\documentclass[11pt]{article}
\pdfoutput=1
\usepackage{jheppub}
\usepackage{amsmath,amssymb,amsfonts,graphicx,slashed,xcolor,subcaption}

\hypersetup{
    pdfstartview={FitH},    
    pdftitle={title},    
    pdfauthor={authors},     
}

\newcommand{\be}{\begin{equation}}
\newcommand{\ee}{\end{equation}}
\newcommand{\bea}{\begin{eqnarray}}
\newcommand{\eea}{\end{eqnarray}}
\newcommand{\beas}{\begin{eqnarray*}}
\newcommand{\eeas}{\end{eqnarray*}}
\newcommand{\ba}{\begin{array}}
\newcommand{\ea}{\end{array}}
\newcommand{\tr}{{\rm tr}}

\newcommand{\cDsl}{{{\cal D}\kern-.65em /\,}}
\newcommand{\cDslsm}{{{\cal D}\kern-.5em /\,}}
\newcommand{\nabslsm}{\nabla\kern-.55em /}
\newcommand{\pasl}{\pa\kern-.55em /}
\newcommand{\psl}{p\kern-.45em /}
\newcommand{\Dsl}{D\kern-.65em /}
\newcommand{\Asl}{A\kern-.55em /}
\newcommand{\nabsl}{\nabla\kern-.65em /\kern+.2em}
\newcommand{\qsl}{q\kern-.5em /}
\newcommand{\ksl}{k\kern-.5em /}
\newcommand{\rsl}{r\kern-.5em /}

\newcommand{\cDslLCsq}{{\stackrel{\circ}{\cDsl^{\kern2pt 2}}}}

\newcommand\cc[1]{#1^{^{\kern-6pt \circ}}\kern2pt}

\def\tr{{\rm tr}}

\newcommand{\pa}{\partial}

\newcommand{\beq}{\begin{equation}}
\newcommand{\eeq}{\end{equation}}
\newcommand{\beqn}{\begin{eqnarray}}
\newcommand{\eeqn}{\end{eqnarray}}

\def\dalemb#1#2{{\vbox{\hrule height .#2pt
\hbox{\vrule width.#2pt height#1pt \kern#1pt
\vrule width.#2pt}
\hrule height.#2pt}}}

\newcommand{\bs}{\boldsymbol}

\newcommand{\source[1]}{{\lambda}^{{\scriptscriptstyle(#1)}}}
\newcommand{\sourcet[1]}{{\tilde\lambda}^{{\scriptscriptstyle(#1)}}}
\newcommand{\sourceh[1]}{{\hat\lambda}^{{\scriptscriptstyle(#1)}}}

\title{Nonlocal multi-trace sources and bulk entanglement \\ in holographic conformal field theories}

\author[a]{Felix M. Haehl}
\author[a]{\!, Eric Mintun}
\author[a]{\!, Jason Pollack}
\author[b]{\!, Antony J. Speranza}
\author[a]{\!, \qquad \qquad \qquad Mark Van Raamsdonk}

\affiliation[\,a]{Department of Physics and Astronomy, University of British Columbia,\\
6224 Agricultural Road, Vancouver, B.C.\ V6T 1Z1, Canada.}
\affiliation[\,b]{Perimeter Institute for Theoretical Physics, Waterloo, ON N2L 2Y5, Canada}

\emailAdd{f.m.haehl@gmail.com}
\emailAdd{eric.mintun@gmail.com}
\emailAdd{jasonpollack@gmail.com}
\emailAdd{asperanza@gmail.com}
\emailAdd{mav@phas.ubc.ca}

\abstract{
We consider CFT states defined by adding nonlocal multi-trace sources to the Euclidean path integral defining the vacuum state. For holographic theories, we argue that these states correspond to states in the gravitational theory with a good semiclassical description but with a more general structure of bulk entanglement than states defined from single-trace sources. We show that at leading order in large $N$, the entanglement entropies for any such state are precisely the same as those of another state defined by appropriate single-trace effective sources; thus, if the leading order entanglement entropies are geometrical for the single-trace states of a CFT, they are geometrical for all the multi-trace states as well. Next, we consider the perturbative calculation of $1/N$ corrections to the CFT entanglement entropies, demonstrating that these show qualitatively different features, including non-analyticity in the sources and/or divergences in the naive perturbative expansion. These features are consistent with the expectation that the $1/N$ corrections include contributions from bulk entanglement on the gravity side. Finally, we investigate the dynamical constraints on the bulk geometry and the quantum state of the bulk fields which must be satisfied so that the entropies can be reproduced via the quantum-corrected Ryu-Takayanagi formula.
}


\begin{document}

\maketitle

\parskip=10pt

\section{Introduction}

In recent years, it has become clear that the emergence of gravitational physics from certain non-gravitational quantum systems with large $N$,\footnote{ Here, $N$ is a parameter related to the number of degrees of freedom in the theory; for example, the rank of a gauge group, or some power of the central charge in a CFT.} as suggested by the AdS/CFT correspondence \cite{Maldacena:1997re,Witten:1998qj,Aharony:1999ti}, can be made particularly transparent by considering the structure of entanglement in the non-gravitational system (see, e.g., \cite{Maldacena:2001kr, Ryu:2006bv, Swingle:2009bg, VanRaamsdonk:2009ar, VanRaamsdonk:2010pw}). For states with a dual gravitational description, entanglement entropies for spatial subsystems (considered at leading order in $N$) are related to the areas of extremal surfaces in the corresponding spacetime \cite{Ryu:2006bv,Hubeny:2007xt}.

Vacuum entanglement entropies of ball-shaped regions in any conformal field theory are equivalent to the areas of extremal surfaces in an auxiliary AdS spacetime. For arbitrary first-order perturbations to the vacuum state, the ball entanglement entropies can always be captured by extremal surface areas in some first-order perturbation of this AdS spacetime. Furthermore, the Entanglement First Law \cite{Bhattacharya:2012mi} implies that the perturbed spacetime geometry must satisfy Einstein's equations linearized about AdS \cite{Lashkari:2013koa,Faulkner:2013ica}.\footnote{ In \cite{Lewkowycz:2018sgn} this argument was generalized to linearized equations about more general background states.}

In \cite{Faulkner:2017tkh}, these results were extended to second-order perturbations: for a class of CFT states produced by adding local sources to the Euclidean path integral that describes the vacuum state, the ball entanglement entropies up to second order in the sources defining the state can be captured geometrically by a second-order perturbation to AdS. Again, the structure of CFT entanglement (in particular, its relation to the one-point functions of local operators) implies that these perturbations must satisfy local gravitational equations, which now include nonlinear contributions.\footnote{  The second-order calculation is sensitive to two central charge parameters in the CFT, characterizing respectively the vacuum entanglement of balls and the stress tensor two-point function. If these $a$- and $c$-type central charges are equal, the gravitational equations are those of pure Einstein gravity, otherwise the auxiliary gravitational theory involves higher derivatives \cite{Haehl:2017sot,Haehl:2015rza}.}

The class of states considered in \cite{Faulkner:2017tkh} is still significantly smaller than the set of states in the gravity picture that would be expected to have a good semiclassical description. As described in \cite{Botta-Cantcheff:2015sav,Faulkner:2017tkh,Marolf:2017kvq}, defining states via single-trace sources in the Euclidean path integral corresponds to considering coherent states of the bulk fields. As we recall in \S\ref{sec:pathintegrals}, coherent states have a very constrained entanglement structure: for free field theories, the entanglement entropy for any spatial subsystem is the same as for the vacuum state, since the coherent state can be obtained by the action of local unitary operations on the vacuum.

In this paper, our goal is to study the entanglement structure of holographic CFT states which correspond to general states of the bulk gravitational theory with a good semiclassical description. We argue that to construct states with a more general structure of bulk entanglement, we can use the Euclidean path integral with sources for nonlocal double-trace and general multi-trace operators of the form
\be
\label{nonlocal}
\int dx_1 \cdots dx_n\, \source[n]_{\alpha_1 \cdots \alpha_n} (x_1, \dots, x_n) \,{\cal O}_{\alpha_1}(x_1) \cdots {\cal O}_{\alpha_n}(x_n) \; .
\ee
Here, ${\cal O}_\alpha$ are low-dimension operators in the CFT associated with the light bulk fields. The sourced operators could include components of the stress-energy tensor or some other operators with spin. We describe a path integral construction for both pure states and for general mixed states of the bulk fields.

As an example, using only single and double-trace sources leads to bulk states which are general Gaussian states (e.g. squeezed states) in the free field limit.\footnote{  General Gaussian states are created from the vacuum by acting with the exponential of a quadratic combination of creation and annihilation operators. They may also be characterized as states whose Wigner distribution in phase space is positive-definite.} These states are completely characterized by their one- and two-point functions. Including nonlocal sources with three or more operators introduces non-Gaussianities.

On the gravity side, the expectation is that the entanglement structure for these general states should be reproduced using the quantum RT ($qRT$) formula \cite{Faulkner:2013ana, Engelhardt:2014gca},
\be
\label{QRT}
\Delta S_A^{CFT} = {1 \over 4 G} \Delta {\rm  Area}(\tilde{A}) + \Delta S^{bulk}_{\Sigma} \; .
\ee
Here, the surface $\tilde{A}$ is a bulk surface homologous to $A$ that extremizes the full expression on the right hand side of (\ref{QRT}), and the second term is the vacuum-subtracted bulk entanglement entropy of the bulk fields in the region $\Sigma$ bounded by this surface. For bulk states that are not coherent, the second term should play a significant role in reproducing the CFT entanglement. In this paper, we would like to understand better how this works in detail.

In particular, an interesting physical question is whether matter gravitates in the same way independent of its entanglement structure. For example, does a coherent state of fields with a certain stress-energy tensor have the same gravitational effects as matter with the same stress energy tensor but in some squeezed state, or in a state where it is highly entangled with some distant system? Our results below set up a framework for investigating these questions in the context of AdS/CFT.

\subsection{Summary of results}

We now provide an overview of our findings.

\subsubsection*{Leading order entanglement is geometric}

Our first main result is that for any state created by general nonlocal sources of the form (\ref{nonlocal}), the CFT entanglement entropies for arbitrary spatial regions at leading order in the $1/N$ expansion are the same as for another state with only a single-trace source $\lambda_{eff}(x)$, determined in terms of the sources $\source[n]$ via a self-consistency equation (see \eqref{STsource} and \eqref{eq:lnZmult} for explicit expressions). Thus, any results demonstrating that leading order CFT entanglement entropies can be represented geometrically for states defined via local sources immediately extend to the much more general class of states defined by general sources (\ref{nonlocal}).

An interesting aspect of this result is that the effective single-trace source depends non-trivially on all the multi-trace sources in general, but vanishes if the single-trace sources are set to zero. This means that double- and higher-trace sources on their own change the quantum state and entanglement structure of the bulk fields but cannot produce any backreaction. But when single-trace sources are turned on in addition, the final backreacted geometry depends on which multi-trace sources are already present.

Our result presents a generalization of the prescription \cite{Witten:2001ua,Berkooz:2002ug} that concerns the holographic description of local multi-trace sources. In particular, one way of writing our composite effective single-trace source is using expression \eqref{STsource}, which is formally equivalent to the prescription of \cite{Witten:2001ua}, where the effects of local multi-trace sources are captured by a modified boundary condition in the AdS/CFT dictionary.

\subsubsection*{$1/N$ corrections to entanglement: non-analyticities as a signal of bulk quantum effects}

We next proceed to consider entanglement entropy at subleading orders in the $1/N$ expansion.
In AdS/CFT, it is expected that these subleading contributions to CFT entanglement are captured by bulk quantum effects summarized by the quantum RT ($qRT$) formula \eqref{QRT}.

One of our main observations is that contributions to the CFT entanglement entropy at these subleading orders exhibit qualitatively different features from the ${\cal O}(N^2)$ entropies. We find that in the naive perturbative expansion in sources $\source[n]$, divergences can appear at specific orders in the sources if the support of the sources $\source[n]$ extends beyond a certain strip of analyticity in Euclidean time, see for example \eqref{eq:divSk}.\footnote{ Similar observations regarding the breakdown of naive perturbation theory have been made recently in  \cite{Sarosi:2017rsq,Lashkari:2018tjh}.} We argue that these divergences are an artifact of the full expression for the entropies having non-analytic contributions. The presence of such terms in the entropies is consistent with the expectation that the dual gravitational description involves non-geometric contributions (such as bulk entanglement).

\subsubsection*{Reproducing subleading entanglement entropies via gravity}

As a generalization of the results in \cite{Faulkner:2017tkh}, we would like to show directly that for states in a holographic CFT generated by the general sources (\ref{nonlocal}), CFT entanglement entropies can be captured by the quantum RT formula for an appropriately chosen bulk quantum state. Assuming this works, we would like to understand what constraints the bulk state must satisfy, i.e., whether it is possible to show that some quantum version of the gravitational constraints must be satisfied. This was done for general linear perturbations in \cite{Swingle:2014uza}, and very explicitly for a different class of (one-particle) states in \cite{Belin:2018juv}; but as we will see, there are qualitatively new features in the case of non-linear multi-trace perturbations.\footnote{ See also \cite{Goel:2018ubv,Cottrell:2018ash,Miyaji:2018atq,Ecker:2019ocp} for similar recent constructions.}

\subsubsection*{Emergence of bulk locality and quantum Einstein equation}

A particular motivation for our investigations is to understand in more detail how bulk locality emerges from CFT physics and whether there are any nonlocal quantum gravitational effects that correct this. A fascinating aspect of the classical calculations showing that geometries which represent CFT entanglement entropy must satisfy Einstein's equations is that the locality of these equations is not assumed from the start, but rather comes out as part of the derivation. The starting point, an expression for the CFT entanglement entropy in terms of CFT one-point functions, leads to a very nonlocal set of constraints in the gravity theory, relating the areas of extremal surfaces to the asymptotic metric in the region bounded by these surfaces. It is only after an application of Stokes' theorem that this set of constraints reduces to a local equation.

When bulk quantum corrections are considered, the classical RT formula is replaced by the quantum RT formula (\ref{QRT}), introducing the bulk entanglement term. For general first-order perturbations, this term can be written as the integral of a local expression; in the derivation of constraints on bulk geometry starting from the qRT formula, this entanglement term gives rise to the expectation value of the stress tensor as a source for the linearized Einstein's equations \cite{Swingle:2014uza}. This allows us to derive a quantum version of Einstein's equation, see \eqref{eq:qEinstein}. For CFT states corresponding to bulk coherent states, we have argued that $\Delta S^{bulk}_{\Sigma} = 0$, so the bulk entanglement term doesn't contribute.\footnote{ Indeed, it was noted in \cite{Faulkner:2017tkh} that for states defined by single-trace sources, the CFT entanglement entropy to second order in the sources could be reproduced via the classical RT formula even at finite $N$.}

Outside of these special cases, there is no reason for the bulk entanglement to cancel or be represented by a local integral. Thus, it is very interesting to understand if and how the assumption that CFT entanglement entropy is correctly captured by a gravity calculation can still lead to a local bulk constraint. We will see below that the equivalence between bulk relative entropy (which can be understood as the nonlocal part of the bulk entanglement entropy) and CFT relative entropy \cite{Jafferis:2015del} plays an important role in this emergence of locality. We also give a direct argument that the CFT relative entropy for our multi-trace states is equal to the relative entropy of a corresponding bulk state. Our argument provides a more explicit alternative to the one in \cite{Jafferis:2015del} and appears to hold beyond the ${\cal O}(N^0)$ considered in \cite{Jafferis:2015del}.

\subsection{Outline}

The outline of this paper is as follows.

In \S\ref{sec:pathintegrals}, we describe the class of states that we will consider. We explain how incorporating multi-trace sources into the Euclidean path integral defining the states allows us to control the bulk entanglement structure, for example allowing us to study states in which matter in one region is entangled with matter in a distant region or even with matter in some disconnected spacetime.

In \S\ref{sec:EE}, we consider the calculation of entanglement entropy in perturbation theory in the sources using the replica method. We discuss the $N$-scaling of various terms in the perturbative calculation, explaining a ``diagrammatic'' method to understand which power of $N$ governs a particular contribution. In this representation, the order $N^2$ contributions to entanglement entropy come exclusively from tree diagrams; diagrams with loops only contribute at lower orders in $1/N$.

In \S\ref{sec:Nsquared}, we show that for any state defined by nonlocal sources, all the order $N^2$ contributions to the CFT entanglement entropy for spatial regions are the same as for another state defined by an effective single-trace source.
Our result implies that in any CFT for which the order $N^2$ entanglement entropies of states defined by single-trace sources are represented geometrically, the order $N^2$ entanglement entropies for all multi-trace states are also represented geometrically, via the classical RT formula applied to the geometry associated with the effective single-trace source. To study the situation where bulk entanglement is important, we need to go to order $N^0$.

In \S\ref{sec:Subleading}, we consider contributions to the entanglement entropy at subleading orders in $1/N$. Here, we have a new qualitative effect: in the replica method calculation of entanglement entropy, we can have loop diagrams which are sensitive to the topology of the replica manifold. We argue that these diagrams can give rise to contributions to the perturbative expansion that are non-analytic in the sources, and that such contributions indicate that the bulk interpretation of the entanglement is no longer purely geometrical. Making use of a toy model, we show that such non-analyticities can give rise to divergences in the naive perturbative expansion.

In section \S\ref{sec:quantum2}, we consider explicit calculations of entanglement entropy and relative entropy for ball-shaped regions, reviewing a method to calculate these entropies at fixed perturbative orders in the sources which avoids the replica trick. We show directly that some of these perturbative contributions are divergent, and interpret the divergences as symptoms of the non-analyticities discussed in section \S\ref{sec:Subleading}. This discussion is complemented by Appendix \ref{sec:replica}, where we find the same divergences using the replica trick, and Appendix \ref{app:kthorder}, which discusses the divergences at higher orders in perturbation theory.

In \S\ref{sec:quantum1}, we take steps toward understanding directly when the CFT entanglement entropies including $1/N$ corrections can be reproduced gravitationally via the quantum RT formula. We show that this equality is equivalent to a collection of more basic statements, including a quantum version of the bulk gravitational equations, the equivalence of bulk and boundary relative entropies (as suggested in \cite{Jafferis:2015del}), and a quantum version of the Hollands-Wald geometrical identity that played a central role in \cite{Faulkner:2013ica, Faulkner:2017tkh}.\footnote{ See also the recent paper \cite{Lewkowycz:2018sgn}, which takes a different but related approach to relating the quantum RT formula and the gravitational equations.} Checking the validity of the quantum RT formula reduces to demonstrating that these other statements hold.

Alternatively, assuming the validity of the quantum RT formula leads to a particular form of the bulk gravitational constraints. Our framework indicates that the equivalence of bulk and boundary relative entropies is crucial in order to remove potentially nonlocal terms in these quantum gravitational equations. We give a direct general argument that for the class of states described in \S\ref{sec:pathintegrals}, this equivalence of relative entropies holds. We discuss what remains to be shown for these higher order perturbations in order to demonstrate the locality of the quantum gravitational constraints and/or to identify potential nonlocal contributions.

We end in \S\ref{sec:discussion} with a discussion, including various possible future directions for investigation.

\section{Path integral states}
\label{sec:pathintegrals}

In this section we discuss a class of CFT states prepared via a Euclidean path integral construction that we argue produces general (perturbative) bulk states with a good semiclassical description.

\subsection{General Setup}

A class of holographic CFT states that is expected to have a nice semiclassical description are states $| \Psi_\lambda \rangle$ defined via the Euclidean path integral as
\be
\label{PIstate}
\langle \phi_0 | \Psi_\lambda \rangle = \int^{\phi(\tau=0) = \phi_0}_{\tau<0} [d \phi(\tau)] e^{-S_E - \int_{-\infty}^0 d \tau \int dx \lambda_\alpha(x, \tau) {\cal O}_\alpha(x,\tau)} \; ,
\ee
where ${\cal O}_\alpha(x,\tau)$ correspond to low-dimension operators dual to light fields in the bulk. We have defined $| \Psi_\lambda \rangle$ by writing its wave functional, i.e.~the overlap with states $| \phi_0 \rangle$ satisfying $\hat{\phi} | \phi_0 \rangle = \phi_0 | \phi_0 \rangle$. Specifying the state requires specifying an entire family of couplings $\{\lambda_\alpha\}$, which we have abbreviated as $\lambda$ in writing $\Psi_\lambda$. States of the form \eqref{PIstate} can be understood as giving rise to coherent states of the perturbative bulk fields \cite{Botta-Cantcheff:2015sav,Marolf:2017kvq}, and there is a standard holographic prescription that can be used to compute the corresponding Lorentzian geometries produced by such states \cite{Skenderis:2008dg}. Note that in order to define excited states of the original theory, the sources $\lambda_\alpha(x, \tau)$ are taken to vanish as $\tau \to 0$. Otherwise, such a path integral defines some state of a different perturbed theory.

At linear order in the sources $\lambda_\alpha(x, \tau)$, the bulk state corresponding to (\ref{PIstate}) can be written as $|0 \rangle + c_{\alpha,n} a^\dagger_{\alpha,n} |0 \rangle + \dots$, where $n$ labels all the mode operators associated with a bulk field labeled by $\alpha$. This follows since the insertion of a light single-trace operator into the path integral gives rise to a state with a single particle on top of the AdS vacuum. The coefficients $c_{\alpha,n}$ can be determined explicitly as an integral transform of the sources $\lambda_\alpha(x, \tau)$ \cite{Marolf:2017kvq}. This transformation is formally invertible, so we have a one-to-one correspondence between sources and bulk perturbations at the linearized level.

Looking at the states (\ref{PIstate}) at second order in perturbation theory, we expect the bulk state to be corrected by terms $c_{\alpha \alpha', n \; n'} a^\dagger_{\alpha,n} a^\dagger_{\alpha',n'}|0 \rangle$, but the coefficients of these are again fixed by the sources $\lambda_\alpha(x, \tau)$ (or alternatively by the first order coefficients $c_{\alpha,n}$). This implies that the form of the bulk state at second order in perturbation theory is very constrained in CFT states of the form (\ref{PIstate}). These constraints amount to saying that in the free-field limit on the bulk side, the bulk state should be coherent: the exponentiated operator in the path integral leads to an exponentiated creation operator $e^{c_{\alpha,n} a^\dagger_{\alpha,n}} |0 \rangle$ in the bulk description.

\subsubsection*{Multi-trace states}

From the bulk point of view, states with more general $c_{\alpha \alpha', n \; n'}$ are still completely reasonable low-energy states. For example, they may correspond to squeezed states of the bulk modes, or states with entanglement between distant particles. In order to represent these in the CFT, we can consider more general states\footnote{  When $x$ and $x'$ are coincident, ${\cal O}_\alpha(x,\tau) {\cal O}_\alpha(x', \tau')$ is shorthand for a low-dimension double-trace operator whose insertion in the Euclidean path integral creates a two-particle state.}
{\small
\be
\label{PIstate2}
\boxed{
\begin{split}
&\qquad\qquad\qquad\qquad\qquad\quad\langle \phi_0 | \Psi_\lambda \rangle = \int^{\phi(\tau=0) = \phi_0}_{\tau<0} [d \phi(\tau)] \; \exp\Big[{-S_E - S_{\{\source[i]\}}} \Big] \; ,\\
&
S_{\{\source[i]\}} = \sum_n \int_{-\infty}^0 d \tau_1 \cdots  d \tau_n \int dx_1 \cdots dx_n\,  \source[n]_{\alpha_1 \cdots \alpha_n} (x_1,\tau_1, \dots, x_n,\tau_n) \,{\cal O}_{\alpha_1}(x_1,\tau_1) \cdots {\cal O}_{\alpha_n}(x_n,\tau_n) \,.
\end{split}
}
\ee
}\normalsize
Again, we take the sources to vanish as $\tau_i \to 0$. We will later often drop the operator labels $\alpha_i$. By analogy with \eqref{PIstate} above, the subscript in $| \Psi_\lambda \rangle$ should be understood to denote the collection of sources $\{\source[i]\} = \{\source[1]_{\alpha_1},\source[2]_{\alpha_1 \alpha_2},\,\ldots,\source[n]_{\alpha_1\ldots\alpha_n}\}$.

\begin{figure}
\begin{center}
\includegraphics[width=.6\textwidth]{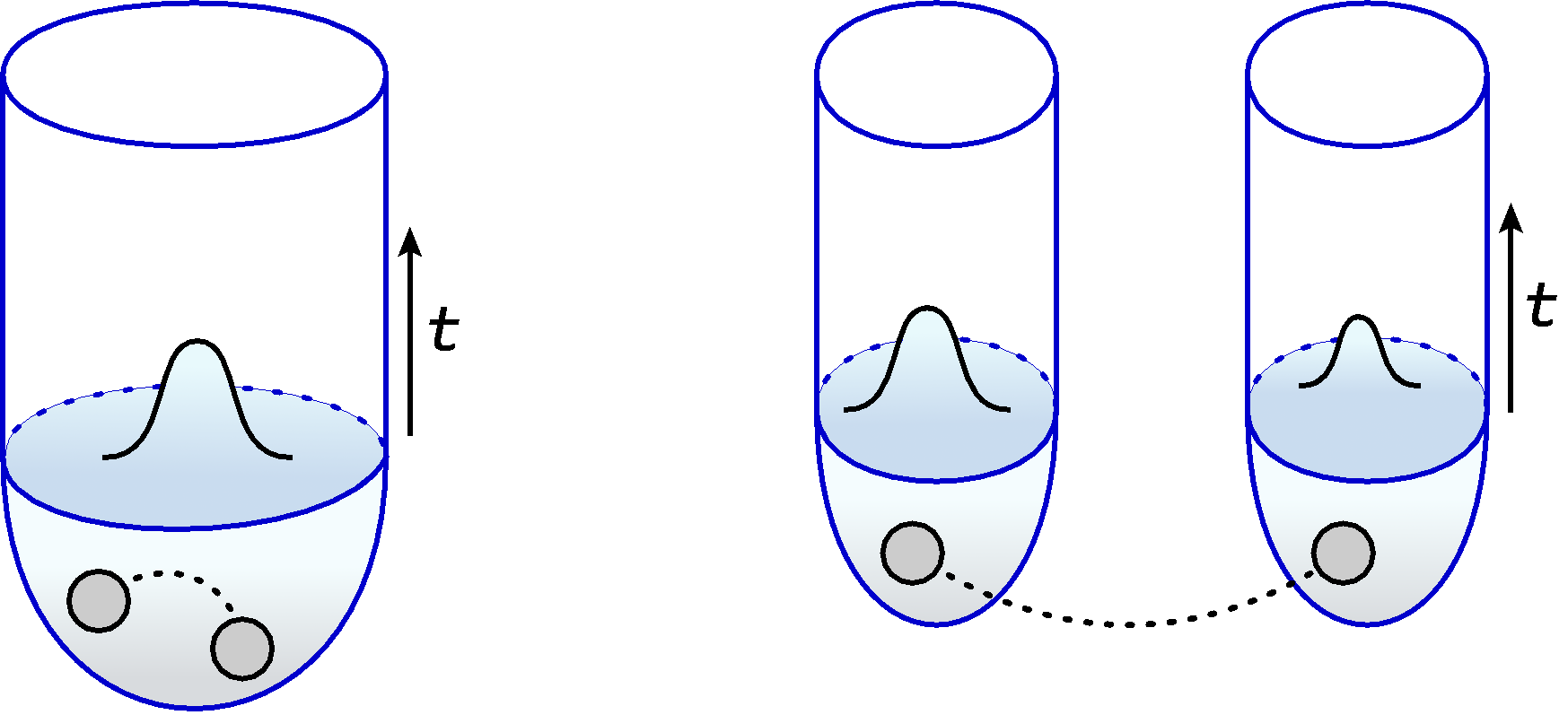}
\end{center}
\caption{Schematic illustration of the setup. The Euclidean path integral (for example, over a half-sphere) prepares the ground state for subsequent Lorentzian evolution in real time $t$. In the presence of double-trace sources (illustrated with a dotted line connecting operator insertions) we create Gaussian excitations on top of the vacuum. The case where double trace sources create correlations between disconnected systems is similar to the path integral preparation of the thermofield double state.}
\label{fig:setup}
\end{figure}

For example, in the limit where the bulk theory is free, states with single and nonlocal double-trace sources give rise to general Gaussian states of the bulk effective theory,
\be
e^{c_{\alpha,n} a^\dagger_{\alpha,n} + c_{\alpha \alpha', n \; n'} a^\dagger_{\alpha,n} a^\dagger_{\alpha',n'}} |0 \rangle \; .
\ee
These states are fully determined by their two-point functions. See the left panel of Fig.\ \ref{fig:setup} for an illustration.

In general, the multi-trace sources in (\ref{PIstate2}) can correspond to local perturbations to the CFT Lagrangian by irrelevant operators. However, if we choose the sources to vanish in the limit when some of the operators become coincident, we expect the perturbative behavior of observables for the multi-trace states to be no worse than for single-trace states with the same operators.

We shall argue that the multi-trace path integral states provide a useful generalization of coherent states: they are very general, but at the same time physically interesting and intuitive as they simply correspond to more general excitations of low-energy modes in semiclassical holography. This distinguishes them from other possible generalizations of \eqref{PIstate} such as quantum superpositions of coherent states, which would not correspond to low-energy excitations around a simple gravitational background geometry.\footnote{ We thank O.\ Parrikar for raising this point.}

\subsection{Path integral states coupling disconnected spacetimes}

The multi-trace states above allow a more general structure of bulk entanglement than for states with local sources, for which the bulk entanglement is expected to be similar to the bulk vacuum state. To illustrate this clearly, we can consider the case where the theory is defined on a disconnected spacetime, for example two copies of some base spacetime $M$, with a CFT living on each one. In this case, we can add sources that couple the two separate CFTs
\be
\langle \phi_1 \phi_2 | \Psi_\lambda \rangle = {1 \over Z_\lambda} \int [d \phi_1] [d \phi_2] \;e^{-S_1 - S_2 - \int dx_1 dx_2\, \source[2](x_1,x_2) {\cal O}(x_1) {\cal O}(x_2) },
\ee
so that the individual CFTs are each now in a mixed state. The bulk interpretation is that the matter in one spacetime is entangled with matter in the other spacetime. This is similar to the path integral construction of the thermofield double state, where instead of sources coupling operators on the two parts of the path integral, the Euclidean path integral is defined on a space that is geometrically connected. This is illustrated in the second part of Fig.\ \ref{fig:setup}.

Similar sources coupling two non-interacting theories have appeared recently in \cite{Gao:2016bin,Maldacena:2017axo}. There, the sources were Lorentzian sources coupling two CFTs in a thermofield double state, and the effect of the sources was to make the wormholes traversible. In the present context, the sources produce an entangled state of decoupled CFTs, so the entanglement entropy between the CFTs is time-independent in the Lorenzian picture.\footnote{ See also \cite{Aharony:2006hz} for a similar construction.}

\subsection{Path integrals for a general density matrix of the bulk effective theory}

Given any collection of CFTs in a state prepared using arbitrary multi-trace sources which in general couple operators in the various CFTs, we can obtain a single path integral expression for the density matrix of one of the CFTs by ``integrating out'' the remaining ones. To see this, consider again an example with two CFTs, where
\be
\langle \phi_1 \phi_2 | \Psi_\lambda \rangle = {1 \over Z_\lambda} \int [d \phi_1] [d \phi_2]\; e^{-S_1 - S_2 - \int \lambda_1^{(1)}{\cal O}_1-\int \lambda_2^{(1)}{\cal O}_2 - \int \lambda_{12}^{(2)} {\cal O}_1{\cal O}_2+\ldots },
\ee
The case with multiple CFTs is completely analogous.

If we evaluate the density matrix for the first CFT, we obtain a path integral where fields in the first CFT are integrated over the full Euclidean space with a cut at $\tau=0$ and while fields in the second CFT are integrated over the full space. Working perturbatively in the sources, each term (including those coming from the normalization) has some CFT${}_1$ operators inserted in the CFT${}_1$ path integral, and some CFT${}_2$ operators inserted in the CFT${}_2$ path integral. The latter path integral can be performed to give a Euclidean $n$-point function in the second CFT, which is just some particular function of the coordinates where the operators were inserted. We are left with a term in the first CFT path integral with various operators inserted, and with a coefficient involving the various $\lambda$s and the $n$-point function coming from the second CFT. All together, we have a series of such terms. Formally, we can write the resulting series as the exponential of some other series. Thus, the final expression for the density matrix in the first CFT takes the form
\be
\label{effsource}
\langle \phi_- | \rho_1 | \phi_+ \rangle = {1 \over Z_\lambda} \int_{\phi(\tau\rightarrow0_- ) = \phi_-}^{\phi(\tau\rightarrow0_+ ) = \phi_+} [d \phi] \;e^{- S_1 - \int d x_1 \source[1]_{eff}(x_1) {\cal O}(x_1) - \int d x_1 d x_2 \source[2]_{eff}(x_1,x_2) {\cal O}(x_1){\cal O}(x_2)  + \dots}
\ee
where in general we will have an infinite series of terms.

At first sight, this expression looks the same as the one we would get by using sources to create a pure state in a single CFT and then calculating the density matrix. However, in the single CFT case, the sources in (\ref{effsource}) could only be nonzero when all the coordinates are in the same half space ($ \tau < 0$ or $\tau > 0$). In the more general case where we start with multiple CFTs, we will have sources coupling the $\tau < 0$ region to the $\tau > 0$ region, and the resulting density matrix will correspond to a mixed state of the CFT.

From now on, we will avoid talking about situations with multiple CFTs, and simply consider general density matrices for a single CFT of the form (\ref{effsource}), allowing general sources consistent with Hermiticity. It is plausible that such an expression corresponds to the most general (perturbative) density matrix in the bulk effective field theory. To see this, note that such a density matrix takes the form
\be
\label{genrho}
\rho = \sum_{n_L,n_R} c_{\alpha_1 \cdots \alpha_{n_L} \beta_1 \cdots \beta_{n_R}} a^\dagger_{\alpha_1} \cdots a^\dagger_{\alpha_{n_L}} |0 \rangle \langle 0 | a_{\beta_1} \cdots a_{\beta_{n_R}} \; .
\ee
In the expression (\ref{effsource}), we can split the sources into separate terms where a specific number of operators appear in the $\tau > 0$ region and a specific number of operators appear in the $\tau < 0 $ region. For each term in (\ref{genrho}), we then have a corresponding source; for example, the term in (\ref{genrho}) with $n_L$ and $n_R$ ladder operators on the left and right can be associated with the source
\be
\int dx_1^- \cdots dx_{n_L}^-  dx_1^+ \cdots dx_{n_R}^+ \source[n_L + n_R]_{eff}(x_1^- \dots x_{n_L}^-,  x_1^+ \cdots x_{n_R}^+) {\cal O}(x_1^-) \cdots {\cal O}(x_{n_L}^-)  {\cal O}(x_1^+) \cdots {\cal O}(x_{n_R}^+)
\ee
We make this association because the leading effect of this source is to alter the coefficient $c_{\alpha_1 \cdots \alpha_{n_L} \beta_1 \cdots \beta_{n_R}}$ in the expansion of the density matrix. This source will also have effects at higher order.

In the following sections, we will consider the calculation of spatial entanglement entropy in general states of the form (\ref{effsource}), working perturbatively in the various sources, to see whether we can always understand these using the quantum RT formula.

\subsection{$N$ scaling}
\label{sec:Nscaling}

Before concluding this section, we comment on the scaling of various quantities with $N$ in a large $N$ theory. We recall that for a large $N$ gauge theory in the  't Hooft limit, the Lagrangian is normalized as
\be
{\cal L} \sim {1 \over g^2} \tr(F^2 + \dots) = {N \over \lambda} \tr(F^2 + \dots) \,.
\ee
Here, the trace scales like $N^1$, so the action is of order $N^2$. More generally, if we normalize single-trace operators as ${\cal O} \sim {1 \over N} \tr(\cdots)$, then the connected $n$-point functions of normalized operators scale like $1/N^{2n-2}$.

We would now like to understand the possible $N$-scaling for our various sources. In order to have a well-defined large $N$ limit, we need actions of the form
\be
\label{largeN}
S \sim N^2 f({\cal O}_i)\,,
\ee
where $f$ has coefficients of order $N^0$. Comparing this to our expression (\ref{PIstate2}), we see that the various  sources $\lambda^{(n)}_{\alpha_1\ldots \alpha_n}$ (either single-trace or multi-trace) must scale like $N^2$ or slower in order for the large $N$ limit to be well defined.  In all future discussions of $N$-scaling, we will assume sources $\lambda^{(n)}_{\alpha_1\ldots \alpha_n}$ have this maximal scaling of $N^2$.

\section{Entanglement entropy}
\label{sec:EE}

In this section, we discuss the calculation of entanglement entropies for path integral states with general multi-trace sources. We will specifically be interested in understanding how the various contributions scale with $N$ in the $1/N$ expansion and identifying which contributions have the leading-order $N^2$ behavior. We begin with a brief review of the general replica method for calculating entanglement entropies and then consider carrying out such a calculation perturbatively in the sources to determine which are the leading contributions in the $1/N$ expansion.\footnote{ In section 6, we will review an alternative method for these perturbative calculations which does not use the replica trick.}

\subsection{Replica method}

We will now briefly review the replica method for computing subsystem entanglement entropies for states defined via Euclidean path integrals with sources. We begin with a mixed state defined as
\be
\label{genmixed}
\langle \phi_- | \rho | \phi_+ \rangle = {1 \over Z_\lambda} \int_{\phi(\tau\rightarrow0_- ) = \phi_-}^{\phi(\tau\rightarrow0_+ ) = \phi_+} [D \phi] \;e^{- S - \int d x_1 \source[1](x_1) {\cal O}(x_1) - \int d x_1 d x_2 \source[2](x_1,x_2) {\cal O}(x_1){\cal O}(x_2)  + \dots}
\ee
where the integral is over Euclidean space minus a cut at $\tau = 0$.

The density matrix for a spatial subsystem $A$, defined as $\rho_A = \text{tr}_{A^c} \rho$, is obtained from the full density matrix (\ref{genmixed}) by making the identification $\phi_- = \phi_+$ on the complementary subsystem $A^c$ and integrating over $\phi_\pm|_{A^c}$. This leaves a path integral over the Euclidean space minus the region $A$ on the $\tau = 0$ slice,
\be
   \langle \phi_-^A | \rho_A | \phi_+^A \rangle =  \frac{1}{Z_\lambda} \int_{\phi(\tau \to 0^+,\,x\in A) = \phi_+^A}^{\phi(\tau \to 0^-,\,x\in A) = \phi_-^A} [D\phi]\; e^{- S - \int d x_1 \source[1](x_1) {\cal O}(x_1) - \int d x_1 d x_2 \source[2](x_1,x_2) {\cal O}(x_1){\cal O}(x_2)  + \dots} \,,
\ee
The entanglement entropy associated with this reduced density matrix can be computed as
\be
  S_A \equiv - \text{tr}_A \left( \rho_A  \, \log \rho_A \right)
  = - \frac{d}{d q} \log \text{tr}_A \left( \rho_A^q \right) \big{|}_{q=1}
  \,.
\ee
where the derivative in the expression on the right requires analytic continuation in the power $q$.\footnote{ Alternatively, the entanglement entropy may be expressed in terms of $\text{tr}_A \left( \rho_A^q \right)$ for integer powers via an integral formula such as equation (\ref{Sint}) below.}
The powers of the reduced density matrix appearing here can be computed as
\be
\label{eq:trrho}
  \text{tr}_A\left( \rho_A^q\right) =  \int [D\phi_1^A] \cdots [D\phi_q^A] \;   \langle \phi_q^A | \rho_{A} | \phi_{q-1}^A \rangle \cdots \langle \phi_2^A | \rho_{A} | \phi_1^A \rangle \, \langle \phi_1^A | \rho_{A} | \phi_q^A \rangle
   = \frac{Z_q}{Z^q},
\ee
where $Z_q$ is a path integral over the ``replica manifold'' $M_q$ obtained by gluing $q$ copies of the CFT spacetime $M_1$ in a cyclic manner along the region $A$ such that the field $\phi$ continuously transitions from one sheet to the next:
\be
 Z_q = \int_{M_q} [D\phi] \; e^{- S - \int d x_1 \source[1](x_1) {\cal O}(x_1) - \int d x_1 d x_2 \source[2](x_1,x_2) {\cal O}(x_1){\cal O}(x_2)  + \dots} \, ,
\ee
where the sources have been replicated on each sheet of $M_q$.

This reduces the computation of $S_A$ to evaluating the Euclidean path integrals $Z \equiv Z_1$ and $Z_q$ over the original spacetime $M_1$ and the replica manifold $M_q$ (in two-dimensional CFTs $M_q$ is a $\mathbb{Z}_q$ symmetric Riemann surface).

\subsection{Order $N^2$ contributions to entanglement}

Consider a Euclidean holographic CFT perturbed by a general nonlocal multi-trace deformation
\be
\label{deformed}
S_E \to S_E + \sum_n \int dx_1 \cdots dx_n \;\source[n](x_1, \dots, x_n) {\cal O}(x_1) \cdots {\cal O}(x_n)  \equiv S_{\{\source[i]\}}\; .
\ee
where the sources $\source[n]$ are each of order $N^2$ and the operators are normalized so that their connected $N$-point functions are order $1/N^{2(n-1)}$. Here, we are suppressing an index that labels the type of operator under consideration.

Consider the calculation of the corresponding partition function $\log Z[\{\source[i]\}]$ perturbatively in the sources. We would like to understand the terms in the perturbative expansion of this expression that contribute at order $N^2$, i.e., at leading order in the $1/N$ expansion. Suppose we have a contribution with $V_n$ factors of the source $\source[n]$. This will have $n_{tot} = \sum_n n V_n$ operators appearing in the path integral. The resulting $n_{tot}$-point function will have a connected contribution, but will also have various disconnected contributions which factorize into lower-point functions. In order to contribute to $\log Z[\{\source[i]\}]$, even these ``disconnected'' contributions must be connected in the sense that the various lower-point correlators are coupled together via the sources. Terms in $Z[\{\source[i]\}]$ which completely factorize into two or more parts (not connected by sources) arise from exponentiating these connected terms in $\log Z[\{\source[i]\}]$.

To understand the $N$ scaling of the various possible contributions, we will introduce a diagrammatic representation of the various contributions. We represent an $n$-trace source $\source[n]$ by $n$ circles (i.e., operator insertions) connected by dashed lines via an $n$-prong vertex (filled red circle, representing the source):
{\small
\be
\begin{split}
&\includegraphics[width=.65\textwidth]{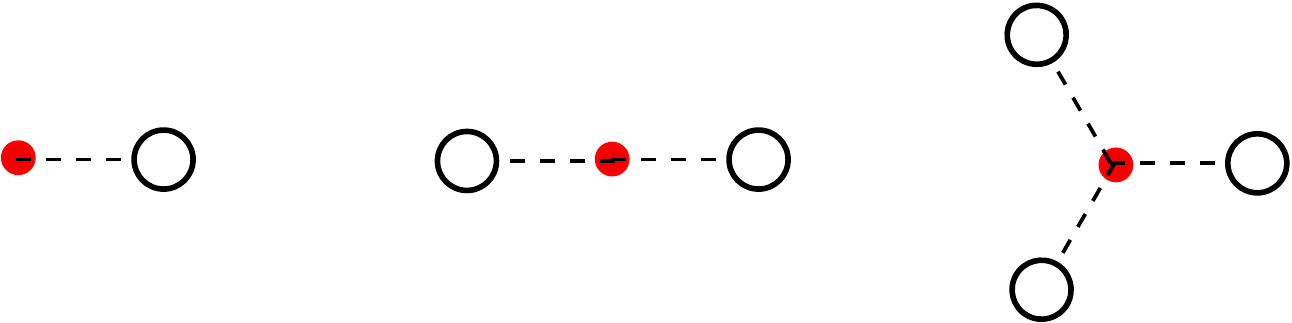}\\
&\;\;\; \int  \source[1] {\cal O}
\qquad\qquad \quad\iint  \source[2] {\cal O}{\cal O}
\qquad\qquad\qquad \iiint  \source[3] {\cal O}{\cal O}{\cal O}
\end{split}
\ee
}\normalsize
Further, a connected $k$-trace correlator is represented as a different type of ``vertex'' (shown in blue below) joining $k$ of the operators via solid segments.

For instance, the following is a relevant tree graph contributing to the 11-point function of order ${\cal O}((\source[1])^6 \source[2] \source[3])$:
\be
\begin{split}
 \includegraphics[width=.4\textwidth]{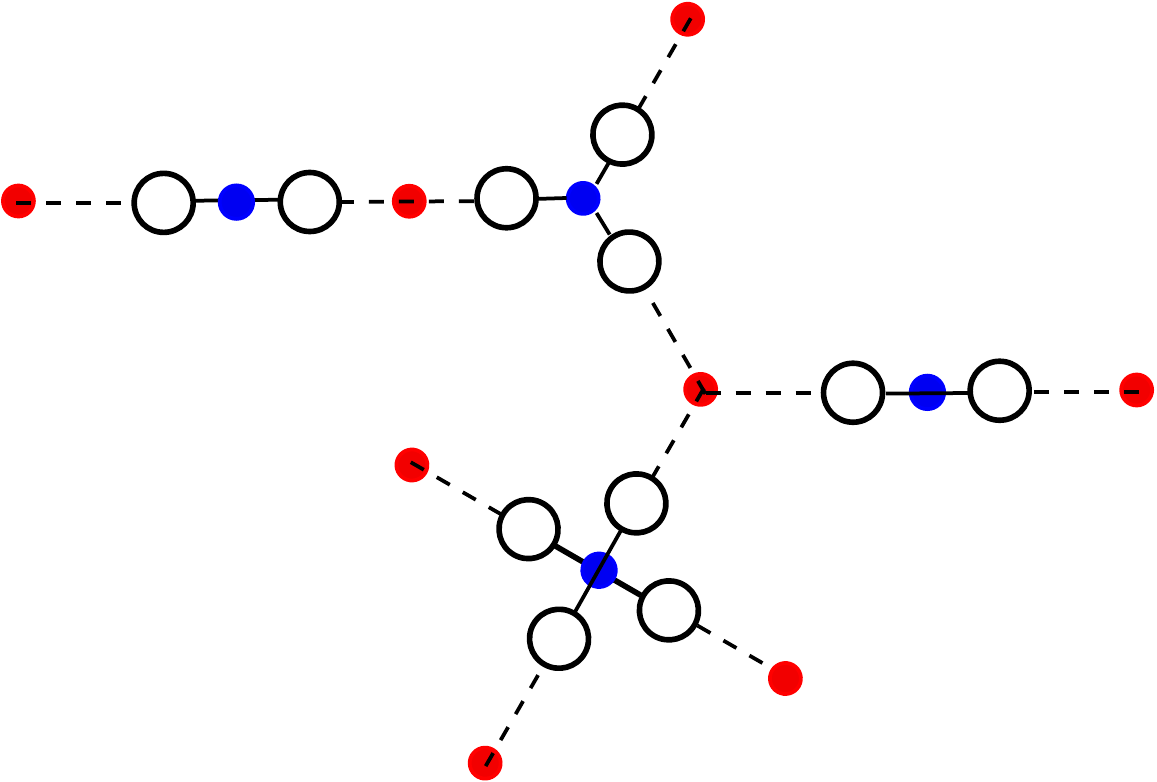}
\end{split}
\ee
This diagram corresponds to a factorization into two 2-point functions, one 3-point function, and one 4-point function. We see that each ``edge'' of the graph (connecting a pair of vertices) includes a single dashed segment, a single operator insertion (not considered to be a vertex), and a single solid segment.

Now, consider a term in the contribution with $V_n$ factors of the source $\source[n]$ in which the various operators are contracted up via $N_k$ connected $k$-point vacuum correlators (potentially for various different $k$). Since the number of edges equals the number of operators, and each operator appears in a single correlator, we have
\be
\# \; {\rm edges} = \# \; {\rm operators} = \sum_k k N_k  \; .
\ee
The number of vertices is
\be
\# \; {\rm vertices} = \sum_n V_n + \sum_k  N_k \; .
\ee
By the Euler relation, we have that the number of loops in the graph is
{\small
\be
L \equiv \# \; {\rm loops} = 1 +( \# \; {\rm edges}) - (\# \; {\rm vertices} )= 1 + \sum_k k N_k - (\sum_n V_n + \sum_k  N_k) = 1 + \sum_k (k-1) N_k - \sum_n V_n
\ee
}\normalsize
Finally, recalling from Sec.~\ref{sec:Nscaling} that the sources are each at most of order $N^2$ and the $k$-point connected correlators are of order $1/N^{2(k-1)}$, we see that the leading $N$-scaling for the contribution associated with such a graph is
\be
{N^{2 \sum_n V_n} \over  N^{\sum_k 2 (k-1) N_k} } = N^{2 \sum_n V_n - \sum_k 2 (k-1) N_k}  = N^{2 - 2 L} \; .
\ee
Thus, the leading power of $N$ from the contribution associated with a specific diagram will be simply $N^{2 - 2L}$, where $L$ is the number of loops in the diagram. In particular, {\it order $N^2$ contributions to $\log Z[\{\source[i]\}]$ correspond to fully connected tree graphs.}

It is useful to note that the terminal nodes of any non-vanishing tree diagram must correspond to insertions of a single-trace source. Thus, while the multi-traces sources affect the partition function at order $N^2$ in the presence of single-trace sources, multi-trace sources on their own do not affect the partition functions (or entanglement entropies) at order $N^2$.

\subsubsection*{$S(A) = S(A^c)$ at order $N^2$ for path integral states}

The partition function calculations we have just discussed feed into the replica method calculation of entanglement entropies for subsystems of the CFT. Thus, perturbative contributions to entanglement entropy at order $N^2$ come from tree diagrams. We would now like to argue that perturbative order $N^2$ contributions to CFT entanglement entropy vanish when considering the entropy of the whole CFT and that the order $N^2$ entanglement entropy for a subsystem $A$, computed perturbatively in the sources, is always the same as for the complementary subsystem $A^c$.

First, consider the case where we are calculating the  entropy of the full CFT. As we have discussed, this can be nonzero when the multi-trace sources couple the $\tau < 0$ and $\tau > 0$ regions of the path integral defining the density matrix, e.g.~in the case where we traced out another CFT. However, we will now see that for these path integral states, treated perturbatively in the sources, the entropy can only be order $N^0$.

First note that from \eqref{eq:trrho} we have
\be
\log(\tr(\rho^q)) = \log(Z_q) - q\log(Z_1)\,,
\ee
where $Z_q$ is the path integral on the usual $q$-sheeted surface obtained by gluing together $q$ copies of Euclidean space across a cut corresponding to the region under consideration. However, in this case, we are considering the  entropy of the full CFT state, so the cut is the entire $\tau = 0$ slice. This means that the multi-sheeted surface used in computing $Z_q$ consists of $q$ disconnected sheets, obtained by gluing the $\tau > 0$ region of the first copy of Euclidean space with the $\tau < 0$ region of the second copy, and so forth. Denoting the space with $n$ disconnected copies of the original Euclidean space as $M_q$, we now have that
\be
\label{RenyiZ}
\log(Z_q) - q\log(Z_1) = Z_{M_q}(\hat{\lambda}) - Z_{M_q}(\tilde{\lambda})\,,
\ee
where the second term is computed with sources $\tilde{\lambda}$ that are simply the original sources repeated on each sheet, while the first term is computed with sources $\hat{\lambda}$ defined from $\tilde{\lambda}$ by incrementing by one the sheet number of every coordinate with a $\tau$ value less than zero. For example, a nonzero source $\source[m+n](x_1^+, \dots x_m^+,x_1^-, \dots, x_n^-)$ coupling $m$ operators with $\tau>0$ to $n$ operators with $\tau < 0$ gives rise to nonzero sources $\sourcet[m+n]((x_1^+)_{(k)}, \dots (x_m^+)_{(k)},(x_1^-)_{(k)}, \dots, (x_n^-)_{(k)})$ and $\sourceh[m+n]((x_1^+)_{(k)}, \dots (x_m^+)_{(k)},(x_1^-)_{(k+1)}, \dots, (x_n^-)_{(k+1)})$.

Now, the sources $\hat{\lambda}$ and $\tilde{\lambda}$ are the same if and only if the original sources do not couple operators with $\tau > 0$ to operators with $\tau < 0$. Otherwise, the expression (\ref{RenyiZ}) is generally nonzero. However, we will now see that at order $N^2$, all contributions to the expression (\ref{RenyiZ}) cancel between the first and second terms.

The key point is that these contributions have a connected tree structure, as described in the previous section. If we consider any contribution to the first term of (\ref{RenyiZ}) with this topology, there are $(q-1)$ equivalent contributions to the first term obtained by acting with the $\mathbb{Z}_q$ symmetry that cyclically permutes the sheets. There are also $q$ equivalent contributions in the second term of (\ref{RenyiZ}), appearing with the opposite sign. These are obtained by replacing all coordinates in the sources and operators with coordinates at the same location but all on the same sheet (there are $n$ choices for which one).

Conversely, if we start with any non-zero contribution to the second term with a connected tree structure, then all sources and operators must be on the same sheet, so there are $(q-1)$ equivalent contributions to the second term obtained by the $\mathbb{Z}_q$ symmetry. But there are also $n$ equivalent contributions to the first term (but with the opposite sign). These are obtained by starting with some single-trace source corresponding to a terminal node of the tree, and, moving away from this on the tree, replacing any multi-trace source $\sourcet[n]$ we encounter with a corresponding non-zero multi-trace source $\sourceh[n]$, such that the coordinate of the operator we have already encountered on our path through the tree left fixed. Translations by $\mathbb{Z}_q$ give the other $q$ equivalent contributions.

As an example, the expression
\be\label{eq:ex11}
\sourcet[1](x^{(1)}_{1+}) \sourcet[2](x^{(1)}_{2+}, x^{(1)}_{3-}) \sourcet[1](x^{(1)}_{4+}) \langle {\cal O}(x^{(1)}_{1+}) {\cal O} (x^{(1)}_{2+}) \rangle \langle {\cal O}(x^{(1)}_{3-}) {\cal O} (x^{(1)}_{4+}) \rangle
\ee
is replaced with the equivalent expression
\be\label{eq:ex12}
\sourceh[1](x^{(1)}_{1+}) \sourceh[2](x^{(1)}_{2+}, x^{(2)}_{3-}) \sourceh[1](x^{(2)}_{4+}) \langle {\cal O}(x^{(1)}_{1+}) {\cal O} (x^{(1)}_{2+}) \rangle \langle {\cal O}(x^{(2)}_{3-}) {\cal O} (x^{(2)}_{4+}) \rangle \;.
\ee
This is depicted in the following graphs, where each rectangle represents a copy of the CFT spacetime and the $\tau=0$ spatial slice runs horizontally through the middle:
\be
\begin{split}
 \includegraphics[width=.9\textwidth]{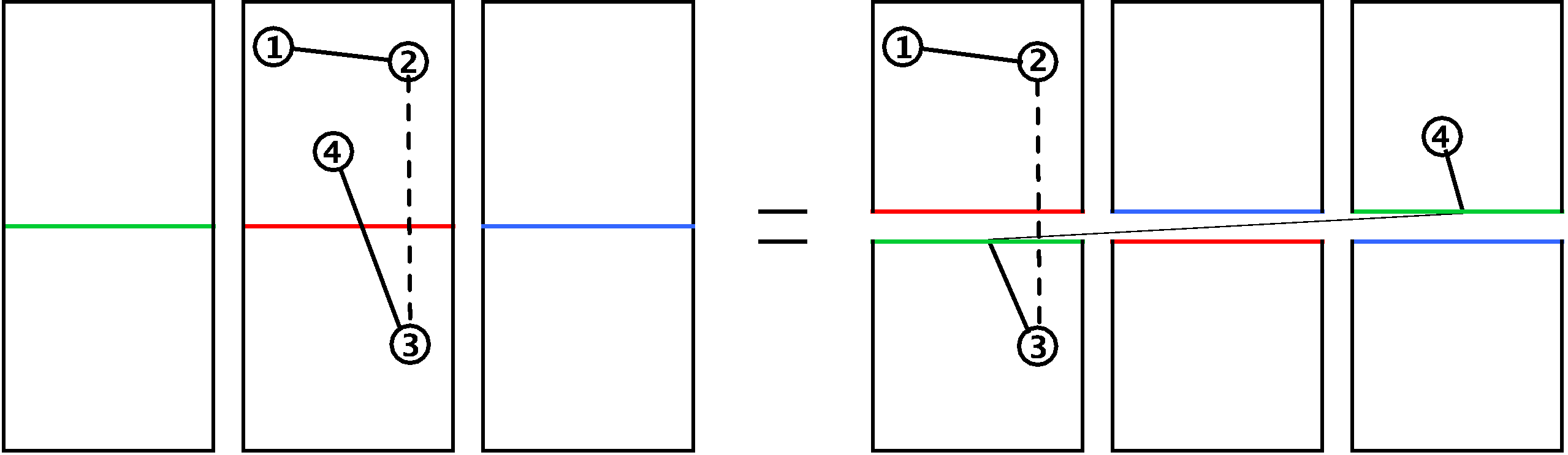}
\end{split}
\nonumber
\ee
The left hand side shows the contribution to $q\log (Z_1)$ as in \eqref{eq:ex11} for $q=3$ identical sheets. The right hand shows the corresponding equal contribution to $\log(Z_q)$ as in \eqref{eq:ex12}. Equally colored $\tau=0$ slices show the cyclic replica identifications (the copy with red edges is labelled as sheet 1, and the one with green edges as sheet 2 in the equations above).

Thus, we have shown a one-to-one correspondence between nonzero order $N^2$ contributions to the first and second terms in (\ref{RenyiZ}), so the ${\cal O}(N^2)$ perturbative contributions to the Renyi entropies and thus the  entanglement entropy must vanish.

This shows in particular that we cannot produce ${\cal O}(N^2)$ entanglement between two CFTs perturbatively using only the multi-trace path integral states we are considering.  Further, it shows that for a subsystem $A$ of a single CFT, we must have $S(A) = S(A^c)$ at ${\cal O}(N^2)$, since by the triangle inequality for entanglement entropies,\footnote{ Note that the UV divergences cancel in the expression on the left side.}
\be
|S(A) - S(A^c)| \le S(A \cup A^c) \sim {\cal O}(N^0)\; .
\ee
The same conclusion may also be reached by a diagrammatic argument similar to the one above, which shows that for every ${\cal O}(N^2)$ perturbative contribution to $S(A)$, we have an equivalent contribution to $S(A^c)$.

\subsubsection*{No bulk entanglement at order $N^2$}

These calculations also suggest that for states of the type we are considering, the CFT entropy at order $N^2$ should be interpreted purely geometrically on the gravity side (e.g. through the classical RT formula). If it were possible to have ${\cal O}(N^2)$ bulk entanglement across some surface accounting for part of an  ${\cal O}(N^2)$ CFT subsystem entanglement, then we should also be able to have ${\cal O}(N^2)$ bulk entanglement between matter in two disconnected AdS spacetimes. But we have seen that this does not occur (perturbatively, for the path integral states we are considering).

\section{Reproducing order $N^2$ entanglement with single-trace states}
\label{sec:Nsquared}

In this section, we will consider the CFT entanglement entropy at order $N^2$ for states defined by general multi-trace sources. We will show that the entanglement entropy at this order is identical to that in another state we construct with only single-trace sources, where the effective single-trace sources are determined from the original sources by a certain self-consistency equation. Thus, all results demonstrating the geometrical character of the ${\cal O}(N^2)$ entanglement for single-trace states automatically carry over to this much more general class of states.

Our effective single-trace sources are very similar to those appearing in \cite{Witten:2001ua,Berkooz:2002ug}, where the AdS/CFT dictionary was discussed in the case of {\it local} multi-trace sources. (Though, in  that context, the sources were used to deform the theory as opposed to the state.)

\subsection{Nonlocal multi-trace deformations}
\label{sec:NonLocalMulti}

Consider a Euclidean holographic CFT perturbed by a general nonlocal multi-trace deformation:
\be
\label{deformed2}
S_E \to S_E + W[{\cal O}] \,, \qquad W[{\cal O}] = \sum_n  \int dx_1 \cdots dx_n\, \source[n](x_1, \dots, x_n) \, {\cal O}(x_1) \cdots {\cal O}(x_n)  \; .
\ee
where we are suppressing an index that labels the type of operator under consideration. Note that, following Sec.~\ref{sec:Nscaling}, the sources $\source[n]$ are each of order $N^2$ and the operators are normalized so that their connected $N$-point functions are order $1/N^{2(n-1)}$. Finally, we will assume that $\source[n](x_1,\ldots,x_n)$ is completely symmetric under permutations of insertion points; this assumption is merely to simplify notation, and can be dropped without any complications.

First, consider the calculation of the one-point function of ${\cal O}$,
\be
\label{eq:1ptO}
\langle {\cal O}(x) \rangle_{\{\source[i]\}} = {\delta \over \delta \source[1](x)} \log Z (\{\source[i]\})\,,
\ee
perturbatively in all the multi-trace sources appearing in \eqref{deformed2}. We claim the following:\\

{\it To leading order in $1/N$, the one-point function in the presence of multi-local multi-trace sources can be computed equivalently in a state with only an effective single-trace source:
\be
\label{STsource}
\boxed{
\langle {\cal O}(x) \rangle_{\{\source[i]\}}  = {\delta \over \delta \lambda_{eff}(x)} \log Z (\lambda_{eff})
 \qquad\text{with}\qquad
 \lambda_{eff}(x) = \frac{\delta W[\langle {\cal O}\rangle_{\lambda}]}{\delta \langle{\cal O}(x)\rangle_\lambda} \bigg{|}_{\lambda = \lambda_{eff}}\,.
 }
\ee
}

{\it Derivation:}
Let us understand the perturbative contributions to the one point function at leading order in large $N$ to $\langle {\cal O}(x) \rangle_{\{\source[i]\}}$. To this end, we consider expanding the exponential in the definition of the multi-trace state, obtaining an infinite series of terms each involving some vacuum correlation function.

In the way that we are normalizing the operators, the leading contribution to the one-point function is order $N^0$. We need to understand which of the infinite number of correlators are of order $N^0$. Suppose we have a contribution with $V_n$ factors of the source $\source[n]$, and suppose that all the operator insertions from expanding the exponent are contracted up via $N_k$ connected $k$-point vacuum correlators. These correlators must be connected to each other by the various nonlocal sources, since the normalization factor $Z^{-1}$ in the derivative of the logarithm \eqref{eq:1ptO} automatically cancels out all the disconnected terms.

We can represent the various contributions diagrammatically, as we did in the previous section. We represent the original operator we are taking the expectation value of by a filled circle. The operators coming from the sources in the exponent are represented by empty circles. A multi-trace insertion involving $k$ operators is denoted by the $k$ operators being connected by dashed lines with $k$ ends.\footnote{ In this case, we suppress the dashed line in the case of a single-trace source.} For example:
{\small
\be
\begin{split}
&\includegraphics[width=.9\textwidth]{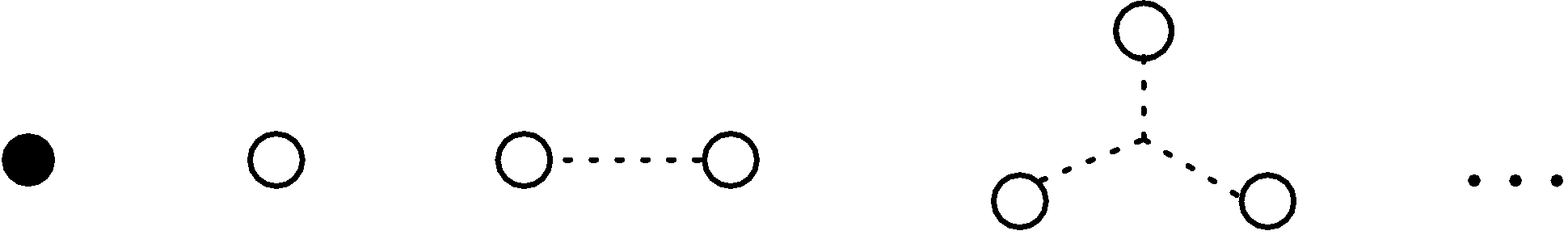}\\
&{\cal O}(x) \qquad\;\;\; \int  \source[1] {\cal O}
\qquad\qquad \iint  \source[2] {\cal O}{\cal O}
\qquad\qquad\qquad\quad \iiint  \source[3] {\cal O}{\cal O}{\cal O}
\end{split}
\ee
}\normalsize
Finally, connected vacuum $n$-point correlators for some set of operators are represented by the corresponding circles being connected by solid edges to an $n$-point vertex. As an example, we have for the one-point function in a state with only single-trace sources:
\be\label{eq:graph2}
\langle {\cal O} \rangle_{\source[1]} \;\;=
\begin{gathered} \quad \includegraphics[width=.7\textwidth]{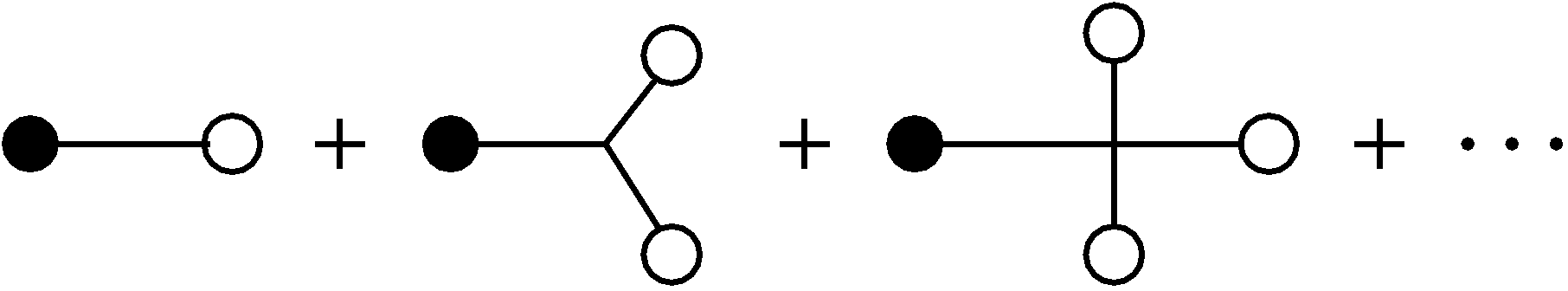}
\end{gathered}
\ee

All contributions at leading order in $1/N$ correspond to diagrams where the vacuum correlators are connected up by the multi-trace sources in a tree graph. The argument is almost exactly the same as in the diagrammatic partition function calculation of \S\ref{sec:EE}. For instance, an 11-point function relevant for $\langle {\cal O}\rangle_{\{\source[i]\}}$ at leading order in $1/N$ is given by the following tree:
\be
\begin{split}
 \includegraphics[width=.4\textwidth]{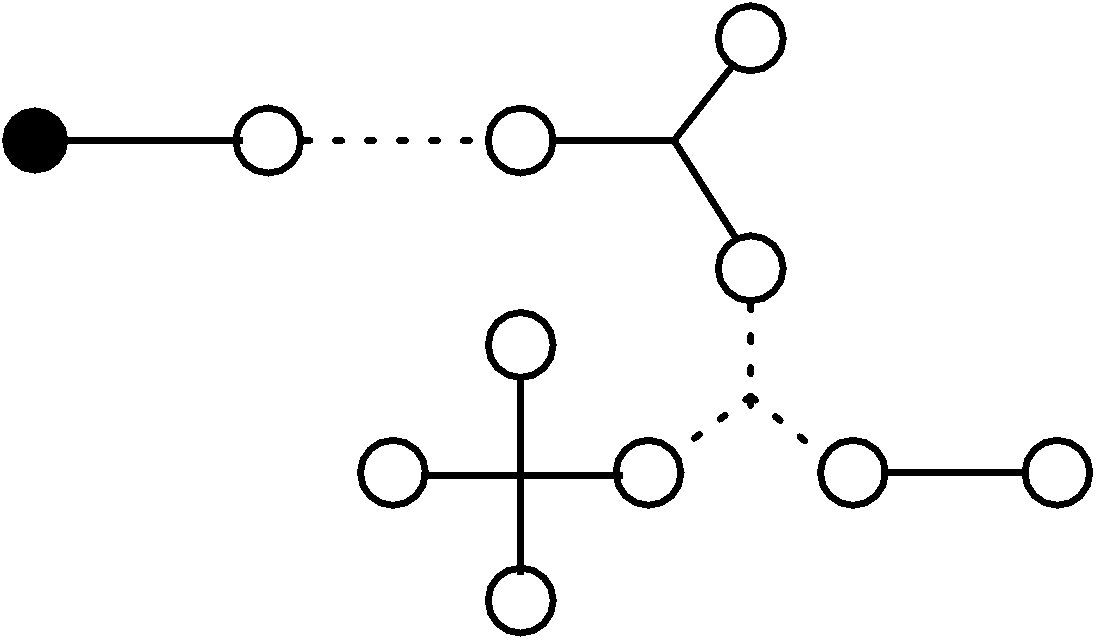}
\end{split}
\ee
It is fifth order in $\source[1]$, first order in both $\source[2]$ and $\source[3]$, and it factorizes into two 2-point functions, one 3-point function, and one 4-point function.

As a result of this, it is possible to represent the complete set of diagrams for the order $N^0$ contributions to the one-point function with general sources in terms of an effective operator insertion:
\be
\langle {\cal O} \rangle_{\{\source[i]\}} \;\;=
\begin{gathered} \quad \includegraphics[width=.7\textwidth]{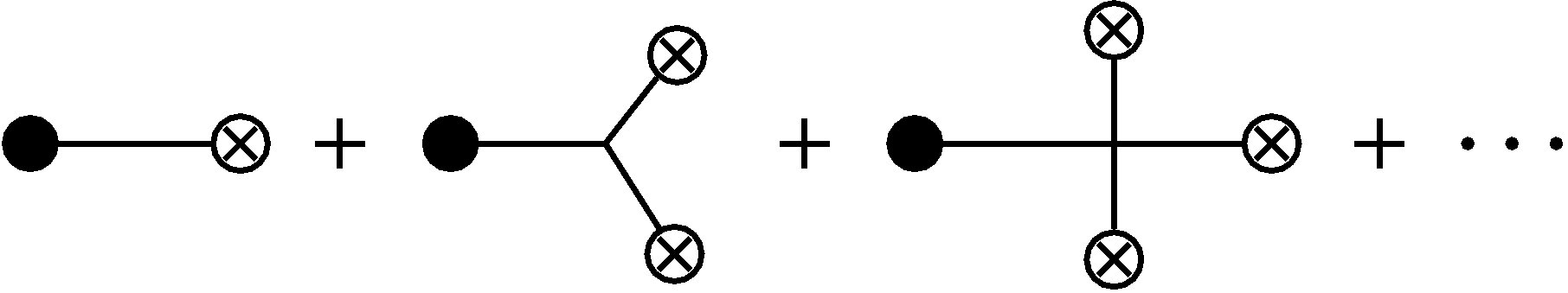}
\end{gathered}
\ee
i.e., it takes the same form as \eqref{eq:graph2}, but with an effective single-trace source, represented by a crossed circle defined as an open circle (one-point function) attached to all possible higher order trees:
\be
\label{eq:lambdaeffgraph}
 \includegraphics[width=.7\textwidth]{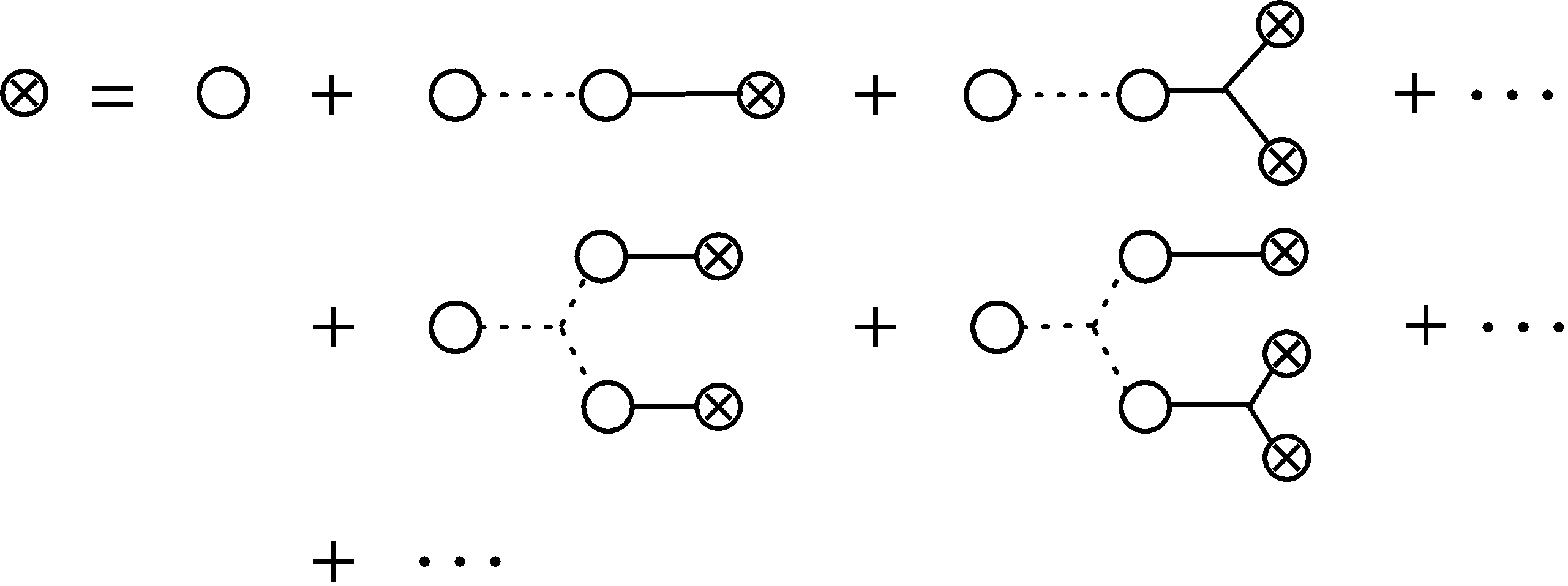}
\ee
The diagrammatics illustrates that the effective source is given by
\bea
\lambda_{eff}(x) &=& \source[1](x) +2 \int \source[2](x,x_1) \langle {\cal O}(x_1) \rangle_{\lambda_{eff}} +  3 \int \source[3](x,x_1,x_2) \langle {\cal O}(x_1) \rangle_{\lambda_{eff}} \langle {\cal O}(x_2) \rangle_{\lambda_{eff}} \cr
&=& \left. {\delta W[{\cal O}] \over \delta {\cal O}(x)}\right|_{{\cal O}(x) \rightarrow \langle {\cal O}(x) \rangle_{\lambda_{eff}}} \cr
 &=& \frac{\delta W[\langle {\cal O}\rangle_{\lambda}]}{\delta \langle{\cal O}(x)\rangle_\lambda} \bigg{|}_{\lambda = \lambda_{eff}}.
\eea

This completes the proof of the claim (\ref{STsource}).

\subsection{Partition function from effective single-trace source states}

The order $N^2$ result for the one-point functions implies that
\be
\label{OnePt}
 {\delta \log Z(\{\source[i]\}) \over \delta \source[1](x)} = \langle {\cal O}(x) \rangle_{\{\source[i]\}}  = \langle {\cal O}(x) \rangle_{\lambda_{eff}}  = {\delta \log Z(\lambda_{eff}) \over \delta \lambda_{eff}(x)} \; .
\ee
From this, we can derive an expression for the leading large $N$ contribution to the logarithm of the partition function in the theory with multi-trace sources, in terms of quantities computed in the theory with an effective source. Since we have seen that the ${\cal O}(N^2)$ partition function for $\source[1] = 0$ is the same as the unperturbed partition function, the full partition function $\log Z(\{\source[i]\})$ can be defined by this requirement and the first equality above. We now show that (to leading order in large $N$), these requirements are satisfied by
\be
\label{eq:lnZmult}
\boxed{
\log Z(\{\source[i]\}) = \log Z(\lambda_{eff})  - \sum_{n \ge 2} (n-1) \int \source[n](x_1, \dots, x_n) \langle {\cal O}(x_1) \rangle_{\lambda_{eff}} \cdots \langle {\cal O}(x_n) \rangle_{\lambda_{eff}} \; .
}
\ee
To see that this reduces to the unperturbed partition function when $\source[1] = 0$, we note that the effective source also vanishes in this case, and all one-point functions vanish in the absence of sources. To check the first equality in (\ref{OnePt}), first recall that
\be
\label{lameff}
\lambda_{eff}(x) = \source[1](x) + \sum_{n \ge 2} n \int \source[n](x, x_1 \dots, x_{n-1}) \langle {\cal O}(x_1) \rangle_{\lambda_{eff}} \cdots \langle {\cal O}(x_{n-1}) \rangle_{\lambda_{eff}},
\ee
so that
{\small
\be
\label{eq:deltalambdaeffdeltalambdaone}
{\delta \lambda_{eff}(x') \over \delta \source[1](x)} = \delta(x - x') + \sum_{n \ge 2} n (n-1) \int \source[n](x', x_1 \dots, x_{n-1}) \langle {\cal O}(x_1) \rangle_{\lambda_{eff}} \cdots \langle {\cal O}(x_{n-2}) \rangle_{\lambda_{eff}} {\delta \langle {\cal O}(x_{n-1}) \rangle_{\lambda_{eff}} \over \delta \source[1](x)}\,.
\ee
}
Then
\be
\begin{split}
&{\delta  \over \delta \source[1](x)} \left(\log Z(\lambda_{eff}) - \sum_{n \ge 2} (n-1) \int \source[n](x_1, \dots, x_n) \langle {\cal O}(x_1) \rangle_{\lambda_{eff}} \cdots \langle {\cal O}(x_n) \rangle_{\lambda_{eff}}\right) \\
&\quad = \int d x' \;{\delta \log Z(\lambda_{eff}) \over \delta \lambda_{eff}(x')} {\delta \lambda_{eff}(x') \over \delta \source[1](x)} \\
&\qquad \qquad  - \sum_{n \ge 2} n (n-1) \int \source[n](x', x_1 \dots, x_{n-1}) \langle {\cal O}(x') \rangle_{\lambda_{eff}} \cdots \langle {\cal O}(x_{n-2}) \rangle_{\lambda_{eff}} {\delta \langle {\cal O}(x_{n-1}) \rangle_{\lambda_{eff}} \over \delta \source[1](x)} \\
&\quad = \int d x'\; \langle {\cal O}(x') \rangle_{\lambda_{eff}} {\delta \lambda_{eff}(x') \over \delta \source[1](x)} \\
&\qquad  \qquad  - \sum_{n \ge 2} n (n-1) \int \source[n](x', x_1 \dots, x_{n-1}) \langle {\cal O}(x') \rangle_{\lambda_{eff}} \cdots \langle {\cal O}(x_{n-2}) \rangle_{\lambda_{eff}} {\delta \langle {\cal O}(x_{n-1}) \rangle_{\lambda_{eff}} \over \delta \source[1](x)} \\
&\quad = \int d x'\; \langle {\cal O}(x') \rangle_{\lambda_{eff}} \delta(x-x') \\
&\quad = \langle {\cal O}(x) \rangle_{\lambda_{eff}} = \langle {\cal O}(x) \rangle_{\{\source[i]\}} \; .
\end{split}
\ee
as desired. In the third line, we have used the last equality in \eqref{OnePt}. The cancellation in the second-to-last line comes from inserting \eqref{eq:deltalambdaeffdeltalambdaone}.

Diagrammatically, one can understand the result \eqref{eq:lnZmult} as follows. The left hand side of that equation to ${\cal O}(N^2)$ is a sum over all possible tree graphs consisting of any number of (empty) circles connected by dotted lines (integrated multi-trace sources) or solid lines (correlation functions). The first term on the right hand side reproduces most of these diagrams; namely, it contains all such tree graphs with the restriction that none of the outermost nodes are connected to the rest of the tree by a dotted line. This can be seen from the pictorial representation of the effective source in \eqref{eq:lambdaeffgraph}. The remaining terms on the right hand side of \eqref{eq:lnZmult} serve to provide precisely the missing graphs which involve dotted lines as ends of branches.

In Appendix \ref{sec:HigherPoint} we demonstrate how to calculate higher-point correlation functions using the effective single-trace coupling.

\subsubsection*{Calculating entanglement entropy using the replica trick}

Now that we have a prescription for calculating the CFT partition function with multi-trace sources in terms of states with an effective single-trace source, we can use it to calculate the entanglement entropy using the replica trick. However, there are differences from the usual story. First, we need to take into account the extra terms \eqref{eq:lnZmult}. Second, the effective source for the case of a multi-sheeted replica manifold is not simply a replicated version of the effective source that appears in the single-copy case. The effective source is determined by (\ref{lameff}); in that expression, the expectation value of an operator with a replicated source is not the same as the expectation value of an operator with source $\lambda_{eff}$ in the single-copy case. One way to understand this is that the sources from other sheets contribute to the one-point function at a given point via the CFT two-point function between the sheets.

Let's now consider the calculation. We begin by recalling that the entanglement entropy for a region $A$ may be calculated as
\be
\label{Replica}
S_A = \left. -{d \over dq} \log(\tr_A(\rho_A^q)) \right|_{q=1} = -\left. {d \over dq} \big(\log Z_q(\{\source[i]\}) - q \log Z_1(\{\source[i]\})\, \big) \right|_{q=1},
\ee
where $Z_q(\{\source[i]\})$ is the partition function calculated on a multi-sheeted surface defined by gluing $q$ copies of Euclidean space via the region $A$. According to (\ref{eq:lnZmult}), the partition function here is equal (to leading order in $1/N$) to the partition function with an effective single-trace source $\lambda_{eff}^{(q)}(x)$ defined analogous to (\ref{lameff}), plus the second term in (\ref{eq:lnZmult}) depending explicitly on the multi-trace couplings.  

After making this substitution in (\ref{Replica}), we can divide contributions to the entropy into three terms.  One term comes directly from the $q$-derivative of the second term in  (\ref{eq:lnZmult}), which we will call $\Delta S_A^{(1)}$.  The other two terms come from the $q$-derivative of $\log Z_q(\lambda_{eff}^{(q)})$: there is one part due to the $q$-dependence of $\lambda_{eff}^{(q)}$ and a remaining part which is present for a source which does not depend on $q$.  The former we will call $\Delta S_A^{(2)}$, while the latter is the contribution that gives precisely the area of the Ryu-Takayanagi surface in the geometry produced by $\lambda_{eff} \equiv \lambda_{eff}^{(q = 1)}$:
\be
\label{eq:SAcompl}
 \left. -\frac{d}{dq} \big( \log Z_q(\lambda_{eff} ) - q \log Z_1( \{ \source[i]\}) \, \big) \right|_{q=1} = S_A(\lambda_{eff}) \,.
\ee

 Let's now understand the additional term due to the $q$ dependence of $\lambda_{eff}^{(q)}$. This contribution is
\be
\Delta S_A^{(2)} = -\int d^d x  {\delta \log Z_{q=1}(\lambda_{eff}^{(q=1)}) \over \delta \lambda_{eff}^{(q=1)}(x)} {d \lambda_{eff}^{(q)}(x) \over dq} \; .
\ee
From (\ref{lameff}), we have
\be
\label{dlameff}
{d \lambda^{(q)}_{eff}(x) \over dq}=  \sum_{n \ge 2} n(n-1) \int \source[n](x, x_1 \dots, x_{n-1}) \langle {\cal O}(x_1) \rangle_{\lambda_{eff}} \cdots \langle {\cal O}(x_{n-2}) \rangle_{\lambda_{eff}}{d \langle {\cal O}(x_{n-1}) \rangle_{\lambda_{eff}} \over dq}.
\ee
Thus, using the last relation in (\ref{OnePt}), we have
\be
\Delta S_A^{(2)} =  -\sum_{n \ge 2}  n (n-1) \int \source[n](x_1, x_2 \dots, x_{n}) \langle {\cal O}(x_1) \rangle_{\lambda_{eff}} \cdots \langle {\cal O}(x_{n-1}) \rangle_{\lambda_{eff}} {d \langle {\cal O}(x_{n}) \rangle_{\lambda_{eff}} \over dq}\,.
\ee
But we now see that this precisely cancels $\Delta S_A^{(1)}$, i.e., the $q$-derivative of the second term in (\ref{eq:lnZmult}). Therefore, the terms in \eqref{eq:SAcompl} are complete and all other contributions cancel.

{\it We conclude that the order $N^2$ entanglement for the states defined by multi-trace sources is given by the Ryu-Takayanagi formula applied to the geometry defined by the effective single-trace source:}
\be
\label{eq:EEeff}
\boxed{
S_A(\{\source[i]\}) = S_A(\lambda_{eff}) +{\cal O}(N^0)\,.
}
\ee

\subsubsection*{Comments}

Our field-theoretic proof that the entanglement entropy at leading order in large $N$ is captured by an effective single-trace source relies on the Euclidean replica trick. Here, the number of replicas $q$ is an integer, which we analytically continue to compute the limit $q \rightarrow 1$. For non-integer $q$, it is necessary to invoke some type of analytic continuation, and the meaning of the analytically continued expressions is not clear from the field theory point of view. Note, however, that non-integer values of $q$ are very natural and easy to implement from a bulk perspective \cite{Lewkowycz:2013nqa,Dong:2016fnf}. In that case the replica manifold is constructed as a spacetime with a conical excess of $2\pi q$, which can reasonably take non-integer values. This can be taken as evidence that the somewhat formal manipulations that we have employed in our derivation are sensible, at least in the case of holographic theories.

While the order $N^2$ entanglement entropies are the same for our original multi-trace state and the state defined by the effective single-trace source, we emphasize that the same is not true for the Renyi entropies.\footnote{ We thank Xi Dong for asking this question.} The $q$-th R\'{e}nyi entropy is given by
\be
\label{qrenyi}
S^{(q)}_A = {1 \over 1-q} \log(\tr_A(\rho_A^q)) = {1 \over 1-q} \big(\log Z_q(\{\source[i]\}) - q \log Z_1(\{\source[i]\})\big) \,.
\ee
The explicit expression (\ref{eq:lnZmult}) shows that partition functions for general integer values of $q$ do not agree with those for the effective single-trace state, and so the Renyi entropies will be different in general. This is in accord with the expectation that while the order $N^2$ entanglement entropy captures properties of the classical geometry dual to a holographic state, the Renyi entropies for $q \ne 1$ contain more fine-grained information that goes beyond the classical geometry \cite{Dong:2018lsk}.

An interesting consequence of the observations in this section and the previous section is that for a holographic theory, the dual geometry, to the extent that it can be reconstructed from the CFT one-point functions and order $N^2$ entanglement entropies, is not affected by any of the multi-trace sources when there are no single-trace sources present but do affect the geometry when there are single-trace sources. This is immediate from our diagrammatic representation of contributions to the partition function and correlators, since there are non-vanishing tree diagrams built from single and multi-trace sources, but no nonvanishing tree diagrams with only multi-trace sources.\footnote{ In contrast, if we construct a state with multi-trace Euclidean sources and then add a Lorentzian source, it can be checked that the resulting geometry is the same as it would have been without the multi-trace sources.}

\subsection{Bulk interpretation in AdS/CFT}

The results of this section have been purely field-theoretic, but they have a natural interpretation for holographic theories.

Consider for simplicity the case where we are sourcing a scalar operator ${\cal O}$. For holographic theories, there is a corresponding bulk scalar field $\phi$ and we can interpret the ${\cal O}(N^2)$ part of the partition function $\log Z(\{\source[i]\})^{-1}$ as being equal to a gravitational action $S_{bulk} = S_{grav} + S_\phi$. The bulk scalar field has the usual asymptotic expansion
\be
   \phi(x,z\sim 0) \sim z^{d-\Delta} \left( \alpha(x) + {\cal O}(z^2) \right) + z^\Delta \left( \beta(x) + {\cal O}(z^2) \right) \,,
\ee
where $\Delta$ is the dimension of the operator.

For the geometry dual to a state defined by single trace sources, $\beta(x)$ is proportional to the expectation value $\langle {\cal O}(x) \rangle_{\{\source[i]\}}$, while $\alpha(x)$ is proportional to the source.

For a state defined via multi-trace sources, we have seen that one point functions and entanglement entropies at order $N^2$ are the same as for another state defined with an effective single trace source. Thus, at the classical level, these two states have the same dual geometry. The value of $\alpha(x)$ in this geometry is equal to the source for the single-trace state, so by definition, it is equal to the effective source for our multi-trace state. The value of $\beta(x)$ is equal to the one point function of ${\cal O}(x)$ in the single-trace state, which is also equal to the one-point function for the original multi-trace state.

With these identifications, the defining relation \eqref{STsource} for the effective source gives the relation
\be
\label{eq:WittenGen}
  \alpha(x) = \frac{\delta W[\beta]}{\delta \beta(x)} \, ,
\ee
which can be interpreted as a modified boundary condition that can be used to determine the classical dual geometry for states defined by multi-trace sources. There are two immediate consistency checks of this statement: for single-trace sources ($W[{\cal O}] = \int dx\, \source[1](x) {\cal O}(x)$) this is consistent with the expectation that $\alpha(x)$ should be proportional to the source. Similarly, for {\it local} double-trace sources the prescription \eqref{eq:WittenGen} reproduces the known result \cite{Witten:2001ua,Berkooz:2002ug}. Note that the deformation $W[{\cal O}]$ defined in \eqref{deformed2} corresponds to adding a boundary term $W[\phi(x,z \sim 0)]$ to the standard Euclidean bulk action, which is consistent with the modification to the boundary condition.

In summary, to calculate the order $N^2$ entanglement entropy or one-point functions holographically for a state defined by nonlocal multi-trace sources in the Euclidean path integral, we should find a solution of the bulk equations of motion with boundary conditions \eqref{eq:WittenGen}, and then use the standard AdS/CFT dictionary (e.g. the classical RT formula for ${\cal O}(N^2)$ entanglement) in this geometry.

\section{$1/N$ corrections and bulk entanglement}
\label{sec:Subleading}

In this section, we discuss $1/N$ corrections to the CFT entanglement entropy. We will see these exhibit qualitatively new features, including a breakdown of naive perturbation theory in the sources, which are consistent with the expectation that the bulk interpretation is no longer purely classical.

\subsection{Replica method topology}

In our diagrammatic representation of perturbative contributions to the partition function (\S\ref{sec:EE}), we have seen that the order $N^2$ contributions must come from tree diagrams.

\begin{figure}
\begin{center}
\includegraphics[width=.5\textwidth]{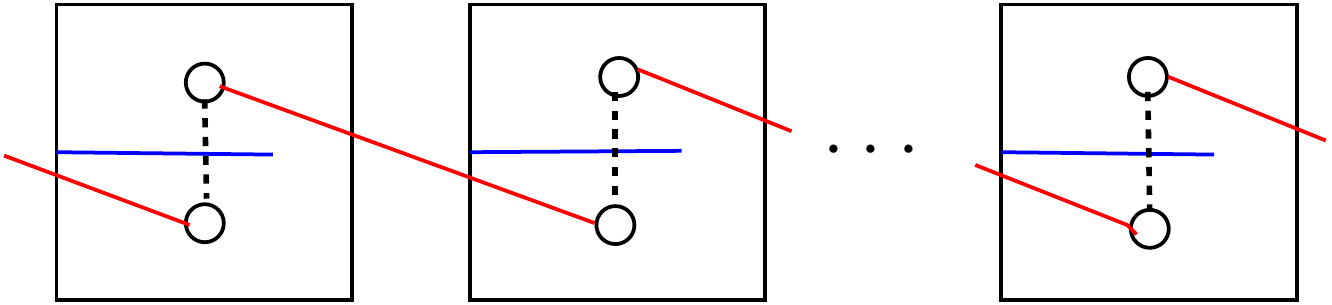}
\end{center}
\caption{Order $\lambda^n N^0$ contributions to $\log Z_n[\{\source[2]\}]$ that are non-contractible on the replica manifold. These contributions survive in the limit when the subsystem is the whole CFT, leading to non-analytic $\lambda \log \lambda$ behavior.}
\label{fig:wrapping2}
\end{figure}

These have the topological feature that they are ``contractible'' on the replica manifold (when the latter is connected); we will argue below that this property corresponds to perturbative contributions to the entanglement entropy that are analytic in the sources. On the other hand, when we consider $1/N$ corrections to the entanglement entropy, we have contributions corresponding to diagrams with ``loops'', such as the one in Fig. \ref{fig:wrapping2}. In this case, the loop can wind around a non-trivial cycle in the replica manifold, and we will see that this can give rise to contributions to the entanglement entropy that are non-analytic, involving logarithms of the sources.

To motivate the connection between the diagram topology and analyticity properties, consider the contributions to the replica-manifold partition function as a function of some parameter $\lambda$ associated with the sources. In the perturbative expression for $\log(Z_n)$ we may have terms with specific fixed powers of $\lambda$, e.g. $\lambda^4$, but we may also have terms scaling as $\lambda^n$ (or $\lambda^{n+2}$, etc...). In the limit $n \to 1$ used in calculating entanglement entropy, the latter terms give contributions involving $\log(\lambda)$. For a specific replica number, e.g. $n=4$, both the $\lambda^4$ terms and the $\lambda^n$ terms would appear as $\lambda^4$, but these arise from different types of diagrams.

Terms of the first type (with a fixed power of $\lambda$) come from diagrams that appear for any value of the replica number, for example where the operators all live on a few adjacent replicas and are not sensitive to the full geometry of the replica manifold. These typically map to equivalent contributions under the $Z_n$ replica symmetry, so the contribution comes with an overall factor of $n$. After calculating the derivative $d/dn \log(Z_n)$ to compute entanglement entropy, we end up with a contribution that has the same fixed power of $\lambda$ that we started with. On the other hand, contributions that give rise to $\lambda^n$ terms are those that are sensitive to the topology of the full replica manifold, for example, contributions with operators on each sheet for which the correlators connect operators on all the sheets in a loop, as in Fig. \ref{fig:wrapping2}. For these contributions, the derivative $d/dn \log(Z_n)$ acting on $\lambda^n$ terms in $\log(Z_n)$ gives rise (at least naively) to $\lambda \log(\lambda)$ contributions in the limit $n \to 1$.

In a perturbative gravity calculation of the area of an extremal surface in a geometry produced by specified Euclidean sources, it is difficult to see how such logarithmic terms
could arise. Thus, the fact that logarithmic contributions to entanglement entropy show up at order $N^0$ may be indicative of the fact that the corresponding gravitational quantity is no longer purely geometrical, but rather involves contributions associated with bulk entanglement.

To see more explicitly the appearance of these logarithmic contributions and understand how they are reproduced through a calculation on the gravity side, we now consider a simple example.

\subsection{Example: full CFT entanglement entropy for a mixed state}

Consider states defined from the Euclidean path integral with only a double-trace source. We will consider states\footnote{ To avoid an excess of notation, here and in the remainder of the paper we will neglect writing matrix elements or overlaps on density matrices or states, repectively, defined from path integrals, and we will omit boundary conditions when this would not cause confusion. In case of confusion, the reader should refer back to \S\ref{sec:pathintegrals}, e.g. \eqref{PIstate2} and \eqref{effsource}, for our conventions.}
\be
\label{DTstate}
\rho = {1 \over Z_\lambda} \int [d \phi] e^{- S - \int d^d x_1 d^d x_2 \;\source[2](x_1, x_2) {\cal O}(x_1) {\cal O}(x_2)}
\ee
which will be mixed states if the source couples operators with $\tau > 0$ to operators with $\tau < 0$.

By the general discussion above, entanglement entropies for such a state cannot have any order $N^2$ contributions, since there are no tree diagrams that can be built from only double-trace insertions. So the leading contributions will be at order $N^0$. We consider first the von Neumann entropy of the full state.

Note first that a single insertion of ${\cal O}(x)$ in the Euclidean path integral that defines the vacuum state will produce some linear combination of independent states each orthogonal to the vacuum state (assuming that the one-point function of ${\cal O}$ vanishes):\footnote{ Here, the expression on the left side is shorthand for the state whose wave functional is defined by the path integral on the left.}
\be
\label{onepart}
{1 \over Z_0^{1 \over 2}} \int [d \phi] e^{-S_{EUC}} {\cal O}(x) = \sum_\alpha c_\alpha (x)| \alpha \rangle \; .
\ee
Here, we can consider the CFT defined on a spatial sphere so that the sum will be discrete; otherwise, the sum over $\alpha$ will be replaced by an integral. Taking the inner product of two such states for different $x$'s, we get
\be
\label{norm}
\sum_\alpha c_\alpha (x_2) c^*_\alpha (x_1) = {1 \over Z_0} \int [d \phi] e^{-S_{EUC}} {\cal O}(x_1^+){\cal O}(x_2) = \langle {\cal O}(x_1^+){\cal O}(x_2) \rangle \equiv  \Delta(x_1^+, x_2) \; .
\ee
Here, the point $x_1^+$ is obtained from $x_1$ by $\tau \to -\tau$.

To leading order in $\source[2]$, the state of the CFT deformed by a double-trace perturbation as in (\ref{DTstate}) is\footnote{  Here, the factor $N_\lambda$ is defined to be the coefficient of $| 0 \rangle \langle 0 |$ in the normalized state. This includes contributions at order $\source[2]$ coming from the contribution of the identity operator to the OPE in the states defined by two insertions of ${\cal O}$ in each CFT. The remaining terms at order $\source[2]$ are orthogonal to the vacuum.}
\be
\rho = N_\lambda \left( | 0 \rangle \langle 0| + \int dx_1 dx_2 \;\source[2](x_1^-,x_2^+) c_\alpha (x_1^-) c_\beta^* (x_2^+) |\alpha \rangle \langle \beta | + \dots \right) \; ,
\ee
where the omitted terms include terms at second and higher order in $\source[2]$ as well as order $\source[2]$ terms of the form $|0 \rangle \langle \alpha,\beta|$ or $| \alpha,\beta\rangle \langle 0|$ that will only modify the eigenvalues of $\rho$ at second order.

To proceed, we compute the leading order entanglement entropy for any state of the form
\be
\rho = N_\epsilon ( | 0 \rangle \langle 0 | + \epsilon A_{ij} |i \rangle \langle j | + \dots) \; .
\ee
Since the density matrix must have unit trace, the normalization factor is
\be
N_\epsilon^{-1} = (1 + \epsilon \tr(A) + {\cal O}(\epsilon^2)) \; .
\ee
From these expressions for the density matrix and normalization factor, we can read off perturbative expressions for the eigenvalues and calculate the entanglement entropy. The result is
\be
\label{logcont}
S =  - \tr (\epsilon A \log(\epsilon A)) + \tr(\epsilon A) +  {\cal O}(\epsilon^2)
\ee
In our case, we have that
\be
(\epsilon A)_{\alpha \beta} = \int dx^- dx^+ \;\source[2](x^-,x^+) c_\alpha (x^-) c_\beta^* (x^+) \; .
\ee
Using (\ref{norm}), we can write the final answer for the entanglement entropy entirely in terms of $\lambda$ and $\Delta$, since we have:
\be
\label{wrapping}
\tr((\epsilon A)^n) = \int \prod_{i=1}^n \left\{dx_i^- dx_i^+ \source[2](x_i^-,x_i^+) \Delta(x_i^+,x_{i+1}^-) \right\} \; .
\ee
The result in (\ref{logcont}) can be expressed in terms of the quantities in (\ref{wrapping}) via analytic continuation or by integral formulae such as
\be
\label{Sint}
- \tr (\epsilon A \log(\epsilon A)) = \int_0^\infty da \sum_{a=0}^\infty {(-a)^n \over (n+1)!} (\tr((\epsilon A)^n) -1) \; .
\ee
The key point is that the result has a non-analytic $\lambda \log \lambda$ behavior; we will now see how this can arise diagrammatically from the replica method.

\subsubsection*{Replica method}

For a subsystem $A$ (which could be the whole CFT), consider the replica manifold $M^{(n)}_A$ and the associated partition function $Z_n$. On this space, the leading contribution to $\log(Z_n)$ is a contribution at order $\source[2]$ given by
\be
\begin{split}
\log(Z_n)\big{|}_{{\cal O}(\source[2])} &= - \sum_i \int dx_1^{(i)} d x_2^{(i)} \source[2](x_1^{(i)}, x_2^{(i)}) \langle {\cal O}(x_1^{(i)}) {\cal O}(x_2^{(i)}) \rangle_{M^{(n)}_A} \\
&= - n \int dx_1^{(1)} d x_2^{(1)} \source[2](x_1^{(1)}, x_2^{(1)}) \langle {\cal O}(x_1^{(1)}) {\cal O}(x_2^{(1)}) \rangle_{M^{(n)}_A} \; .
\end{split}
\ee
However, this does not necessarily give the leading contribution to the entanglement entropy. We also have a contribution at order $(\source[2])^n$ with one source inserted on each sheet of the replica manifold, given by
\be
(-1)^n \int \prod_{i=1}^n dx_1^{(i)} d x_2^{(i)} \source[2](x_1^{(i)}, x_2^{(i)}) \langle \prod_{i=1}^n {\cal O}(x_1^{(i)}) {\cal O}(x_2^{(i)}) \rangle_{M^{(n)}_A}.
\ee
The contributions to this at leading order in the $1/N$ expansion involve products of $n$ 2-point functions. A contribution that is ``non-contractible'' in the sense described above is
\be
\label{nonC}
(-1)^n  \int \prod_{i=1}^n dx_1^{(i)} d x_2^{(i)} \source[2](x_1^{(i)}, x_2^{(i)}) \prod_{i=1}^n \langle {\cal O}(x_1^{(i)}) {\cal O}(x_2^{(i+1)}) \rangle_{M^{(n)}_A} \; ;
\ee
this corresponds to the diagram shown in Fig. \ref{fig:wrapping2}. In the limit where $A$ becomes the whole space, this is non-vanishing so long as the source couples operators for $\tau < 0$ to operators with $\tau > 0$, and gives a contribution which is exactly of the form (\ref{wrapping}). In the calculation of the entanglement entropy via (\ref{Replica}), this leads to logarithmic contributions as in (\ref{logcont}).

When $A$ is not the whole space, it is less clear that the contributions (\ref{nonC}) lead to logarithmic terms in the entanglement entropy; we will discuss this further below.

\subsubsection*{Gravity calculation}

Now let's understand how the result (\ref{logcont}) can arise from a gravity calculation. We would like to show that the CFT entanglement entropy can be reproduced by the quantum RT formula using an appropriate bulk state. In this case, the area term vanishes since we are calculating the entanglement entropy of the entire CFT. Then the quantum RT formula requires that the CFT entanglement equals the bulk entanglement.

For a holographic theory, inserting the operator ${\cal O}(x)$ in the path integral for a single CFT will define a single particle bulk state
\be
\label{onepart2}
 | \Psi_{{\cal O}(x)} \rangle  = \sum_\alpha {\cal C}_i (x)  a^\dagger_{i} | 0 \rangle + \cdots \; .
\ee
orthogonal to the vacuum state, where the dots indicate subleading terms arising due to bulk interactions. The normalization condition requires that
\be
\label{norm2}
\sum_\alpha {\cal C}_i (x_2) {\cal C}^\dagger_i (x_1) = {1 \over Z} \int [d \phi] e^{-S_{EUC}} {\cal O}(x_1^+){\cal O}(x_2) = \langle {\cal O}(x_1^+){\cal O}(x_2) \rangle \equiv  \Delta(x_1^+, x_2) \; .
\ee
Then we can write the bulk state for the double-trace perturbation as
\be
\rho_{\rm bulk} = N_\lambda(| 0 \rangle \langle 0 | + \int dx_1 dx_2 \source[2](x_1,x_2) {\cal C}_i (x_1) {\cal C}_j (x_2) a^\dagger_{i}| 0\rangle \langle 0| a_{j} + {\cal O}((\source[2])^2) \; .
\ee
At this point, the discussion is entirely parallel to our CFT discussion, and we are guaranteed to get the same result, since all dependence on the coefficients ${\cal C}$ is replaced by dependence on $\Delta$ as in the CFT calculation. Thus, the non-analyticities present in the field theory result also appear in the gravity result since both arise from quantum entanglement.

\subsection{Non-analyticity in sources for subsystem entanglement entropies?}
\label{sec:nonanalytic}

We have seen that contributions to the partition function corresponding to the diagram in Fig.~\ref{fig:wrapping2} give rise to $\lambda \log \lambda$ terms in the entanglement entropy in the limit where the subsystem is the whole space. We have shown that these contributions correspond to bulk entanglement entropy on the gravity side.

In the case where we are computing the entanglement entropy of a subsystem, it seems reasonable to expect that the same diagrams should also be associated with the
 bulk entanglement entropy on the gravity side. It is interesting to ask whether they also show the non-analytic $\lambda \log(\lambda)$ behavior. Naively, this should be the behavior since we are applying $d/dn \log(Z_n)|_{n=1}$ to a contribution of order $\lambda^n$ and taking the limit $n \to 1$.

Whether or not the perturbative expansion of the entropy contains nonanalytic terms
can be seen starting with the definition in terms of the density matrix,
\be
S = -\tr (\rho \log \rho) = -\tr((\rho_0 + \lambda \delta \rho)\log(\rho_0+\lambda\delta \rho)).
\ee
Nonanalytic terms arise because the logarithm has a finite radius of convergence when expanding
around any positive number.  This means that the perturbative expansion will be valid
only if all eigenvalues of $\lambda\rho_0^{-1} \delta\rho$ are less than $1$.  It is easy to
find situations where this fails to hold.  First, if $\rho_0$ does not have full rank,
then any $\delta\rho$ whose kernel does not contain the kernel of $\rho_0$ will fail to
satisfy the requirements for convergence of the logarithm.  In particular, this is the case
where $\rho_0$ is a pure state, as occurs when considering the full CFT and perturbing around
the vacuum state.

On the other hand, for a subregion of a CFT, $\rho_0$ has full rank, and
the eigenvalues of $\lambda\rho_0^{-1}\delta \rho$ must be analyzed to determine the
analyticity of the perturbative expansion.  Had the subspace been finite-dimensional, it is clear
that these eigenvalues would eventually become small enough as $\lambda$ is taken to zero,
and hence the perturbative expansion in $\lambda$ would be valid.  However, for infinite-dimensional
spaces, as is typical of quantum field theories, it can occur that for any finite value
of $\lambda$, there are infinitely many eigenvalues of $\lambda \rho_0^{-1}\delta\rho$
that are larger than 1.  Hence, at no value of $\lambda$ is the expansion valid, and the
perturbative series must be resummed in the subspace in which the perturbative expansion fails.
This generically leads to nonanalyticities in the entropy as a function of $\lambda$.

A simple toy example serves to illustrate this point.  Take $\rho_0$ to be the thermal density
matrix of a single harmonic oscillator at inverse temperature $\beta$. Its eigenvalues
are $p_n = e^{-\beta n}$, up to an overall normalization.  Then choose the perturbation
$\delta \rho$ to be a state at a higher temperature, with eigenvalues $q_n = e^{-(\beta-a)n}$.
Regardless of the size of $\lambda$, eventually the perturbation will become larger than
the original density matrix elements, since the smaller inverse temperature causes the
probabilities of the perturbation to die off more slowly at high energies.

The entropy for this toy system is given by
\be
S = \log Z -\frac1Z\sum_n(p_n + \lambda q_n)\log(p_n + \lambda q_n),
\ee
with $Z$ the partition function ensuring normalization of the density matrix.
This sum will have a contribution
\be
\label{eq:toyfull}
s(a;\beta) = -\sum_{n\geq 0} e^{-\beta n} \log(1 + \lambda e^{a n})\,.
\ee
which is manifestly finite; however, the order $\lambda^k$ term obtained in a naive perturbative expansion of the logarithm is divergent for $a > \beta/k$.
This signals a breakdown in perturbation theory coming from nonanalytic contributions in
$\lambda$, which can be extracted by approximating the sum by an integral,\footnote{ One can
argue using the Euler-Maclaurin formula that the difference between the sum and
integral is  analytic in $\lambda$, hence the integral suffices to capture the nonanalytic
behavior of the sum.}
\be
s(a;\beta) \approx {\cal I} \equiv -\int_0^\infty  dx \; e^{-\beta x} \log(1 + \lambda e^{a x}).
\ee
The series representation of this integral includes a non-analytic piece $\lambda^{\beta/a}$ for non-integer $\beta/a$, or $\lambda^{\beta/a} \log \lambda$ for integer $\beta/a$. For example, with $a=\beta/2$, the naive perturbative expansion (obtained by expanding the integrand in $\lambda$, performing the integral term by term, and then setting $a=\beta/2$)
has a divergent order $\lambda^2$ term,
\be
{\cal I}_{naive} = -\frac{1}{\beta} \left[ 2 \lambda + \infty \lambda^2 - {2 \over 3} \lambda^3 + {1 \over 4} \lambda^4 + \dots \right].
\ee
In the correct expansion of the full result (obtained by first performing the integral and then expanding in $\lambda$) we have in this case
\be
{\cal I}_{correct} = -\frac{1}{\beta} \left[ 2 \lambda + (\log \lambda - 1/2) \lambda^2 - {2 \over 3} \lambda^3 + {1 \over 4} \lambda^4 + \dots \right],
\ee
so the naively divergent term is replaced by $\lambda^2(\log \lambda - 1/2)$, while all the other terms are the same.

In summary, diagrams of the type shown in Fig.~\ref{fig:wrapping2} may indeed lead to non-analytic contributions to the entanglement entropy even in the case where we are considering a subsystem, where the unperturbed density matrix is full rank. We have so far only motivated this by considering a toy example. However, it is interesting to note that in this example, the non-analyticity in $\lambda$ is reflected in the fact that contributions at specific orders in $\lambda$ in the naive perturbative expansion are divergent for certain parameter ranges. We will find precisely this behavior in explicit perturbative CFT calculations of the ${\cal O}(N^0)$ entanglement entropy for a subsystem in the following section.

\section{Calculations of entanglement entropy via perturbation theory in sources}
\label{sec:quantum2}

We now turn to a more detailed discussion of the calculation of entanglement and relative entropies in multi-trace states of the form \eqref{deformed}. As we shall demonstrate, there are divergences at specific orders in the sources in the naive perturbative expansion. We will interpret these as artifacts of the non-analyticities observed in \S\ref{sec:nonanalytic}. We will illustrate the issue with a simple example.

\subsection{Perturbative method}
\label{sec:faulknermethod}

We begin by reviewing a perturbative method for computing subsystem entanglement entropies and/or relative entropies in cases where the unperturbed density matrix is known explicitly, for example in the case of a ball-shaped subsystem. Consider general (mixed) states of the form
\be
\label{rhoexp}
\begin{split}
\rho
& = \rho_0 +  \delta \rho \equiv \rho_0 + \rho_0^{\frac{1}{2}} \, \delta \rho_b \, \rho_0^{\frac{1}{2}} \,,
\end{split}
\ee
where $\delta \rho_b$ is a ``bare'' perturbation, which we introduce for convenience.\footnote{ One motivation for this definition is that it is the smallness of $\delta \rho_b$ rather than $\delta \rho$ that most directly controls the validity of the perturbative expansion of $\log \rho$.} Here, we have in mind that $\rho$ is the density matrix for a ball-shaped subsystem, $\rho_0$ is the vacuum density matrix for the region, and $\delta \rho_b$ represents the perturbations to the vacuum state arising from turning on sources in the path integral.

The entanglement entropy for $\rho$ can be expressed as the expectation value of the modular Hamiltonian $K = -\log \rho$ in the state $\rho$. A general formula for the modular Hamiltonian for a state of the form (\ref{rhoexp}) was derived in \cite{Sarosi:2017rsq} (this builds on earlier developments of \cite{Rosenhaus:2014woa,Rosenhaus:2014zza,Faulkner:2014jva,Sarosi:2016oks}, and was recently discussed more rigorously in \cite{Lashkari:2018oke,Lashkari:2018tjh})\footnote{ See \cite{Ugajin:2018rwd} for a generalization of the perturbative method to R\'{e}nyi entropies.}:
\be
\label{logrho}
\begin{split}
K =& K_0 + \delta K\,, \qquad \qquad K_0 \equiv -\log \rho_0 \,, \\
\delta K &\equiv \sum_{k=1}^\infty (-1)^k \int_{-\infty}^\infty ds_1\cdots ds_k\, \mathcal{K}_k(s_1,\ldots,s_k) \, \prod_{r=1}^k \left( \rho_0^{-\frac{i s_r}{2\pi}  } \, \delta \rho_b  \, \rho_0^{\frac{i s_r}{2\pi}} \right),
\end{split}
\ee
with kernels
\begin{equation}
 \mathcal{K}_1(s_1) = \frac{1}{4} \, \frac{1}{( \cosh \frac{s_1}{2} )^2} \,,\qquad
  {\cal K}_k (s_1,\ldots,s_k) = \frac{(2\pi)^2}{(4\pi)^{k+1}} \frac{i^{k-1}}{\cosh \frac{s_1}{2} \; \cosh \frac{s_k}{2} \; \prod_{r=2}^k \sinh \frac{s_r-s_{r-1}}{2} } \,.
\end{equation}
It will be important in what follows that the kernel ${\cal K}_k$ has poles in the complex $s_r$-planes at $2 \pi i$ separations. We can therefore move the $s_r$ integration contours up and down in the imaginary direction within a strip of width $2\pi i$, while still avoiding poles. Attempting to shift the contours further naively leads to singularities.

Using the definitions above we can decompose the entanglement entropy as
\be
\label{SKeq}
S = S_0 + \tr(\delta \rho K_0 ) + \tr((\rho_0 + \delta \rho) \delta K)
\ee
where the second term, linear in the state perturbation, is the change in expectation value of the unperturbed modular Hamiltonian, and the third term is equal to the relative entropy $S(\rho||\rho_0) = \tr(\rho \log \rho) - \tr (\rho \log \rho_0)$.

\begin{figure}
\begin{center}
\includegraphics[width=.5\textwidth]{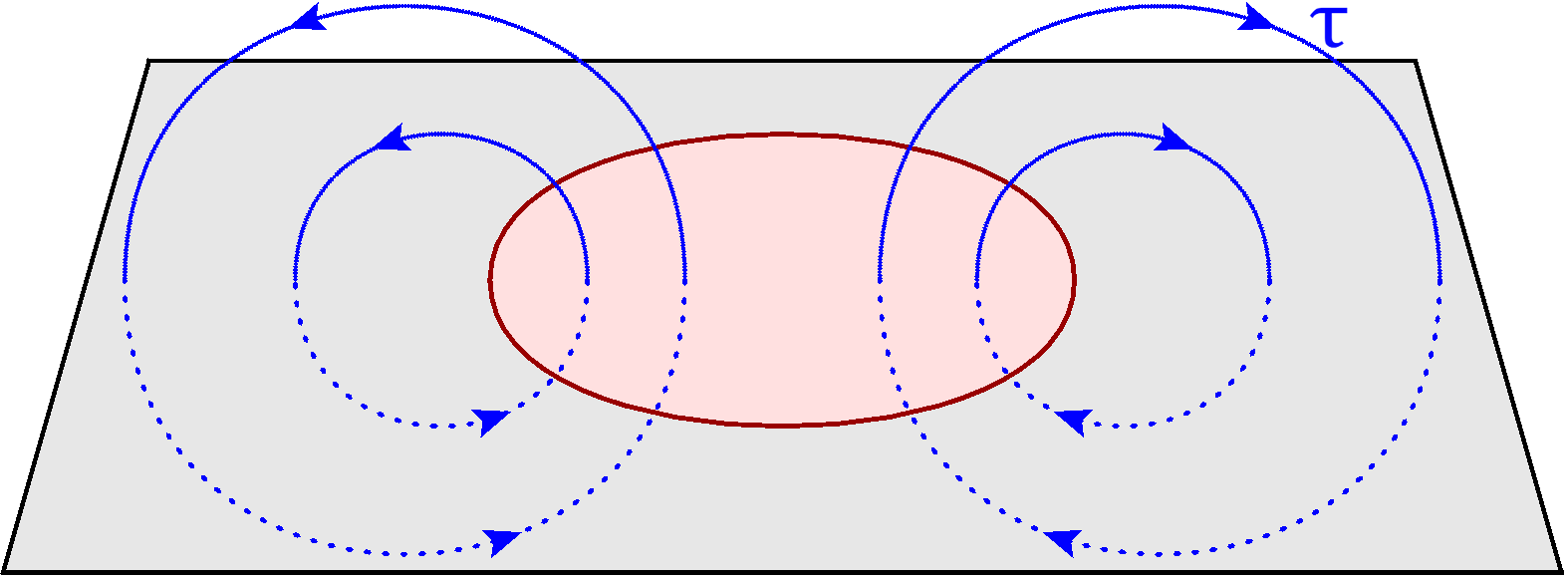}
\end{center}
\caption{Geometrical flow in Euclidean space associated with Euclidean time evolution using the modular Hamiltonian for the shaded ball shaped region.}
\label{fig:Eucflow}
\end{figure}

For ball-shaped regions in a CFT, the modular Hamiltonian is ``local'', i.e. linear in the local stress tensor operator. The unperturbed density matrix can be understood as the thermal state (for some fixed temperature) with respect to this Hamiltonian $K_0$. Real-time evolution with $K_0$, e.g. that induced by the conjugation by imaginary powers of $\rho_0$ in (\ref{logrho}), generates a geometrical flow of operators within the domain of dependence region (casual diamond) of the ball.  Euclidean time evolution with respect to $K_0$ generates an angular flow in Euclidean space, as shown in Fig.\ \ref{fig:Eucflow}, which circulates about the boundary of the ball-shaped region. For a ball of radius $R$ centered at $x=0$, this flow is generated by the conformal Killing vector field
\be
(\zeta^0,\, \zeta^i) = \frac{\pi}{R}   \left( R^2 + (x^0)^2 - \vec{x}^2, \; 2 x^0 x^i \right)  \,.
\ee
As explained in detail in section 3.2 of \cite{Faulkner:2017tkh}, we can choose an angular coordinate $\tau$ along this flow such that $\tau$ runs from $-\pi$ at the $\tau > 0$ surface of the ball-shaped region to $\pi$ at the $\tau<0$ surface, and define coordinates $\tilde{x}$ transverse to this flow direction.

\begin{figure*}
    \centering
    \begin{subfigure}[t]{0.33\textwidth}
        \centering
        \includegraphics[scale=0.5]{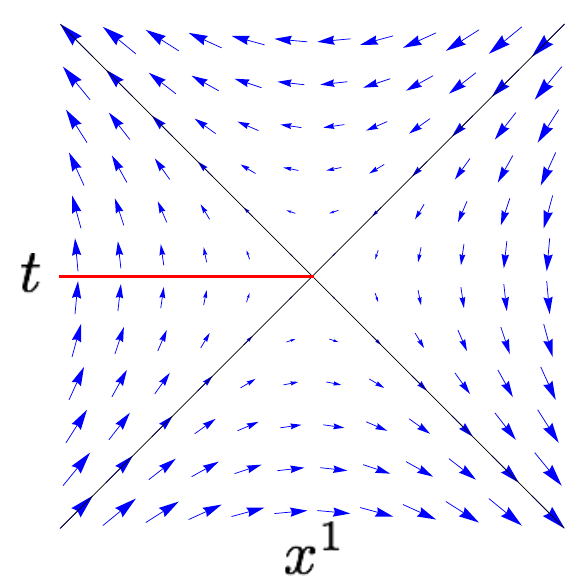}
        \caption{}
    \end{subfigure}%
    ~
    \begin{subfigure}[t]{0.33\textwidth}
        \centering
        \includegraphics[scale=0.5]{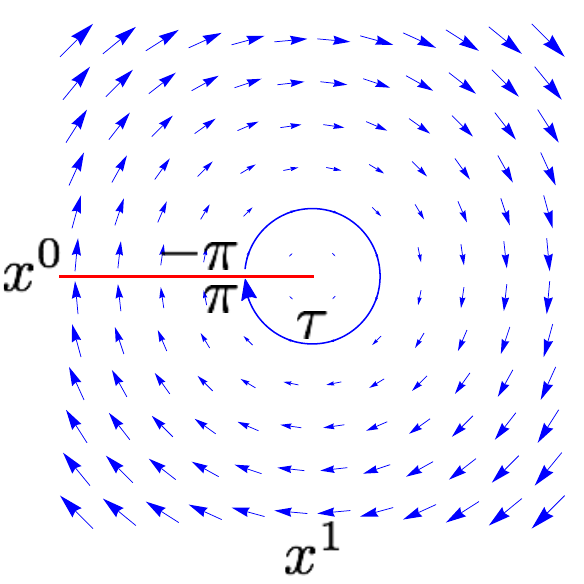}
        \caption{}
    \end{subfigure}%
    ~
    \begin{subfigure}[t]{0.33\textwidth}
        \centering
        \includegraphics[scale=0.5]{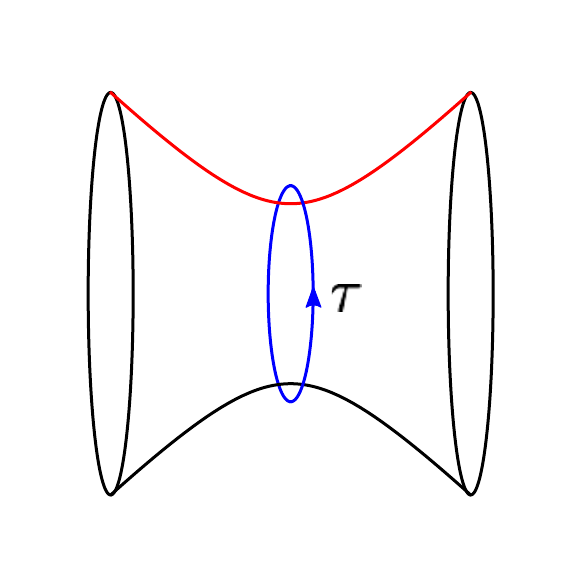}
        \caption{}
        \label{fig:angularTime:c}
    \end{subfigure}
    \caption{(a) Time evolution for a state defined on the half plane $x^1<0$ is given by the usual Rindler time. (b) The associated Euclidean time $\tau$ is the angle around the entangling surface.  (c) The Euclidean plane may be conformally transformed to $\mathbb H^{d-1} \times S^1$, where $\tau$ is now along the $S^1$.}
    \label{fig:angularTime}
\end{figure*}

By a conformal transformation, the domain of dependence region of our ball can be mapped to a half-space, so that the evolution via the modular Hamiltonian correspond to evolution with respect to Rindler time. By a different conformal transformation, we can also map this region to hyperbolic space times time; in the latter picture, the modular evolution maps to ordinary time evoluion. The flow in these cases is depicted in Fig.\ \ref{fig:angularTime}.

\subsubsection*{Perturbation theory in the sources}

Let us now consider specifically the case where the perturbation $\delta \rho$ arises from adding multi-trace sources to the  path integral, as in \S\ref{sec:pathintegrals}. The perturbation $\delta \rho$ then has contributions with various powers of the sources. At $k$-th order in sources, these take the schematic form
{
\be
 \langle \phi_- | \delta^{(k)} \rho | \phi_+ \rangle = \frac{(-)^k}{k!} \int_{ \phi_-}^{\phi_+} [d \phi] \;e^{-S_E}\; \int \prod_{a=1}^k \sum_{n_a} \source[n_a](x_1, \cdots, x_{n_a}) \,{\cal O}(x_1) \cdots {\cal O}(x_{n_a})  - \text{traces}
\ee
}\normalsize
where the integral is over Euclidean space with the ball of interest at $\tau=0$ removed and we impose specific field configurations $\phi_{\pm}$ that appear as boundary conditions at the location of the ball for $\tau = \pm \epsilon$. 
We also subtract ``traces'', which arise from combining lower order terms in the expansion of the exponential of sources into an order $\lambda^k$ contribution. These are also needed to maintain the normalization condition $\text{tr}(\rho) = 1$. We can equivalently write the perturbation in operator language as
\be
\label{rhopert}
\delta^{(k)} \rho_b = \frac{(-)^k}{k!}\int \prod_{a=1}^k \sum_{n_a} \source[n_a](x_1, \cdots, x_{n_a}) \, {\cal T}\left\{ {\cal O}(x_1) \cdots {\cal O}(x_{n_a})\right\} - \text{traces}\,.
\ee
where the time-ordering here is with respect to the angular coordinate $\tau$ defined above (the Euclidean ``modular time''), and the Heisenberg-picture operators are defined via
\be
{\cal O}(\tau, \tilde{x}) \propto e^{\tau K_0} {\cal O}(0, \tilde{x}) e^{-\tau K_0} \; .
\ee
up to conformal factors given explicitly in \cite{Faulkner:2017tkh}.

In the expression (\ref{logrho}) for $\delta K$, the operators appearing in $\delta \rho_b$ are subjected to additional real-time modular flow, so finally, the various terms in the perturbative expansion of the entanglement entropy (\ref{SKeq}) are each expressed in terms of correlators of the form
\be
\label{Euccor}
\tr \left(e^{-2 \pi K_0} {\cal O} (\tau_1 + i s_{n_1},\tilde{x}_1) \cdots {\cal O}( \tau_M + i s_{n_M},\tilde{x}_M) \right) \; .
\ee
integrated against the sources and the kernels appearing in (\ref{logrho}). These are thermal correlators with respect to the modular Hamiltonian, with operators inserted at various complex (modular) times.\footnote{ Alternatively, by a conformal transformation discussed in \cite{Casini:2011kv}, these map to usual thermal correlators for a CFT on hyperbolic space.}

\subsubsection*{Divergences from out-of-time-ordered Euclidean correlators}

While the operators appearing in the various perturbations (\ref{rhopert}) are ordered with respect to the Euclidean modular time $\tau$, those in $\eqref{Euccor}$ may come from terms in (\ref{SKeq}) with multiple $\delta \rho_i$ insertions, and thus are not necessarily time-ordered.

Out-of-time-ordered Euclidean correlators are problematic in quantum field theory since they involve backwards Euclidean time evolution $e^{+ \tau K_0}$, which exponentially enhances UV modes, leading to divergences. As an illustration, for a single quantum harmonic oscillator degree of freedom $x(\tau)$, the Euclidean out-of-time-ordered two-point function gives
\be
{1 \over Z} \tr \left[e^{-\beta H} x(-a) x(0) \right] = {1 \over Z} \tr \left[ e^{-(\beta + a) H} \,\hat{x}\, e^{a H}\, \hat{x} \right] = \frac{1}{2m\omega} {e^{-(\beta + a) \omega} + e^{a \omega} \over 1 - e^{- \beta \omega}}   \; .
\ee
This is well-behaved, but becomes large for large $\omega$. In the context of quantum field theory, we have similar contributions, but these are summed over frequencies, leading to divergences. For example, considering a free massive scalar field theory on a circle of length $L$, we find that
\be
{1 \over Z} \tr \left[e^{-\beta H} \phi(x=0,\tau = -a) \phi(x=0,\tau = 0) \right] = {1 \over 4 L} \sum_{p=-\infty}^\infty {1 \over \omega_p} {e^{-(\beta + a) \omega_p} + e^{a \omega_p} \over 1 - e^{- \beta \omega_p}}   \; .
\ee
where $\omega_p = \sqrt{p^2 + m^2}$. For positive values of $a$ when the correlator is not in time order, this expression diverges due to the exponentially increasing factor $e^{a \omega_p}$ in the sum over spatial momentum.\footnote{ In the time-ordered case with $a < 0$, both terms in the numerator are suppressed for large $\omega$ and we get the usual thermal correlator.} We expect similar divergences to arise any time we have correlators with local operator insertions followed by backwards Euclidean time evolution.

In the context of our entropy calculations, it is sometimes possible to avoid these divergences by shifting the integration contours for the $s$ variables in (\ref{logrho}). The integrals there are invariant if we shift the contour $s_i \in [-\infty, \infty]$ to $s_i \in [-\infty, \infty] + ia$ so long as the shift does not cause the contour to cross a pole in the kernels ${\cal K}$ where $s_i$ appears; this restricts $a$ to an interval of width $2 \pi$. A shift in $s_i$ by $ia$ leads to a shift in the modular times of all operators appearing in the corresponding $\delta \rho$ by $\tau \to \tau - a$. For specific terms of the form (\ref{Euccor}), this is sometimes sufficient to place all the operators in Euclidean time order, but generally this is not possible; in those cases, the contribution at these specific orders in the sources will be divergent, as we saw in the toy calculation in the previous section.

\subsubsection*{Example: second order in a double-trace source}

As an explicit example, we consider the terms at second order in a double-trace source as in (\ref{DTstate}). The first-order shift in the density matrix arising from these sources is
\be
\label{eq:deltarho}
 \delta^{(1)} \rho_b = - \int dx dy \; \source[2](x,y) \, \left[ {\cal {T}} \left\{ {\cal O}(x) {\cal O}(y) \right\} - \langle {\cal T} {\cal O}(x) {\cal O}(y) \rangle \right] \; .
\ee
The simplest situation where divergences can appear is in contributions to entanglement entropy at second order in this perturbation. This is also the leading contribution to the relative entropy (third term in \eqref{SKeq}), and we can write it explicitly using the definitions above as
{\small
\begin{equation}
\begin{split}
\delta^{(2)} S(\rho||\rho_0)
&=  \tr( \delta^{(1)} \rho \; \delta^{(1)} K)  + \tr( \rho_0 \; \delta^{(2)} K) \cr
&= \int ds \, {\cal K}_1(s) \, \big{\langle} \delta \rho_b \, \rho_0^{\frac{1}{2}-\frac{is}{2\pi}} \delta \rho_b \, \rho_0^{\frac{is}{2\pi}-\frac{1}{2}} \big{\rangle}- \int ds_1 ds_2 \, {\cal K}_2(s_1,s_2) \, \big{\langle} \rho_0^{-\frac{is_1}{2\pi}} \, \delta\rho_b \, \rho_0^{\frac{i(s_1-s_2)}{2\pi}} \, \delta \rho_b \, \rho_0^{\frac{is_2}{2\pi}} \big{\rangle} \\
&= \int dx_1dy_1  dx_2dy_2 \, \source[2](x_1,y_1)  \source[2](x_2,y_2) \, \int ds \; {\cal K}_{1}(s)\,
\left( \frac{1}{2} -\frac{i s}{2\pi}\right) \\
&\qquad \times\Big[ \big{\langle} {\cal T} \{ {\cal O}(\tau_{x_1}+is,\tilde{x}_1) {\cal O}(\tau_{y_1} + i s,\tilde{y}_1)\}\, \,({\cal T} \{{\cal O}(\tau_{x_2}-\pi,\tilde{x}_2){\cal O}(\tau_{y_2}-\pi,\tilde{y}_2) \} \big{\rangle} \\
&\qquad\qquad - \big{\langle} {\cal T} {\cal O}(\tau_{x_1}, \tilde{x}_1) {\cal O}(\tau_{y_1} , \tilde{y}_1 ) \big{\rangle} \big{\langle} {\cal T} {\cal O}(\tau_{x_2}, \tilde{x}_2) {\cal O}(\tau_{y_2}, \tilde{y}_2 ) \big{\rangle} \Big] \; ,
\end{split}
\label{eq:SrelkOO}
\end{equation}
}\normalsize
where all correlators are thermal with respect to the modular Hamiltonian. For simplicity, we can take the source to be of the bilocal form
\be\label{eq:sourcesEx}
\source[2](x,y) = \delta\left(\tau_x-\frac{a}{2}\right)\, \delta\left(\tau_y + \frac{a}{2}\right) \, \delta(\tilde{x}) \, \delta(\tilde{y}) \, ;
\ee
general sources can be taken as linear combinations of similar terms. For this source, the leading contribution to relative entropy simplifies to
{\small
\begin{equation}
\begin{split}
&\delta^{(2)} S(\rho||\rho_0)  \\
&\quad =   \int_{-\infty}^\infty ds \; {\cal K}_{1}\left(s\right) \,
 \left(\frac{1}{2}-\frac{i s}{2\pi}\right) \Big[ \big{\langle} {\cal O}\left(\pi+a+is,0\right){\cal O}\left(\pi+is,0\right) \,
{\cal O}\left(a,0\right){\cal O}\left( 0,0\right) \big{\rangle}- \big{\langle} {\cal O}(a) {\cal O}(0) \big{\rangle}^2 \Big],
\end{split}
\label{eq:SrelkOO2Text}
\end{equation}
}\normalsize
where we have choosen the $s$ contour in \eqref{eq:SrelkOO} along the real line in order to achieve the maximum possible separation of the operators from each other. This corresponds to the case $k=2$ in Fig.\ \ref{fig:kcontours} in Appendix \ref{app:kthorder}.

The correlation function in the first term is time-ordered (hence finite) for small enough $a$. However, for $a$ greater than $\pi$, the Euclidean separations are no longer in time order. There is no way to shift the $s$ contour to fix this, since the first two operators and the latter two operators are each separated by more than $\pi$ on the thermal circle of radius $2 \pi$; thus, it's not possible to fit both pairs of operators on the thermal circle in the proper time order.

The divergence in this out-of-time-ordered correlator can be understood by considering a spectral decomposition of the 4-point function into two 2-point functions. This was also observed in a similar context in \cite{Sarosi:2017rsq}. We shall momentarily present an explicit calculation instead.

We can rephrase this as follows: the thermal correlator appearing in the integrand \eqref{eq:SrelkOO2Text} is analytic in a strip of width $2\pi i$. Any attempt at making $a$ exceed $\pi$ makes it impossible to choose a contour for the $s$ integral that achieves Euclidean time ordering. We can therefore understand the divergence of relative entropy as an artifact of trying to analytically continue beyond this strip of analyticity. We will now illustrate this with an explicit calculation.

\subsubsection*{Explicit calculation}

The perturbation (\ref{eq:deltarho}) with the sources (\ref{eq:sourcesEx}) corresponds at first order in the sources to a perturbation
\be
\label{pertstate}
\rho \equiv \rho_\lambda = \rho_\beta + \lambda (\delta \rho - \rho_\beta \tr(\delta \rho)),
\ee
where $\rho$ is a thermal state
\be
\rho_0 \equiv  \rho_\beta = {1 \over Z_\beta} e^{-\beta H}\,, \qquad \qquad Z_\beta = \tr(e^{-\beta H})\,,
\ee
with respect to the modular Hamiltonian $H=K_0$, and the perturbation takes the form
\be
\begin{split}
\delta \rho &= {1 \over Z_\beta} \;e^{-{1 \over 2} (\beta - a) H } {\cal O} e^{- a H } {\cal O} e^{-{1 \over 2}(\beta - a) H}= \frac{1}{Z_\beta} \; \rho_\beta^{\frac{1}{2}} \, {\cal O}\left(\frac{a}{2},{0}\right) \, {\cal O}\left(-\frac{a}{2},{0}\right) \, \rho_\beta^{\frac{1}{2}} \; .
\end{split}
\ee
As we recalled above, this calculation maps via a conformal transformation to an equivalent calculation where $H$ becomes the Hamiltonian for the CFT on hyperbolic space $\mathbb{H}^{d-1}$. For $d=2$, the hyperbolic space is simply the real line, so our calculation becomes equivalent to calculating the relative entropy for a perturbed thermal CFT on flat space.

As a specific example, we will evaluate \eqref{eq:SrelkOO2Text} explicitly for this case. Here, the two-point function for a dimension-$\Delta$ primary operator takes the form
\be\label{eq:2ptOO}
\big{\langle} {\cal O}(\tau) {\cal O}(0) \big{\rangle} = \left[ {\pi^2 \over \beta^2 \sin^2 \left({\pi \tau \over \beta} \right) }\right]^{\Delta} \,.
\ee
We will furthermore set $\beta = 2\pi$ for simplicity. The integral \eqref{eq:SrelkOO2Text} can then be performed explicitly using elementary techniques.\footnote{ We first factorize the 4-point function in \eqref{eq:SrelkOO2Text} into three products of 2-point functions (one of which is cancelled by the normalization). After dropping the term proportional to $s$ (due to symmetry), and replacing the 2-point functions by the explicit expression \eqref{eq:2ptOO}, we perform the change of variables $s = \log \frac{1-y}{y} $. This reduces the integral in \eqref{eq:SrelkOO2Text} to a known expression, see e.g.\ eq.\ (22) of \cite{Schlosser2013}.} We find:
\begin{equation}
\label{eq:appell}
   \delta^{(2)} S(\rho||\rho_0) = \frac{2^{4\Delta-3}\Gamma(2\Delta+1)^2}{\Gamma(4\Delta+2)} \left[ 1 + F_1(2\Delta+1,\,2\Delta,\,2\Delta,\,4\Delta+2;\,1-e^{ia},\,1-e^{-ia} ) \right] \,,
\end{equation}
where $F_1(a,b_1,b_2,c;x,y)$ is the first Appell series. This Appell series has a singularity precisely when $a \rightarrow \frac{\beta}{2} = \pi$. For example, in the case $\Delta = 1/2$, \eqref{eq:appell} reduces to
\be\label{eq:appell2}
 \delta^{(2)} S(\rho||\rho_0)\Big{|}_{\Delta = 1/2} = {1\over 12} \left(1 - {3  \over 2 \sin^2 \left({a  \over 2}\right)} +{3a \over 4 \sin^3 \left({a  \over 2}\right) \cos \left({a  \over 2}\right) }\right) \,.
\ee
The divergence at $a = \pi$ is evident.\footnote{ The result \eqref{eq:appell} reduces to similar trigonometric expressions for many values of $\Delta$; they always diverge as $a \rightarrow \pi$.}

Our calculation above was performed using real time modular flow techniques. Alternatively, one can use the Euclidean replica trick to compute the relative entropy for a perturbed thermal CFT. In Appendix \ref{sec:replica} we use the replica trick as an independent check for the appearance of divergences. As an example, an explicit result for $\Delta = 1/2$ is computed there, and we recover precisely \eqref{eq:appell2} (see \eqref{eq:appell3}); the replica method results for other dimensions are straightforward to obtain from the results of Appendix \ref{sec:replica} and compare to (\ref{eq:appell}), but we have not done this for general $\Delta$.

We further show in Appendix \ref{sec:replica2} that the divergence is not due to spatially coincident operator insertions: it persists if the operators are separated by a finite amount in space.

In Appendix \ref{app:kthorder} we discuss this phenomenon to higher orders. The upshot is that the $k$-th-order relative entropy, $\delta^{(k)} S(\rho ||\rho_0)$ diverges as operator insertions associated with the source $\source[2](x,y)$ are separated by an amount $a$ that exceeds $\frac{\beta}{k} = \frac{2 \pi}{k}$:
\begin{equation}
\label{eq:divSk}
\boxed{
   S(\rho|| \rho_0 ) \big{|}_{\lambda^k} \; \longrightarrow \infty \qquad \text{when} \qquad a \rightarrow \frac{\beta}{k}\,.
   }
\end{equation}
This means that the radius of convergence of the expansion in $\lambda$ is zero.

These conclusions will apply to any nonlocal multi-trace source $\source[i]$, since the finite support of the source in Euclidean time is the cause of the divergence.  However, even for local sources $\source[1]$, the divergence will still occur, albeit at higher orders.  If we deform the state by $\source[1](x) {\cal O}(x)$ and expand to second order, we get
\be
\delta^{(2)} \rho_b = \int dx dy \; \source[1](x) \source[1](y) \, \left[ {\cal {T}} \left( {\cal O}(x) {\cal O}(y) \right) - \langle {\cal T} {\cal O}(x) {\cal O}(y) \rangle \right],
\label{eq:deltarholambda1squared}
\ee
which has the same form as (\ref{eq:deltarho}) above and will hence produce divergences in $\delta^{(4)} S(\rho||\rho_0)$.\footnote{  Unlike for multi-trace sources, the disconnected diagrams in the relative entropy will vanish for perturbations of the form (\ref{eq:deltarholambda1squared}), so to see a divergence explicitly it is necessary to look at the connected four-point function. This is consistent with the fact that the disconnected terms would be ${\cal O}(N^4)$, and thus have to cancel for consistency.}

\subsection{Origin of the divergences}

We have seen that in the calculation of entanglement entropy or relative entropy in perturbation theory in the sources, divergences can appear at fixed orders in perturbation theory. Our discussion in \S\ref{sec:nonanalytic} suggests that these may signal a breakdown of the naive perturbation theory due to non-analyticities (e.g.~logarithmic terms) in the correct perturbative expansion.

Even the specific behavior of the divergences as a function of the parameter $a$ matches with our toy model calculation in \S\ref{sec:nonanalytic}. The behavior in (\ref{eq:divSk}), where the $\lambda^k$ contribution becomes divergent when $a$ increases to $\beta/k$, is exactly the same as in the toy model expression (\ref{eq:toyfull}), in which the $\lambda^k$ term in the naive expansion diverges for $ a \ge \beta/k$.

As further evidence that the divergences we see here are not physical, we can argue directly that the relative entropy calculated in the previous subsection must be a finite quantity.\footnote{ An abstract argument for this was pointed out to us by G.\ Sarosi \cite{Sarosi:2017rsq}: the relative entropy between two states in the same Hilbert space can only diverge if the supports of the spectra are different. In reasonable quantum field theories this is never the case.} The relative entropy is equal to
\be
S(\rho || \rho_0 ) = \Delta \langle H_0 \rangle - \Delta S \,,
\ee
where $H_0 = - \log \rho_\beta$ is the thermal modular Hamiltonian. In the example discussed in the previous subsection, we can calculate directly the change in energy, which gives
\be
\Delta \langle H_0 \rangle = \lambda {d \over d \beta} \Delta_\beta (a)
\ee
and is finite. So the only way for the relative entropy to be infinite is for $\Delta S$ to be $- \infty$. But (for finite volume), the entropy of the thermal state is finite, and the entropy of the perturbed state must be positive, so $\Delta S$ cannot be $- \infty$. However, we can check that the divergence is still present if we use finite volume propagators.\footnote{
The breakdown of perturbation theory that we have seen might be related to the observations in \cite{Lashkari:2018oke,Lashkari:2018tjh}, where it is noted that the naive perturbative expansion in $\delta\rho_b$ is not rigorously well-defined since $\delta \rho_b$ is an unbounded operator.}

\section{CFT entanglement entropy and the quantum RT formula}
\label{sec:quantum1}

We have argued that states defined via the Euclidean path integral with general multi-trace sources correspond to bulk states with a good semiclassical description but a more general structure of entanglement than the single-trace states considered previously. For these states, it is interesting to understand the bulk gravitational physics at the quantum level.

We would like to check, with as few assumptions as possible, that the CFT entanglement entropy for a ball-shaped region can be reproduced by an auxiliary gravity calculation via the quantum RT formula \eqref{QRT}.
An interesting case where this has been accomplished at the quantum level is the paper \cite{Belin:2018juv}, which demonstrates such a matching in the case of a single scalar particle in AdS${}_3$, where the CFT region is taken to be a small interval. 

Alternatively, assuming that there is some bulk state for which the quantum RT formula gives the entanglement entropy of a well-defined CFT state, we would like to see that this bulk state must satisfy some quantum version of the gravitational constraints. As discussed in the introduction, it is interesting to understand to what extent these constraints continue to be local beyond first order in perturbation theory in the case where the bulk state is non-coherent, since the bulk entanglement $\Delta S^{bulk}$ in this case is non-vanishing and cannot be written as the integral of a local expression.

The goal of this section is to review and develop some tools that should be useful in pursuing the directions that we have just outlined, and to provide some helpful intermediate results. We leave a more detailed investigation for future work. For an interesting alternative approach to relating the quantum RT formula and the quantum gravitational constraints, see \cite{Lewkowycz:2018sgn}.

\subsection{Basic identities}

Our arguments will rely on a variety of identities, which we now discuss individually before assembling them to reach various conclusions. We will consider a one-parameter family of CFT states $|\Psi(\lambda) \rangle$, where $|\Psi(0) \rangle$ is the vacuum state. The generic examples we have in mind are states created by perturbative multi-trace sources in the Euclidean path integral. In this case, we can think of each source $\source[n](x_1,\ldots, x_n)$ as some function of $\lambda$ that vanishes for $\lambda = 0$.  Similarly, we consider a one-parameter family of quantum states for some gravitational theory on an asymptotically AdS spacetime, with curvature scale chosen so that CFT entanglement entropies for ball-shaped regions are correctly computed via the RT formula in this spacetime. We will describe the bulk state more explicitly below.

\subsubsection*{The quantum RT formula}

The quantum RT formula (\ref{QRT}) equates the entanglement entropy for a spatial region in a CFT with the ``generalized entropy,'' i.e.\ area plus bulk entropy, of a corresponding region in the gravitational theory, as we discussed in the introduction.\footnote{ For general holographic CFTs, the area can be corrected by higher derivative terms associated with higher derivative terms in the appropriate gravitational action for the dual theory. For simplicity, our discussion in this section will consider the case where Einstein gravity coupled to matter provides the correct description so that the higher derivative contributions are absent. However, much of the discussion goes through in the more general case. } This has been demonstrated in \cite{Faulkner:2013ana} up to order $N^0$ assuming the AdS/CFT correspondence. Subsequently, higher-order quantum corrections have been studied more systematically by also taking into account the necessity to correct the position of the extremal surface in the large $N$ expansion \cite{Engelhardt:2014gca,Dong:2017xht}. These derivations made use of the AdS/CFT correspondence, which we don't want to assume just yet. Thus, for our one-parameter family of CFT/gravity states and a region $A$ in the CFT, we will define the quantity
\be
qRT \equiv - {d \over d \lambda} S^{CFT}_A + {1 \over 4 G_N} {d \over d \lambda} \langle \widehat{\text{Area}}(\tilde{A}) \rangle + {d \over d \lambda} S^{bulk}_{\Sigma} \; .
\ee
The vanishing of $qRT$ is the content of the quantum corrected Ryu-Takayanagi formula.


\subsubsection*{CFT relative entropy, entanglement entropy, and modular energy}

For any one-parameter family of CFT states $|\Psi(\lambda) \rangle$ and any ball-shaped region $A$, we have\footnote{  We define ${\bs \epsilon}_{a_1 \cdots a_k} = \frac{1}{(d+1-k)!} \, \sqrt{-g} \, \varepsilon_{a_1 \cdots a_k b_1 \cdots b_{d+1-k}} \, dx^{b_1} \wedge \ldots \wedge dx^{b_{d+1-k}}$ with the convention that $\varepsilon_{ztx_1 \cdots x_{d-1}} =1$. This induces natural volume forms on lower dimensional surfaces via interior contraction with appropriate normal vectors.
}
\be
\label{RECFT}
{d \over d \lambda} S^{CFT}_A =  {d \over d \lambda} \langle \hat{K}^{vac}_A \rangle  - {d \over d \lambda} S(\rho_A || \rho_A^{vac}) \; ,
\ee
where $\hat{K}^{vac}_A = - \log(\rho_A)$ is the vacuum modular Hamiltonian for the region $A$: this follows directly from the definition of relative entropy. The vacuum modular Hamiltonian for ball-shaped regions is given for any CFT by
\be
\label{boundaryK}
\hat{K}^{vac}_A =\int_A \zeta_A^\mu\, \hat{T}_{\mu \nu} \,{\bs \epsilon}^\nu \; .
\ee
where $\zeta_A^\mu$ is the conformal Killing vector naturally associated with the domain of dependence region for the ball $A$, given explicitly in \cite{Faulkner:2013ica}.

\definecolor{Green}{RGB}{0,170,0}
\definecolor{Blue}{RGB}{0,100,210}
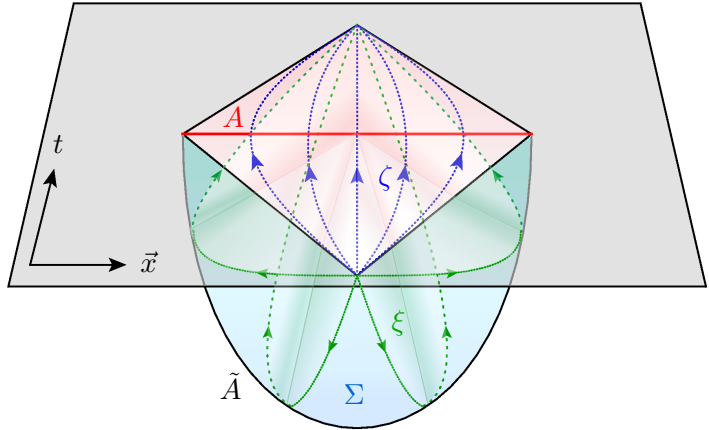
\begin{figure}
\begin{center}
\includegraphics[width=.6\textwidth]{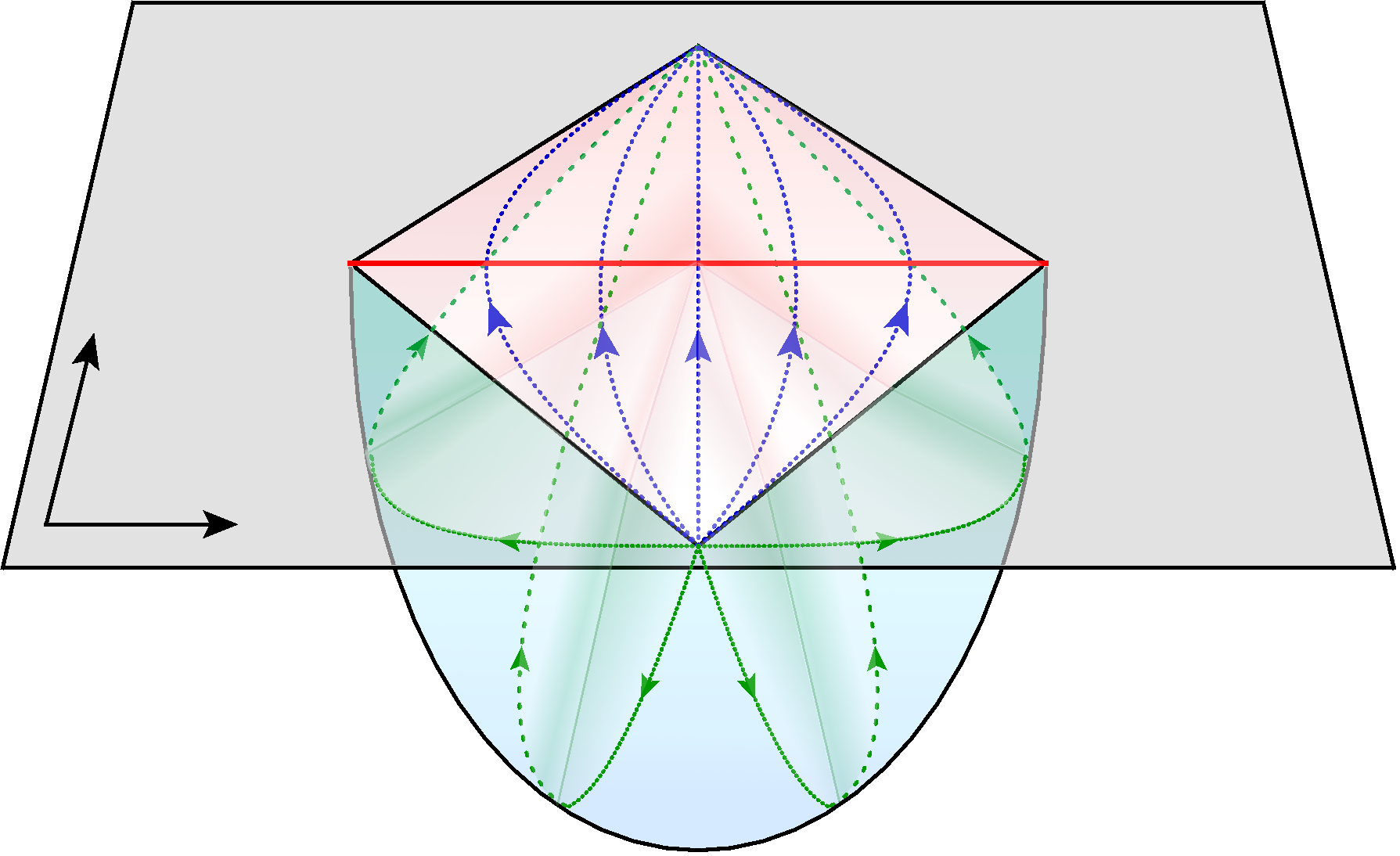}
\put(-248,105){$t$}
\put(-215,60){$\vec{x}$}
\put(-184,115){${\color{red}A}$}
\put(-185,12){$\tilde{A}$}
\put(-138,10){${\color{Blue}\Sigma}$}
\put(-120,38){${\color{Green}\xi}$}
\put(-125,93){${\color{blue}\zeta}$}
\end{center}
\caption{Bulk geometry associated with a ball-shaped region $A$ at the boundary, and Killing boost generating vector fields. The codimension one bulk surface $\Sigma$ is such that $\partial \Sigma = A \cup \tilde{A}$.}
\label{fig:bulk}
\end{figure}

\subsubsection*{Bulk relative entropy, entanglement entropy, and modular energy}

Next, we describe a bulk result analogous to (\ref{RECFT}). First, for {\it any} quantum field theory on AdS, the vacuum modular Hamiltonian for a half-space region $\Sigma$ (bounded by an extremal surface $\tilde{A}$ ending on a ball) is given as\footnote{  This is the AdS analogue of the universal result in Minkowski space that the modular Hamiltonian for a half space is the boost generator.}
\be
\label{bulkMod}
\hat{K}^{vac}_{\Sigma} = \frac{1}{16\pi G_N} \int_{\Sigma} \xi^a\, \hat{T}_{a b} \,{\bs \epsilon}^b \; ,
\ee
where $\xi$ is a Killing vector naturally associated with the bulk domain of dependence of the region $\Sigma$ (see \cite{Faulkner:2013ica} for an explicit expression). In terms of this, we have that (for a quantum field theory on AdS)
\be
\label{REbulk}
{d \over d \lambda} S^{bulk}_{\Sigma} =  {d \over d \lambda} \langle \hat{K}^{vac}_{\Sigma} \rangle  - {d \over d \lambda} S^{bulk}(\rho_{\Sigma} || \rho_{\Sigma}^{vac}) \; .
\ee

In our case, the bulk theory is gravitational, so things are more complicated. For a given one-parameter family of states, the bulk geometry may differ (at least quantum-mechanically) as the parameter is varied. To make contact with the quantum RT formula, we wish to consider the bulk entanglement entropy and relative entropy inside the surface that extremizes the generalized entropy appearing in the quantum RT formula (see also \cite{Engelhardt:2014gca}). We expect that there is a choice of gauge, analogous to the classical ``Hollands-Wald'' gauge \cite{Hollands:2012sf}, for which this extremal surface remains at the same coordinate location so we are dealing with a fixed region $\Sigma$. We assume that it is possible to define a density matrix $\rho_{\Sigma}$ for the bulk fields in this region such that it is possible to compare $\rho_{\Sigma}$ to $\rho_{\Sigma}^{vac}$ and compute a relative entropy. In this case, the definition of relative entropy again leads to the identity (\ref{REbulk}). However, now the expression (\ref{bulkMod}) for the bulk modular Hamiltonian must be improved to include gravitational contributions. We will discuss the correct expression for the bulk modular Hamiltonian below.

\subsubsection*{The JLMS formula}

In \cite{Jafferis:2015del}, it was argued using the AdS/CFT correspondence that there is a direct relation between boundary and bulk modular Hamiltonians, given by
\be \label{JLMS}
-\hat{K}_A + \hat{K}_{\Sigma} + {1 \over 4 G_N} \widehat{\text{Area}}(\tilde{A}) = 0 \,,
\ee
where the area is treated as a quantum operator. We won't assume this relation, but it will be useful for us to define the object
\be
\langle JLMS \rangle \equiv - {d \over d \lambda}\langle \hat{K}^{vac}_A \rangle + {d \over d \lambda}\langle \hat{K}^{vac}_{\Sigma} \rangle + {1 \over 4 G_N}  \frac{d}{d\lambda}\langle \widehat{\text{Area}}(\tilde{A}) \rangle \; .
\ee
To get this we choose the version of \eqref{JLMS} where the modular Hamiltonians are those of regions in the vacuum state, then take the expectation value of both sides in the deformed state $|\Psi(\lambda)\rangle$.

\subsubsection*{Equivalence of boundary and bulk relative entropies}

In \cite{Jafferis:2015del}, it was also argued that the boundary relative entropy for a region $A$ in the CFT is equal to the bulk relative entropy for the corresponding region $\Sigma$ in the associated gravitational theory. This was argued to hold to first subleading order in large $N$, i.e., in situations where the quantum extremal surface in the deformed state is the same as in the original state \cite{Dong:2017xht}. It is unclear whether it should be expected to hold at higher orders in $1/N$. We will argue for this more directly below using a specific construction for bulk states. For now, we define the difference as
\be
RE = {d \over d \lambda} S^{CFT}(\rho_A || \rho_A^{vac}) - {d \over d \lambda} S^{bulk}(\rho_{\Sigma} || \rho_{\Sigma}^{vac}) \; .
\ee

\subsubsection*{Interlude: relations so far}

Based on the definitions so far, we see that by making use of the identities (\ref{RECFT}) and (\ref{REbulk}), we have that
\be
\label{JLMS1}
qRT = \langle JLMS \rangle + RE \; .
\ee
For a given CFT state, we can verify the quantum RT formula by showing that there is a state in some corresponding gravitational theory for which the terms on the right side vanish individually. We will make progress on this below, but first we recall a few more relations that will be useful.

\subsubsection*{The holographic dictionary for the stress-energy tensor: equivalence of CFT modular energy and gravitational boundary energy}

A standard part of the AdS/CFT correspondence is the relation between the CFT stress tensor expectation value and the asymptotic metric. Expressed in Fefferman-Graham coordinates (where the metric operator perturbation about pure AdS is $\hat{h} = \ell^2 z^{d-2} \hat{\Gamma}_{\mu \nu}(z,x) dx^\mu dx^\nu$), this reads
\be
\label{HD}
\langle \hat{T}_{\mu \nu}(x) \rangle = {d \ell^{d-3} \over 16 \pi G_N}  \langle \hat{\Gamma}_{\mu \nu}(z=0,x) \rangle \,.
\ee
As explained in \cite{Lashkari:2013koa}, this is implied by the Ryu-Takayanagi formula, by considering the limit where the boundary region is taken to be an infinitesimal ball. We will use this relation to connect the quantum expectation value of the boundary modular Hamiltonian with a gravitational quantity. We define
\be
ST \equiv   {d \over d \lambda}\langle \hat{{\cal E}}_{grav} \rangle  - {d \over d \lambda} \langle \hat{K}_A^{vac} \rangle \,,
\ee
where
\be
\hat{\cal E}^{grav} =  {d \ell^{d-3} \over 16 \pi G_N} \int_A \zeta_A^\mu \, \hat{\Gamma}_{\mu \nu}(z=0)  \, {\bs \epsilon}^\nu \; .
\ee
Whenever the extrapolate holographic dictionary (\ref{HD}) holds, we have $ST=0$.

\subsubsection*{Quasi-local energy, area, and boundary energy}

Given any theory of gravity on an asymptotically AdS spacetime, it was shown in \cite{Lashkari:2016idm} that there is a natural definition of energy $H_A$ that we can associate to the entanglement wedge of a ball-shaped boundary region $A$. Classically, the energy $H_A$ in our one-parameter family of states $|\Psi(\lambda)\rangle$ is defined covariantly via an integral over $\Sigma$:
\be
\begin{split}
\label{eq:HAomega}
\frac{d}{d\lambda} H_A &= \frac{d}{d\lambda} \int_\Sigma {\bs J}_\xi(g,\phi) - \int_{\partial \Sigma} i_\xi \,{\bs \theta}\big(g, {d g \over d \lambda};\phi, {d \phi \over d \lambda}\big) \\
&= \int_{\Sigma} {\bs \omega}\big({d \over d \lambda} g, {\cal L}_{\xi} g\big)  +\int_{\Sigma} {\bs\omega}_\phi \big({d \over d \lambda} \phi, {\cal L}_{\xi} \phi\big) - \int_{\Sigma}  i_\xi  \left[ {\bf E}(g,\phi) \, {d g \over d \lambda} + {\bf E}^\phi(g,\phi) \, \frac{d\phi}{d\lambda}\right]  \,.
\end{split}
\ee
where ${\bs J}_\xi$ is the $d$-form Noether current associated with $\xi$-diffeomorphisms, and ${\bs \theta}$ is the symplectic potential that arises as the boundary term when varying the bulk Lagrangian \cite{Iyer:1994ys}. Further, ${\bs \omega}$ and ${\bs \omega}_\phi$ are the symplectic current $d$-forms for the gravitational and matter sectors, respectively. For Einstein gravity minimally coupled to a scalar field with mass $m$, the equations of motion ${\bf E}_{ab} \equiv E_{ab} \, {\bs \epsilon}$ and ${\bf E^\phi} \equiv E^\phi \, {\bs \epsilon}$ comprise the Einstein equation and a scalar field equation of motion:
\be
 E_{ab}(g,\phi) = \frac{1}{16\pi G_N} (G_{ab} - \frac{1}{2} T_{ab})\,, \qquad
 E^\phi(g,\phi) = \frac{1}{16\pi G_N} (\Box - m^2) \phi \,,
 \ee
 where $T_{ab}$ is the usual scalar field stress-energy tensor and $G^{ab}$ the Einstein tensor including cosmological constant.
For more details on this formalism, as well as explicit expressions for various terms, we invite the reader to consult \cite{Faulkner:2017tkh} or \cite{Lashkari:2016idm}.

For a one-parameter family of asymptotically-AdS metrics, the energy $H_A$ is related to the quantities that we have defined previously by
\be
\label{HWclassical}
{d \over d \lambda} H_A ={d \over d \lambda}  {\cal E}_{grav}  - {1 \over 4 G_N} {d \over d \lambda}  \text{Area}(\tilde{A})  -  2\int_{\Sigma} \xi^a  \left[ {d \over d \lambda} {E}_{a b}(g,\phi) \right] {\bs \epsilon}^b   \,,
\ee
where $E_{ab}(g)$ is the gravitational equation of motion (see above).
 This identity holds off-shell for Einstein gravity coupled to matter, but also holds for more general theories of gravity if the area here is replaced by a more general (Wald) entropy functional.

Quantum-mechanically, there should be some version of this identity that holds as an operator relation. For our purposes, we only need it at the level of expectation values, so we will define the  ``Hollands-Wald'' \cite{Hollands:2012sf} combination
\be
\label{HW}
HW = {d \over d \lambda} \langle \hat{H}_A  \rangle - {d \over d \lambda}  \langle \hat{{\cal E}}_{grav}  \rangle  + {1 \over 4 G_N} {d \over d \lambda}  \langle \widehat{\text{Area}}(\tilde{A}) \rangle  + 2 \int_{\Sigma} \xi^a \,   \left[{d \over d \lambda} \langle \hat{E}_{ab} \, \rangle \right] \,{\bs \epsilon}^b \; .
\ee
This vanishes in classical gravity as a consequence of Noether's theorem for diffeomorphism invariance \cite{Hollands:2012sf}.
In the present case, we leave open for now the possibility that HW is non-vanishing due to quantum effects. However, we will argue below that, at least perturbatively in our multi-trace path integral states, $HW$ can indeed be shown to vanish.

\subsubsection*{Equivalence of the bulk modular Hamiltonian and the quasi-local energy}

%

Finally, we return to the bulk modular Hamiltonian. In quantum field theory, the expression (\ref{bulkMod}) has the interpretation as the energy in the region $\Sigma$, using the vector field $\xi$ as the associated time. However, it is challenging to generalize this to a gravitation theory where the spacetime is dynamical.

As we have discussed above, the natural analogue of the region $\Sigma$ in a gravitational theory is the entanglement wedge region inside the quantum extremal surface. This provides a covariant ``subregion'' defined for states that may have different classical geometries. At the classical level, there is a natural gravitational energy $H_A$ associated with such a region, as we have discussed above. Naively, we might then expect that the gravitational version of the modular Hamiltonian is a quantum operator version of the gravitational energy $H_A$. Thus, we might expect the vanishing of the expression
\be
KH = {d \over d \lambda} \langle \hat{K}_{\Sigma}^{vac} \rangle - {d \over d \lambda} \langle \hat{H}_{A} \rangle \,.
\ee
However, there are a number of subtleties here that complicate the situation.

Importantly, it is a challenge even to precisely define the Hilbert space on which the modular Hamiltonian would act.\footnote{ The presence of diffeomorphism gauge symmetry implies that the quantum gravitational Hilbert space does not factorize into spatial subregions, so the density matrix and modular Hamiltonian for a subregion is ill-defined \cite{Casini:2013rba}.  By  specifying
an algebra of gauge-invariant observables associated with the subregion, one can define
a density matrix associated with this subalgebra.
This algebraic definition is ambiguous up to a choice of center for the subalgebra,
and it is unlikely that any choice of subalgebra
leads to an entropy that agrees with
the replica trick when the gauge symmetry is
nonabelian (as is true with diffeomorphisms) \cite{Donnelly:2014new}.  Instead, it appears necessary
to extend the Hilbert space by edge mode degrees of freedom on the boundary,
and these degrees of freedom will contribute to the modular Hamiltonian. An additional
complication in gravitational theories is that diffeomorphism symmetry can change the coordinate
location of the entangling surface, and it can be nontrivial to specify the subregion of
interest in a gauge-invariant manner for different geometries.}

When considering linearized graviton and matter field perturbations around a fixed
background, the situation is somewhat improved.  The gravitons can then be
treated as free spin-2 particles, which have a sensible quantum field theory interpretation.\footnote{ It is still necessary to introduce edge modes to match the replica entropy,
but since linearized diffeomorphism symmetry is abelian, their quantization
should be fairly analogous to the abelian gauge field, considered, for example, in
\cite{Donnelly2015E, Donnelly2016W}.}
In this picture, the modular Hamiltonian can be defined through equation
(\ref{bulkMod}), where the linearized graviton stress appears in addition to the matter field
stress tensor.  The graviton stress tensor is ambiguous up to the addition of total derivatives,
which would contribute boundary terms to (\ref{bulkMod}).\footnote{  These should be resolved
in conjunction with the edge mode contribution, so that the entropy coincides with the
replica entropy for a spin-2 field, computed in \cite{Solodukhin2015}.} Such boundary terms were discussed in \cite{Hollands:2012sf}. Including them, more complete expression for the perturbative energy up to quadratic order in the graviton field is known as the canonical energy; this is also the quadratic version of $\hat{H}_{A}$, so the vanishing of $KH$ appears to hold perturbatively at this order.

\subsubsection*{Master identity}

Starting from (\ref{JLMS1}) and making use of the subsequent definitions, we now have that
\be
\langle JLMS\rangle = ST + HW + KH + EoM \; ,
\ee
where the integrated Einstein equations are encoded in
\be
EoM = -2 \int_\Sigma \xi^a  \, \left[ {d \over d \lambda} \langle \hat{E}_{ab}  \rangle \right] \, {\bs \epsilon}^b \,.
\ee
Thus, we have shown that
\be
\label{eq:qRTfinal}
\boxed{
qRT = ST + HW + KH + RE + EoM\, .
}
\ee
The relation \eqref{eq:qRTfinal} is our main formula to analyze. As written, \eqref{eq:qRTfinal} is a tautology, but it provides an interesting relation between statements of physical interest, which may or may not be independently true, depending on the context.

We would like to understand the extent to which the various terms in \eqref{eq:qRTfinal} vanish for multi-trace states without assuming the full holographic dictionary. This amounts a derivation of non-trivial aspects of holography from first principles in this class of states.\footnote{  It is instructive to remind the reader of the derivation of classical second order Einstein equations in \cite{Faulkner:2017tkh,Haehl:2017sot}: there, the classical analog of \eqref{eq:qRTfinal} was used, where all objects are truncated at order $N^2$. At the classical level, one clearly has $ST = HW = KH = 0$. This was used to derive the Einstein equations $EoM=0$ from \eqref{eq:qRTfinal}.
More concretely, the non-trivial calculation performed in \cite{Faulkner:2017tkh} can be thought of as an explicit demonstration that $RE =  HW + EoM$ at order $N^2$ for single-trace states at second order in sources for any conformal field theory. Assuming $ST = HW = KH = 0$ and the Ryu-Takayanagi formula, the fundamental relation \eqref{eq:qRTfinal} then reduces to the vanishing of integrated Einstein equations, $EoM=0$, which amounts to a universal derivation of second-order gravitational dynamics.}

\subsection{Quantum-gravitational constraints}

In this section, we would like to understand the quantum-gravitational constraints on a spacetime which correctly calculates the entanglement entropy of a CFT state via the quantum RT formula.

We will assume that for the one-parameter family of CFT states $|\Psi(\lambda)\rangle$, we have found a one-parameter family of states of a corresponding gravitational theory, described by asymptotically AdS geometry $M(\lambda)$ and bulk quantum state $|\Psi^{bulk}(\lambda) \rangle$ for the fields on $M(\lambda)$. We assume that the gravitational state correctly computes the CFT entanglement entropy via the quantum RT formula \cite{Faulkner:2013ana,Engelhardt:2014gca}. In this case, starting from the basic identity (\ref{eq:qRTfinal}), we have that $qRT = ST = 0$, leaving
\be
\label{eq:ident2}
HW + KH + RE + EoM = 0.
\ee

The simplest expectation is that for the correct choice of state in the gravity picture, $HW$, $KH$, and $RE$ are all zero. This would imply that the usual gravitational equations hold at the level of expectation values, $EoM=0$. We will be able to show this directly for certain terms in the perturbative expansion in sources, which is already very interesting as it corresponds to a derivation of Einstein equations in the presence of quantum effects at specific perturbative orders. However, we don't have an argument in general to show $EoM=0$. This leaves open the interesting possibility that $HW + KH + RE$ could be non-vanishing in general; in this case, the non-vanishing part would represent some corrections to the quantum-gravitational equations.

An interesting point here is that while $EoM$ corresponds to an integrated local quantity, some of the remaining terms are explicitly nonlocal. In particular, the bulk relative entropy can be understood as the nonlocal part of vacuum-subtracted bulk entanglement entropy. Ending up with a completely local constraint requires that these nonlocal terms cancel completely. In the next subsection, we will argue directly for such a cancellation in the term $RE$ by showing directly that bulk relative entropy cancels with boundary relative entropy in this term for a suitable choice of state in the gravitational theory.

It is very interesting to explore carefully any possible residual source of nonlocality, as some authors have suggested the need for some degree of nonlocality in quantum-gravitational physics, for example, in order to resolve the black hole information paradox (see e.g., \cite{Mathur:2002ie,Giddings:2006sj,Giddings:2012gc,Bao:2017who}).

The remainder of this section serves two purposes: $(i)$ to demonstrate that in the multi-trace Euclidean path integral states we can argue for the vanishing of various terms in \eqref{eq:qRTfinal} from first principles; and $(ii)$ to use this argument to derive the ``quantum'' Einstein equations $\langle \hat{E}_{ab} \rangle = 0$ at certain perturbative orders, while also elucidating how one should understand these equations.

\subsubsection{Direct argument for the equality of relative entropies}
\label{sec:RE0}

In this subsection, we argue directly that for holographic CFT states defined via the Euclidean path integral with general perturbative sources, there is a corresponding state in the associated gravitational theory for which the relation $RE=0$ holds, i.e.~for which the CFT relative entropy for a ball-shaped region (relative to its vacuum entropy) is equal to the bulk relative entropy for the fields contained by the quantum extremal surface. In our perturbative framework expectation values in the multi-trace states are linear combinations of vacuum correlators. After defining the bulk state appropriately, the statement $RE=0$ thus reduces to an application of the standard extrapolate dictionary of AdS/CFT in the vacuum state, which we will assume holds true.

For any set of sources $\{\source[n]\}$ in the CFT, we can define a bulk quantum state via a Euclidean path integral in the bulk quantum theory (to the extent that such a path integral is well-defined). We insert sources in this path integral that are closely related to those of the CFT. In particular, for any CFT source, we insert a corresponding source in the bulk path integral whose support in the radial direction is localized near the asymptotic AdS boundary and whose dependence on the remaining coordinates matches that of the CFT sources.

Concretely, recalling the definition  \eqref{PIstate2} of the CFT state, we define a corresponding bulk state by:
\be
\label{PIstateBulk}
\langle \varphi_0 | \Psi_\lambda^{bulk} \rangle = \int^{\phi(\tau=0) = \varphi_0}_{\tau<0} [d \phi(\tau)] \;e^{-S_E^{bulk} - S^{bulk}_{\{\source[n]_{bulk}\}}} \; .
\ee
Here, $\source[n]_{bulk}$ are Euclidean bulk sources for the bulk fields; we take the bulk sources to be localized near the AdS boundary and equal to the CFT sources, up to an appropriate scaling.  For instance, for a scalar field of dimension $\Delta$ on Poincar\'{e} AdS with boundary at $z=0$, if we choose
\be
\source[n]_{bulk}(x_1,z_1;\ldots;x_n,z_n) = \epsilon^{n (d+1-\Delta)} \source[n](x_1,\ldots,x_n) \prod_{i=1}^n \delta(z_i - \epsilon) \, ,
\ee
we find that
\be
\label{ExtDic}
\lim_{\epsilon \to 0} \epsilon^{-m \Delta} \langle \hat{\phi}(x_1, \epsilon) \dots \hat{\phi}(x_m, \epsilon) \rangle_{\{\source[n]_{bulk}\}} = \langle \mathcal{O}(x_1) \dots \mathcal{O}(x_m) \rangle_{\{\source[n]\}}\,,
\ee
assuming the extrapolate dictionary for correlators on the vacuum.\footnote{ More precisely, we are assuming here the Euclidean version of the extrapolate dictionary for vacuum correlators, as developed, for instance, in \cite{Skenderis:2008dg,Christodoulou:2016nej}. Note that an analogous step was involved in the derivation of nonlinear Einstein equations in \cite{Faulkner:2017tkh}: there, the authors start by writing the two-point function that computes second order relative entropy as the asymptotic value of a bulk field in a single-trace state. This is essentially the extrapolate dictionary for vacuum two-point functions. Note, however, that two-point functions are entirely determined by kinematics; in the present context we assume the extrapolate dictionary for higher-point vacuum correlation functions.}

Now, the perturbative calculation of the bulk relative entropy for the region $\Sigma$ is formally identical to the calculation in the CFT (i.e. the last term in (\ref{SKeq}), making use of (\ref{logrho})), with the replacements
\be
{\cal O}_\alpha(x) \to \epsilon^{-\Delta_\alpha} \phi_\alpha(x,\epsilon)\,, \qquad \qquad \rho_0^{CFT} \to \rho_0^{bulk} \; .
\ee
In the expression (\ref{logrho}), the operators $\rho_0^{CFT}$ gave rise to a geometrical flow associated with the conformal Killing vector $\zeta$. The final result for each term in the perturbative expansion was a Euclidean correlator of operators at spatial locations and complex times; see e.g.~\eqref{eq:SrelkOO2Text}. The action of $\rho_0^{bulk}$ on bulk operators is also via a geometrical flow, corresponding to the Killing vector $\xi$ described above. But near the AdS boundary, where all the bulk fields $\phi_\alpha$ appear, the Killing vector $\xi$ coincides with the vector $\zeta_A$, so the action of $\rho_0^{bulk}$ on the fields $\phi_\alpha$ is precisely the same as the action of $\rho_0^{CFT}$ on the operators ${\cal O}_\alpha$. Thus, assuming the validity of the extrapolate dictionary (\ref{ExtDic}), we will find that $S_{rel, bulk} = S_{rel, CFT}$ (and therefore $RE = 0$) order by order in the perturbative expansion.

Referring back to the discussion at the end of the previous subsection, we see that a significant source of potential nonlocality in the quantum gravitational constraints is removed.

An interesting point is that at leading order in $N$, we may replace the full correlators in the perturbative expansion of the CFT relative entropy with just their two-point (or possibly three-point) disconnected pieces. Then, the bulk relative entropy depends only on bulk quantities that are derivable from the CFT using symmetry principles alone.  We should then be able to make the following statement, similar to the results in \cite{Faulkner:2017tkh}: given a CFT state defined by perturbative Euclidean sources $\{\source[i]\}$, we can build an auxiliary AdS geometry $\mathcal{M}$ and auxiliary quantum field theory state of fields $\phi$ such that, to leading order in $N$, $RE = 0$ perturbatively in $\{\source[i]\}$.  Moreover, for the appropriate choice of $\{\source[i]\}$ (e.g., choosing only bilocal sources), the bulk relative entropy will be non-trivial, in the sense it is not just equal to the canonical energy or some other geometric quantity.  The auxiliary bulk $\{ \mathcal{M}, |\Psi_\lambda^{bulk} \rangle \}$ that gives this result is the same one that properly calculates other relevant quantities in the CFT.

Our argument for the equality of relative entropies is valid only to the extent that the bulk path integral can be treated as a quantum field theory path integral. It would be useful to understand better when this breaks down, and investigate the possible residual terms in $RE$.\footnote{ In particular, these may come about due to quantum corrections that affect the location of the extremal surface \cite{Dong:2017xht,Engelhardt:2014gca}. Here, we will mostly be concerned with the first subleading effects in $1/N$, where this observation is not relevant. However, when searching for possible nonlocal contributions to the quantum gravitational equations of motion, these effects might be important.}

\subsubsection{Quantum-corrected Einstein equation: first order in the graviton operator}
\label{sec:qEE}

Assume now that we are in a situation where $RE=0$, e.g. where the argument in the previous section holds, or where we are considering any first order variation in the state.\footnote{ Recall that the leading contribution to relative entropy is second order in the state perturbation.}

We thus have that $HW + KH +  EoM = 0$, or explicitly:
\be
\label{qrtcheck0}
{d \over d \lambda} \langle \hat{K}_{\Sigma}^{vac} \rangle - {d \over d \lambda}  \langle \hat{{\cal E}}_{grav}  \rangle  + {1 \over 4 G_N} {d \over d \lambda}  \langle \widehat{\text{Area}}(\tilde{A}) \rangle = 0 \,.
\ee
The various operators appearing in expectation values can be expanded perturbatively in field operators. For example, the area operator expanded in terms of the metric perturbation field $\hat{h}$ has linear terms, quadratic terms, etc... . At a fixed order in the sources, the terms with different powers of the field operator typically contribute at different orders in $1/N$. In this section, we will see that it is simple to show that the gravitational equations hold at the level of expectation values at any order in sources and $1/N$ for which
\begin{enumerate}
\item
Only terms linear in the metric perturbation contribute to the last two terms in (\ref{qrtcheck0}).
\item
The bulk modular Hamiltonian is the integral of a local quantity, as in (\ref{bulkMod}).
\end{enumerate}
When the first condition is satisfied, we can write \eqref{qrtcheck0} as
\be
\label{qrtcheck0b}
{d \over d \lambda} \langle \hat{K}_{\Sigma}^{vac} \rangle - {d \over d \lambda}  \langle \hat{{\cal E}}_{grav}^{(1)}  \rangle  + {1 \over 4 G_N} {d \over d \lambda}  \langle \widehat{\text{Area}}^{(1)}(\tilde{A}) \rangle = 0 \,.
\ee
Next, we can make use of the purely gravitational operator identity
\be
\label{WaldId}
-  \hat{{\cal E}}_{grav}^{(1)} +  {1 \over 4 G_N}  \widehat{\text{Area}}^{(1)}(\tilde{A})  = - \frac{1}{8\pi G_N}\int_\Sigma \xi^a \,  \hat{G}^{(1)}_{ab} \, {\bs \epsilon}^b \,,
\ee
where superscript ${}^{(1)}$ refers to linearization in the graviton operator.
This follows immediately from the corresponding classical identity which is a linearized version of the classical Hollands-Wald identity (\ref{HWclassical}), aka Noether's theorem for $\xi$-diffeomorphisms. More explicitly, we consider a linearized graviton fluctuation on the AdS background: $\hat{g} = g_0 + \hat{h}$ such that  \eqref{WaldId} should be understood as the linearized version of \eqref{HWclassical} lifted to an operator statement. Note that this linearized identity is purely gravitational: the right-hand side involves the linearized Einstein tensor $\hat{G}^{(1)}_{ab}$ as opposed to the full Einstein equation.

Subtracting the expectation value of \eqref{WaldId} from \eqref{qrtcheck0b}, we find
\be
{d \over d \lambda} \left[ \langle  \hat{K}_{\Sigma}^{vac} \rangle - \frac{1}{8\pi G_N} \int_\Sigma \xi^a \,\big{\langle} \hat{G}_{ab}^{(1)}    \big{\rangle}  \,{\bs \epsilon}^b\right]  = 0 \; .
\ee
Finally, using the expression $(\ref{bulkMod})$, we arrive at the conclusion that the linearized Einstein equations hold at the level of expectation values for the specific orders in perturbation theory where the various conditions above are satisfied.\footnote{ An additional argument is required to pass from the integrated equations to the local equations, but this is the same as in \cite{Faulkner:2013ica}.} If we write the linearized metric operator as $\hat{g} = g_0 + \hat{h}$, these equations read as follows:
\be
\label{eq:qEinstein}
\boxed{
 \frac{d}{d\lambda} \left[ G^{(1)}_{ab}( \langle \hat{h} \rangle)  - \frac{1}{2} \,  \big{\langle}\hat{T}_{ab} \big{\rangle} \right]  = 0\,.
   }
\ee
We will refer to this as the first-order {\it quantum Einstein equation}. In particular, it is first order in $\hat{h}$. However, if the state under consideration is a multi-trace Euclidean path integral state, then this tadpole equation for $\langle \hat{h} \rangle$ makes sense at any order in the sources and is thus fully nonlinear in the state. This equation was derived for general first order perturbations in \cite{Swingle:2014uza}, but we will see that the present derivation applies in some cases not covered by \cite{Swingle:2014uza}.

\subsubsection*{Example: Double-trace sources for scalar primary operators}

We now present an example of a situation where the conditions of the previous section apply. Consider the case of a state created by a double-trace source $\source[2]$ for a scalar primary operator ${\cal O}$. In the bulk, this corresponds to source terms of the form $\iint \source[2]_{bulk} \, \hat{\phi} \hat{\phi}$ as described in \S\ref{sec:RE0}.
Every term in \eqref{qrtcheck0} then corresponds to some correlation functions of the graviton $\hat{h}$ with the scalar field operators $\hat{\phi}$.

It is easy to see that only terms linear in $\hat{h}$ contribute to \eqref{qrtcheck0} at order $N^0$. The $N$ scaling is counted in a way analogous to the situation at the boundary (see \S\ref{sec:Nscaling}).\footnote{ In fact, one could use the HKLL prescription \cite{Hamilton:2005ju,Hamilton:2006az} to write a boundary representation of the bulk operator $\hat{h}$ in terms of a smearing of the boundary stress tensor. All bulk correlation functions then turn into boundary correlation functions to which the discussion in \S\ref{sec:Nscaling} applies.
}
In particular an expectation value $\langle \hat{h} \rangle$ in a double-trace state scales as $N^{-2}$ at any order in $\source[2]_{bulk}$. This contributes to the quantities in \eqref{qrtcheck0} at order $N^0$.
At $k$-th order in the source, the bulk expectation value $\langle \hat{h} \rangle$ thus becomes a $(2k+1)$-point correlation function that scales as $N^0$. Any further graviton operator insertions lead to suppression in $1/N$. We can think of this diagrammatically as follows:
\vspace{.5cm}
\begin{equation*}
\label{eq:graph5}
\langle \hat{h} \rangle \quad=  \qquad
\begin{gathered}
\includegraphics[width=.6\textwidth]{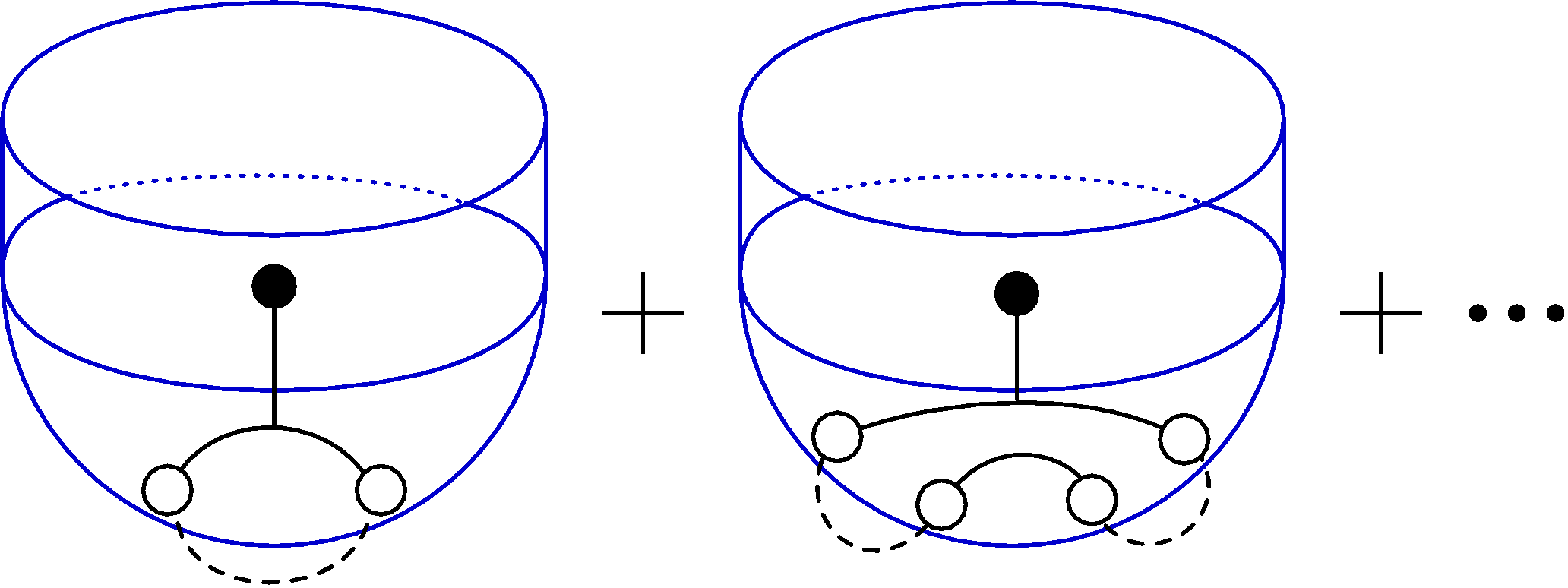}
\end{gathered}
\end{equation*}
which represents the contributions at first and second order in $\source[2]_{bulk}$, respectively.
In this expression, the solid dot stands for the graviton insertion, and the white dots represent operator insertions due to the double-trace bulk sources (the dotted lines). Solid lines represent Wick contractions at leading order in large $N$.  This notation is completely analogous to the one used in \S\ref{sec:Nsquared}. From the analysis in that section, we can immediately infer from the presence of a loop that the above diagram is subleading in the large $N$ expansion compared to classical gravity. Indeed, the above diagram has one loop and correspondingly represents an ${\cal O}(N^0)$ contribution to \eqref{qrtcheck0}. The presence of more than one loop will lead to further suppression in $1/N$. Similarly, any  higher-point functions of $\hat{h}$ will be subleading, as well.

More explicitly, at any order in $\source[2]_{bulk}$ there exist diagrams with only a single loop, such that they contribute at order $N^0$ in \eqref{eq:qEinstein}. An example of such a diagram would be the following contribution to the Einstein tensor term:
\begin{equation*}
\label{eq:graph5}
\left. G^{(1)}_{ab}( \langle \hat{h} \rangle) \right|_{{\cal O}((\source[2]_{bulk})^n)} \quad \ni  \qquad
\begin{gathered}
\includegraphics[width=.25\textwidth]{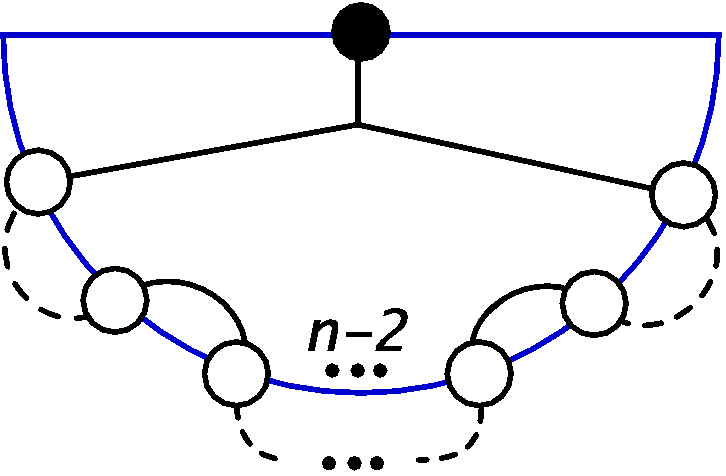}
\end{gathered}
\end{equation*}
In this sense one should understand \eqref{eq:qEinstein} as a tadpole equation for $\langle \hat{h}\rangle$ at any order in the source. For similar remarks in a general AdS/CFT setup, see also \cite{Faulkner:2013ana,Engelhardt:2014gca,Dong:2017xht}.

Using diagrams such as the one given above, one can illustrate the contributions to the quantum Einstein equations at any desired order in $\source[2]_{bulk}$ and establish that they all contribute at the same order in $1/N$. More generally, any such diagram will be of order $(N^2)^{1-L-(I-1)}$, where $L$ is the number of loops, and $I$ is the number of external operator insertions (one for $\hat{h}$ and two for the matter stress tensor). This makes it easy to identify the relevant diagrams at any given order in the large $N$ expansion.

\subsubsection{Second order ``quantum" Einstein equations}

In order to derive from the master identity \eqref{eq:qRTfinal} the quantum equations of motion that include backreaction due to gravitons, we need a state in which we require expansion of \eqref{qrtcheck0}  to at least quadratic order in the graviton operator. We would then write \eqref{qrtcheck0} as
\be
\label{qrtcheck0c}
{d \over d \lambda} \left\{  \frac{1}{16\pi G_N} \int_{\Sigma} \xi^a\, \langle\hat{T}_{a b}(\hat{\phi},\hat{\phi}) \rangle \,{\bs \epsilon}^b -   {{\cal E}}_{grav}^{(1)}(\langle\hat{h}  \rangle)  + {1 \over 4 G_N}  {\text{Area}}^{(1)}(\langle\hat{h}\rangle)+\langle {\text{Area}}^{(2)}(\hat{h},\hat{h}) \rangle \right\} = 0 \,,
\ee
where superscripts indicate the order of expansion in $\hat{h}$ around the background.

An example of a state where \eqref{qrtcheck0c} captures all the relevant terms at leading order in large $N$ is given by the state created by a double-trace source for the stress tensor. We shall consider this example below and will denote the source as $\source[2]_T$ to distinguish it from the primary operator sources used previously. It is straightforward to see that in the presence of a stress tensor source the terms in \eqref{qrtcheck0c} that involve two-point functions of $\hat{h}$ contribute at the same order as the one-point function of $\hat{h}$ and the two-point function of $\hat{\phi}$.

To derive the quantum equation of motion in this setup we would like to combine \eqref{qrtcheck0c} with an appropriate version of $HW=0$ (analogous to \eqref{WaldId}).
The main task is thus to establish $(i)$ the quantum generalization of the Hollands-Wald formula ($HW=0$) at higher orders in operators, and $(ii)$ the equality of relative entropies, $RE=0$. These steps deserve further study and we hope to address them in the future. Here we shall simply explore some consequences that would follow.

We anticipate the following statement as a consequence of expanding \eqref{HW} to quadratic order in operators:
\be
\begin{split}
\label{HW22}
0 = HW^{(2)}
&= {d\over d\lambda} \bigg\{ \langle \hat{H}_A \rangle
-  {\cal E}_{grav}^{(1)}(\langle\hat{h}\rangle)
 + {1 \over 4 G_N} \left[ \text{Area}^{(1)}(\langle\hat{h}\rangle) + \langle {\text{Area}}^{(2)}(\hat{h},\hat{h}) \rangle  \right] \\
&\qquad\quad + \frac{1}{8\pi G_N} \int_{\Sigma} \xi^a \,   \left[  G^{(1)}_{ab}(\langle \hat{h}\rangle) + \big{\langle} G_{ab}^{(2)}(\hat{h},\hat{h})  - \frac{1}{2} \, T_{ab}(\hat{\phi},\hat{\phi})
\big{\rangle} \right] \,{\bs \epsilon}^b \bigg\}\,,
\end{split}
\ee
where we already dropped some terms involving equations of motion for the AdS background solution and the tadpole $\langle \hat{h} \rangle$. We kept all terms that are relevant at order $N^0$ for any number of sources $\source[2]_T$. We are being very cavalier about properly defining any of the terms quadratic in operators.\footnote{ However, note a nice feature of our multi-trace states: due to their normalization,  UV divergences that naively appear for two-point functions of coincident operators are automatically regulated. For instance, $\langle \hat{h}(X)\hat{h}(X)\rangle_{\source[2]} \sim \iint \source[2]_{bulk}(Y,Z) \, \langle \hat{h} (X)\hat{h}(Y) \rangle \langle \hat{h}(X) \hat{h}(Z) \rangle$. The coincident operators end up being separated by large $N$ factorization.} One might be able to verify (or falsify) this relation using arguments similar to those in \cite{Jafferis:2015del}. A more rigorous approach would be to lift the gravitational Noether theorem to a quantum Ward identity and find a way to generalize it to the case \eqref{HW22}.

Using the vanishing of \eqref{HW22} in our basic identity \eqref{eq:ident2} (and further setting $KH=RE=0$), would yield the following equations of motion:
\be
\label{eq:EOM2}
 \frac{d}{d\lambda}   G_{ab}^{(1)} (\langle \hat{h} \rangle)
   = \frac{d}{d\lambda}  \Big{\langle} G_{ab}^{(2)}(\hat{h},\hat{h})  - \frac{1}{2} \, T_{ab}(\hat{\phi},\hat{\phi})\Big{\rangle}     \,.
\ee
These equations of motion describe the first non-trivial backreaction on the classical geometry due to graviton and matter fluctuations in multi-trace states. We will refer to them as \emph{second order ``quantum'' Einstein equations}.

The first term on the right hand side of the equation of motion \eqref{eq:EOM2} is rather interesting: it captures the backreaction of two-point graviton correlations on the dynamics of the geometry. This case contains genuinely new features as it illustrates how gravitons essentially backreact like a complicated set of matter fields. It also goes beyond merely an explicit illustration of the abstract arguments given previously in \cite{Swingle:2014uza} (see also \cite{Dong:2017xht}), where these effects had not been captured.\footnote{ However, this quantum effect can often be too suppressed at large $N$ for our purposes. A situation where it does contribute at order $N^0$ is in Euclidean path integral states with a double-trace source for the stress tensor, $\source[2]_T(x,y)$. One can check that for primary operator sources $\source[2]_{\cal O}(x,y)$ the first term on the right hand side of \eqref{eq:EOM2} is subleading in $1/N$ compared to the other terms.}

\section{Summary and discussion}
\label{sec:discussion}

In this paper we studied CFT states constructed by performing a Euclidean path integral in the presence of nonlocal multi-trace sources, \eqref{PIstate2}. These states provide an interesting generalization of previously studied states associated with single-trace sources because their entanglement structure is genuinely different from that of coherent excitations of the vacuum. We investigated the structure of such states using perturbation theory both in $1/N$ and in the sources.
One of our main motivations was to learn about the extent to which properties of these states can be geometrized using ideas from AdS/CFT. The findings of this work can be distilled into three main statements:

$(i)$ At leading order in large $N$, correlation functions and entanglement entropy in multi-trace states can be equivalently captured by a different state that only contains a composite effective single-trace source (see, for example, \eqref{STsource}). In the context of AdS/CFT, this implies that to leading order in $1/N$ any quantity that can be geometrized in single-trace states under some assumptions about the theory, can also be geometrized in multi-trace states under the same assumptions. The most well-known example of this is the statement that for {\it any} CFT there exists a geometry which captures entanglement entropies of ball-shaped regions up to second order in sources and additionally satisfies the gravitational equations of motion.

$(ii)$ At subleading orders in the $1/N$ expansion, perturbation theory in sources has surprising features. Using the replica trick, real-time methods, and toy models, we argued that the appearance of divergences in the perturbative expansion in sources is generic (see, for example, \eqref{eq:divSk}). We argued that these divergences are not physical, but rather an artifact of non-analytic terms in the perturbative expansion at subleading orders in $1/N$. The fact that non-analytic dependence on sources cannot be expected to have a geometrical interpretation in AdS/CFT motivates the necessity of invoking bulk entanglement in the dual quantum-gravitational description.

$(iii)$ Finally, we studied the constraints on bulk quantum gravity states imposed by the requirement that they capture the entanglement entropies of ball-shaped regions in multi-trace states at the first subleading order in $1/N$. Our master identity \eqref{eq:qRTfinal} provides a relation between different quantum-gravitational constraints that one can study independently in perturbation theory. We gave direct arguments for the emergence of a quantum generalization of Einstein's equation in the context of multi-trace states (see \eqref{eq:qEinstein} and \eqref{eq:EOM2}). This involves demonstrating the existence of a bulk quantum state which captures the entanglement entropies to subleading orders in $1/N$, and deriving the statement that expectation values of fields in this state necessarily satisfy quantum equations of motion.

\subsubsection*{Comments and future directions}

It is interesting to understand how general the states are that we study in this paper. We expect these states to be very generic in the sense that allowing for arbitrary $n$-point sources should allow one to give arbitrary values to all $n$-point correlation functions in the theory. Such generality and the fact that there is a systematic way to study the multi-trace Euclidean path integral states, would imply that these states provide a formidable environment to study the detailed mechanisms of AdS/CFT. As a first step towards establishing the generality of these states, one could extend the work \cite{Marolf:2017kvq} to the present setup and give a precise map between multi-trace sources and bulk excitations.

Our understanding of non-analyticities in the perturbative expansion of entropies in these states is incomplete and calls for further studies. A possible avenue would be to try and improve the perturbative expansion by invoking more rigorous methods such as those developed in \cite{Lashkari:2018oke,Lashkari:2018tjh}. It would also be interesting to see these features appear explicitly from a calculation of entanglement entropy between bulk degrees of freedom.

We provided a rather general and therefore somewhat abstract derivation of the quantum Einstein equations at certain perturbative orders in sources and $1/N$, see \eqref{eq:qEinstein}, \eqref{eq:EOM2}. It would be nice to have a more general and perhaps more explicit calculation (similar to either \cite{Faulkner:2017tkh,Haehl:2017sot} or  \cite{Belin:2018juv}) which shows the emergence of these equations from purely CFT considerations. Here, it is necessary to understand more precisely how to define bulk relative entropy and bulk modular Hamiltonians, taking into account changes in the location of the quantum extremal surface coming from the quantum extremality conditions \cite{Engelhardt:2014gca,Dong:2017xht}. It will also be important to establish more completely the appropriate quantum version of the Hollands-Wald identity. A strong motivation for such future investigations to explore the tantalizing possibility of nonlocal contributions to the gravitational constraints due to bulk entanglement at some order in perturbation theory, or perhaps non-perturbatively.

\section*{Acknowledgements}
We would like to thank Alexandre Belin, Eliot Hijano, Ted Jacobson, Nima Lashkari, Aitor Lewkowycz, Mukund Rangamani, Gabor Sarosi, Brian Swingle, Tomonori Ugajin for enlightening discussions and comments. We are especially grateful to Onkar Parrikar for initial collaboration and useful discussions. FH would like to thank the Mainz Institute for Theoretical Physics, the KITP Santa Barbara, and the Aspen Center for Physics for hospitality and support during the course of this project. Research at the KITP is supported in part by the National Science Foundation under Grant No. NSF PHY-1748958. The Aspen Center for physics is supported by National Science Foundation grant PHY-1607611. The work of FH and MVR is supported in part by the Simons Collaboration {\it It From Qubit} and a Simons Investigator Award (MVR).
JP is supported in part by the Simons Foundation and in part by the Natural Sciences and Engineering Research Council of Canada. Research at Perimeter Institute is supported by the Government of Canada through Industry Canada and by the Province of Ontario through the Ministry of Research \& Innovation.
\appendix

\section{General correlation functions in the effective geometry}
\label{sec:HigherPoint}

In \S\ref{sec:NonLocalMulti} we discussed the relation between partition functions for multi-trace path integral states and effective sources. We note that a connected $n$-point function can be written, using the result \eqref{OnePt}, as
\beas
\label{eq:corrr}
\langle {\cal O}(x_1) \cdots {\cal O}(x_n) \rangle^{conn.}_{\{\source[i]\}} &=&{\delta \over \delta \source[1](x_1)} \cdots {\delta \over \delta \source[1](x_n)} \log Z(\{\source[i]\}) \cr
&=& {d \over d \source[1](x_1)} \cdots {\delta \over \delta \source[1](x_{n-1})} {\delta \over \delta \lambda_{eff}(x_n)} \log Z(\lambda_{eff}) \; .
\eeas
Diagrammatically, we expect that to leading order in $1/N$ this correlation function should be given by the sum of all possible trees with $n$ external operator insertions and any number of internal insertions of multi-trace sources. Any loop in such a diagram will lead to an additional suppression by $1/N^2$. We will now prove this.

The source $\source[1]$ is not a natural object from the point of view of the theory of the effective coupling. The goal here is to express correlators such as \eqref{eq:corrr} entirely in terms of derivatives with respect to $\lambda_{eff}$. For simplicity, let us restrict to only a single-trace and a double-trace source, so that the effective coupling is
\be
\label{eq:lambdaeff}
\lambda_{eff} (x) = \source[1](x) +  \int dy\, \source[2](x,y) \langle \mathcal{O} (y) \rangle_{\lambda_{eff}} \, .
\ee

\subsubsection*{Two-point functions}

Let us first discuss two-point functions in this setup. We start from
\be
\langle \mathcal{O}(x_1) \mathcal{O}(x_2) \rangle_{\{\source[i]\}}^{conn.} = \frac{d}{d \source[1](x_1)} \frac{d}{d \lambda_{eff} (x_2)} \log Z(\lambda_{eff}) \, .
\label{twopointl1leff}
\ee
We can formally remove the unnatural dependence on $\source[1]$ by calculating
\be
\frac{\delta}{\delta \source[1](x_1)} = \int dy\; \frac{\delta \lambda_{eff}(y)}{\delta \source[1](x_1)}\, \frac{\delta}{\delta \lambda_{eff}(y)} \, .
\ee
To obtain this, we first use \eqref{eq:lambdaeff} to show that
\be
\begin{split}
\frac{\delta \source[1](x_1)}{\delta \lambda_{eff}(x_2)} & = \delta (x_1 - x_2) -  \int dy\; \source[2](x_1, y) \frac{\delta \langle \mathcal{O}(y) \rangle_{\lambda_{eff}}}{\delta \lambda_{eff}(x_2)}
\\ & = \delta (x_1 - x_2) - \int dy \;\source[2](x_1, y)  \langle \mathcal{O}(y) \mathcal{O}(x_2) \rangle_{\lambda_{eff}} \, ,
\end{split}
\label{dl1/dleff}
\ee
where $\langle \mathcal{O}(y) \mathcal{O}(x_2) \rangle_{\lambda_{eff}}$ is the two-point correlator in the theory with the effective coupling.  We now need to invert this expression.  This can be done formally by thinking of ${\delta \source[1](x_1) \over \delta \lambda_{eff}(x_2)}$ as a matrix with indices $x_1$ and $x_2$.  The inversion then takes the form
\be
\begin{split}
& \left [ \delta (x_1 - x_2) -  \int dy\; \source[2](x_1, y)  \langle \mathcal{O}(y) \mathcal{O}(x_2) \rangle_{\lambda_{eff}} \right ]^{-1} \\ & \quad \quad = \delta (x_1 - x_2) +  \int dy_1 \;\source[2](x_1, y_1)  \langle \mathcal{O}(y_1) \mathcal{O}(x_2) \rangle_{\lambda_{eff}}
\\ & \quad \quad \quad \  +  \int dy_1 dy_2 dy_3\, \Big ( \source[2](x_1, y_1)  \langle \mathcal{O}(y_1) \mathcal{O}(y_2) \rangle_{\lambda_{eff}}\; \source[2](y_2, y_3)  \langle \mathcal{O}(y_3) \mathcal{O}(x_2) \rangle_{\lambda_{eff}} \Big )
\\ & \quad \quad \quad \  + \dots .
\end{split}
\ee
This inverse is well-defined as long as the second term in \eqref{dl1/dleff} is small compared to the first in a matrix sense.  If $\source[2]$ is perturbatively small, this is naively true. Then \eqref{twopointl1leff} can be written as
{\small
\be
\label{eq:OOconn}
\langle \mathcal{O}(x_1) \mathcal{O}(x_2) \rangle_{\{\source[i]\}}^{conn.} = \int dy_1 \,\langle \mathcal{O}(x_1) \mathcal{O}(y_1) \rangle_{\lambda_{eff}} \left [ \delta (y_1 - x_2) -  \int dy_2 \source[2](y_1, y_2)  \langle \mathcal{O}(y_2) \mathcal{O}(x_2) \rangle_{\lambda_{eff}} \right ]^{-1} \, ,
\ee
}\normalsize
which is expressed entirely in terms of correlators in the effective theory.  If we expand this expression, we get a sum of terms where each term corresponds to a different number of $\source[2]$s separating the two operator insertions. Diagrammatically, we can write \eqref{eq:OOconn} as
\vspace{.5cm}
\begin{equation*}
\label{eq:graph5} \includegraphics[width=.95\textwidth]{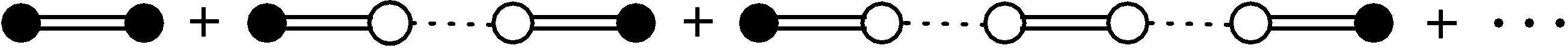}
\end{equation*}
As before, solid circles are the basic operator insertions ${\cal O}(x_1)$ and ${\cal O}(x_2)$, empty circles are operator insertions coming from expanding the exponential in the path integral defining the state, and dashed lines connect the two operators of a bilocal source. A double line, however, is a propagator in the background of the effective source, instead of a vacuum propagator.  To expand it in terms of vacuum propagators and the $\source[1]$ and $\source[2]$ sources, it is necessary to connect an entire tree structure of insertions to each double line.  Thus we recover exactly the sum we expected from analyzing the problem without using an effective source: \eqref{eq:OOconn} is given by the sum of all possible tree diagrams with two solid circles and any number of sources.

\subsubsection*{Higher-point functions}

The expansion of general higher-point correlators in terms of an effective single-trace coupling proceeds similarly to the two-point function case discussed above. The one new qualitative feature is that additional $\source[1]$ derivatives can now act on both $\log Z(\lambda_{eff})$ and on the ${\delta \lambda_{eff}(x_1) \over \delta \source[1](x_2)}$.

As an illustration, start with the definition of $\lambda_{eff}$ as
\be
\lambda_{eff} (x) = \source[1](x) +  \int dy \,\source[2](x,y) \langle \mathcal{O} (y) \rangle_{\lambda_{eff}} +  \int dy_1 dy_2\, \source[3](x,y_1,y_2) \langle \mathcal{O} (y_1) \rangle_{\lambda_{eff}} \langle \mathcal{O} (y_2) \rangle_{\lambda_{eff}} \, .
\ee
Then
\be
\begin{split}
\frac{\delta\source[1](x_1)}{\delta \lambda_{eff}(x_2)} &  = \delta (x_1 - x_2) -  \int dy\, \source[2](x_1, y)  \langle \mathcal{O}(y) \mathcal{O}(x_2) \rangle_{\lambda_{eff}}
\\ & -  \int dy_1 dy_2\, \source[3](x_1,y_1,y_2) \langle \mathcal{O} (y_1) \mathcal{O} (x_2) \rangle_{\lambda_{eff}} \langle \mathcal{O} (y_2) \rangle_{\lambda_{eff}}
\\ & -  \int dy_1 dy_2\, \source[3](x_1,y_1,y_2) \langle \mathcal{O} (y_1)  \rangle_{\lambda_{eff}} \langle \mathcal{O} (y_2) \mathcal{O} (x_2) \rangle_{\lambda_{eff}} \, .
\end{split}
\ee
The desired quantity ${\delta\lambda_{eff}(x_1)\over\delta \source[1](x_2)}$ is the matrix inverse of this.  We are interested in the object
\be
\label{npointl1leff}
\begin{split}
&\langle \mathcal{O}(x_1) \dots \mathcal{O}(x_n) \rangle_{\{\source[i]\}} = \frac{\delta}{\delta \source[1](x_1)} \cdots  \frac{\delta}{\delta \source[1](x_{n-1})} \frac{\delta}{\delta \lambda_{eff} (x_n)} S_{bulk}(\lambda_{eff}) \\
&\qquad\quad= \int dy_1 \dots dy_{n-1}\,\left( \frac{\delta \lambda_{eff}(y_1)}{ \delta \source[1](x_1)} \frac{\delta}{\delta \lambda_{eff}(y_1)} \right)\cdots \frac{\delta \lambda_{eff}(y_{n-1})}{ \delta \source[1](x_{n-1})} \langle \mathcal{O}(y_{n-1}) \mathcal{O}(x_n)\rangle_{\lambda_{eff}}\, .
\end{split}
\ee
The diagrammatics should be understood as follows.  The right-most factor
\be
\frac{\delta \lambda_{eff}(y_{n-1})}{ \delta \source[1](x_{n-1})} \langle \mathcal{O}(y_{n-1}) \mathcal{O}(x_n)\rangle_{\lambda_{eff}}
\ee
is the two point function calculated in \eqref{eq:OOconn}.  It has the tree structure discussed there.  Then, each factor of
\be
\int dy_i \,\frac{\delta \lambda_{eff}(y_i)}{\delta\source[1](x_i)} \frac{\delta}{\delta \lambda_{eff}(y_i)}
\ee
adds to the old tree an entire subtree of correlators and sources connected to a single new operator insertion.  If the $\lambda_{eff}$ derivative hits a ${\delta\lambda_{eff}(x_1)\over \delta \source[1](x_2)}$, this subtree is inserted somewhere on the subtree associated with a previous operator insertion.  If it instead hits $\langle\mathcal{O}(x_n)\rangle_{\lambda_{eff}}$, the subtree is added to the tree associated with the original propagator.  This set of derivatives thus iteratively builds up all possible tree-like decorations of a tree with $n$ operator insertions, as expected.

\section{Divergent relative entropy for perturbed CFT thermal states}

\label{sec:replica}

In this section we provide details of the calculation of the leading order relative entropy for the thermal state of a CFT on $\mathbb{R}^{d-1}$ perturbed by a similar state with operators inserted at $\tau = \pm a$ on the thermal circle $[-\beta/2, \beta/2]$. For $d=2$, this is equivalent by a conformal transformation to the leading order relative entropy for an interval in a CFT perturbed by a bi-local double trace source, as described in \S\ref{sec:quantum2}.

\subsection{Relative entropy in thermal states using replica trick}
\label{sec:replica0}

The perturbation (\ref{eq:deltarho}) with the sources (\ref{eq:sourcesEx}) corresponds at first order in the sources to a perturbation
\be
\label{pertstate}
\rho \equiv \rho_\lambda = \rho_\beta + \lambda (\delta \rho - \rho_\beta \tr(\delta \rho)),
\ee
where $\rho$ is a thermal state
\be
\rho_0 \equiv  \rho_\beta = {1 \over Z_\beta} e^{-\beta H}\,, \qquad \qquad Z_\beta = \tr(e^{-\beta H})\,,
\ee
with respect to the modular Hamiltonian $H=K_0$, and the perturbation takes the form
\be\label{eq:resDeltaHalf}
\begin{split}
\delta \rho &= {1 \over Z_\beta} \;e^{-{1 \over 2} (\beta - a) H } {\cal O} e^{- a H } {\cal O} e^{-{1 \over 2}(\beta - a) H}= \frac{1}{Z_\beta} \; \rho_\beta^{\frac{1}{2}} \, {\cal O}\left(\frac{a}{2},{0}\right) \, {\cal O}\left(-\frac{a}{2},{0}\right) \, \rho_\beta^{\frac{1}{2}} \; .
\end{split}
\ee
for some local operator ${\cal O}$ of dimension $\Delta$.
As we recalled above, this calculation maps via a conformal transformation to an equivalent calculation where $H$ becomes the Hamiltonian for the CFT on hyperbolic space $\mathbb{H}^{d-1}$. For $d=2$, the hyperbolic space is simply the real line, so our calculation becomes equivalent to calculating the relative entropy for a perturbed thermal CFT on flat space. In this case, it is possible to calculate the leading order contribution to relative entropy
\be
S(\rho||\rho_0) \equiv \tr(\rho \log(\rho)) - \tr(\rho \log (\rho_0)),
\ee
very explicitly. We will now perform this calculation (generalizing slightly to perturbations of the CFT on $\mathbb{R}^{d-1}$) using the replica trick


To calculate the second-order relative entropy, we will use the replica trick to consider
\be
\delta^{(2)}S(\rho || \rho_\beta )  = \left[\lim_{n \to 1} {d \over dn} \tr(\rho_\lambda^n) \right]_{\lambda^2} \; .
\ee
We have that
\beas
\tr(\rho_\lambda^n)|_{\lambda^2} &=& \lambda^2 {1 \over 2} n (n-1) \left[ {\rm Str}(\rho_\beta^{n-1} \delta \rho \delta \rho) - 2 \tr(\rho_\beta^{n-1} \delta \rho) \tr(\delta \rho) + \rho(\rho_\beta^n) \tr(\delta \rho)^2 \right],
\eeas
where ${\rm Str}$ indicates a trace symmetrized over the possible orderings.

We define
\be
\Delta_\beta(a) \equiv {1 \over Z_\beta} \tr\left(e^{-(\beta - a)H}{\cal O} e^{-a H} {\cal O}\right) = \frac{1}{Z_\beta} \, \langle {\cal O}(a,0){\cal O}(0,0) \rangle_\beta \; ,
\ee
i.e., the thermal two-point function for operators ${\cal O}$ at the same spatial position separated by distance $a$ on the thermal circle, and
\be
K_\beta(a,b,c) = {1 \over Z_\beta} \tr(e^{-(\beta - a - b - c)H}{\cal O} e^{-a H} {\cal O} e^{-b H} {\cal O} e^{-c H} {\cal O}),
\ee
the similarly-defined four point function. Then we have
\be
\tr(\rho_\lambda^n)|_{\lambda^2} = \lambda^2 {Z_{n \beta} \over Z_\beta^n} \left[ {1 \over 2} n \sum_{k=1}^{n-1} K_{n\beta}(a,(k-1)\beta,a) - n(n-1) \Delta_\beta(a) \Delta_{n \beta}(a) + {1 \over 2} n(n-1) \Delta_\beta(a)^2 \right].
\ee
From this, we obtain
\be
\left[ {d \over d n} \tr(\rho_\lambda^n)|_{n \to 1} \right]_{\lambda^2} = -{1 \over 2} \Delta_{\beta}(a)^2  + {1 \over 2} {d \over d n} \left[n \sum_{k=1}^{n-1} K_{n\beta}(a,(k-1)\beta,a) \right]_{n \to 1},
\ee
For a large-$N$ CFT, we expect that the leading contribution to the four-point function comes from disconnected contributions that can be written in terms of a pair of two-point functions. Specifically, we have
\be
K_{n\beta}(a,(k-1)\beta,a) \sim \Delta_{n \beta}^2(a) + \Delta_{n \beta}^2(k\beta) + \Delta_{n \beta}(k\beta + a) \Delta_{n \beta}(k\beta - a)  \; .
\ee
where the $\sim$ indicates that we are dropping the connected part which comes suppressed by powers of $N^2$. Using this, we get
\be
\label{res1}
\left[ {d \over d n} \tr(\rho_\lambda^n)|_{n \to 1} \right]_{\lambda^2} =  {1 \over 2} {d \over d n} \left[n \sum_{k=1}^{n-1} (\Delta_{n \beta}^2(k\beta) + \Delta_{n \beta}(k\beta + a) \Delta_{n \beta}(k\beta - a))\right]_{n \to 1}.
\ee
Let's now specialize further, to the case where the CFT is defined on $\mathbb{R}^{1,1}$. In this case, the propagator can be written explicitly as
\be
\Delta_\beta(a) = \left[ {\pi^2 \over \beta^2 \sin^2 \left({\pi a \over \beta} \right) }\right]^{\Delta}\,,
\ee
where we have absorbed any overall normalization factor into the definition of $\lambda$. Then from (\ref{res1}), we find for operator dimension $\Delta$:
\be
\begin{split}
\left[ {d \over d n} \log \tr(\rho_\lambda^n)|_{n \to 1} \right]_{\lambda^2}
&= \frac{1}{2} \frac{d}{dn} \left[ n \, \left( \frac{\pi}{n\beta}\right)^{4\Delta} \left( C_{2\Delta} (n;0) + C_{2\Delta}(n;\frac{a\pi}{\beta}) \right) \right]_{n\rightarrow 1} \, ,
\end{split}
\ee
where we have defined
\be
C_p(n;x) = \sum_{k=1}^{n-1} \left[ {1 \over \sin \left({k \pi \over n} + x \right) \sin \left({k \pi \over n} - x \right)} \right]^p \; .
\ee
To compute these sums, we can use the result \cite{Chen:2001}
\be
\label{eq:sinsum}
\sum_{k=1}^{n-1} {1 \over \sin^2 \left({k \pi \over n} \right) - x^2} = {1 \over x^2} - {n \cot(n \arcsin x) \over x \sqrt{1 - x^2}} \; .
\ee
From this, we obtain the generating function
\bea
G_n(x,t) \equiv \sum_{p=1}^\infty t^{2 p} C_p(n;x) &=& \sum_{k=1}^{n-1} {t^2 \over |\sin^2 \left( {k \pi \over n} \right) - \sin^2x | - t^2} \cr
&=&  {t^2 \over \sin^2x + t^2} - t^2 {n \cot[n \arcsin \sqrt{ \sin^2 x + t^2 }] \over \sqrt{\sin^2x + t^2} \sqrt{\cos^2x - t^2}},
\eea
where we have dropped the absolute value in the first line, as its argument turns out to be always positive anyway.
From this, we find that
\bea
\sum_{p=1}^\infty t^{2 p} \lim_{n \to 1} C_p(n;0) &=& 0 \cr
\sum_{p=1}^\infty t^{2 p} \lim_{n \to 1} {d \over d n} C_p(n;0) &=& -1 + {\arcsin(t) \over t \sqrt{1-t^2}}  \; .
\eea
From these generating functionals, we have
\bea
\lim_{n \to 1} C_p(n;0) &=& 0\,, \cr
\lim_{n \to 1} {d \over d n} C_p (n;0) &=& {2^{2 p} \Gamma^2(p+1) \over \Gamma(2 p + 2)}  \; .
\eea
Next, we have
\bea
\sum_{p=1}^\infty t^{2 p} \lim_{n \to 1} C_p \left(n;{\chi }\right) &=& 0\,, \cr
\sum_{p=1}^\infty t^{2 p} \lim_{n \to 1} {d \over d n} C_p\left(n; {\chi }\right) &=& - {t^2 \over t^2 + \sin^2 \chi} + {t^2 \arcsin \sqrt{t^2 + \sin^2 \chi} \over (t^2 + \sin^2 \chi)^{3 \over 2} \sqrt{1 - t^2 - \sin^2 \chi}}  \cr
&=&  {t^2 \over t^2 + \sin^2 \chi}  \left[ -1 + \frac{\text{arcsin}\sqrt{t^2 + \sin^2 \chi}}{\sqrt{t^2 + \sin^2 \chi} \sqrt{ 1- t^2 - \sin^2 \chi }} \right]\,.
\eea
It's probably possible to come up with an explicit formula for the coefficient functions here, but for now, we can give some specific results. For $p=1$, we obtain
\bea
 \lim_{n \to 1} C_1 \left(n;{\chi }\right) &=& 0\,, \cr
 \lim_{n \to 1} {d \over d n} C_1 \left(n;{\chi }\right) &=& { \chi \over \sin^3 \chi \cos \chi}  -{ 1  \over \sin^2 \chi} \; .
\eea
This allows us to give the following explicit result for $\Delta = 1/2$,
\be\label{eq:appell3}
S(\rho_\lambda || \rho_\beta ) \big{|}_{\lambda^2} = \left[ {d \over d n} \log \tr(\rho_\lambda^n)|_{n \to 1} \right]_{\lambda^2} = {\pi^2 \over 2 \beta^2} \left({2 \over 3} - {1  \over \sin^2 \left({a \pi \over \beta}\right)} +{{a \pi \over \beta} \over \sin^3 \left({a \pi \over \beta}\right) \cos \left({a \pi \over \beta}\right) }\right)\,.
\ee
This clearly diverges as $a$ approaches $\beta/2$. This is precisely what we found in \S\ref{sec:quantum2} using modular flow techniques, c.f., \eqref{eq:appell2}.\footnote{ The fact that the relative entropies computed as above have to match \eqref{eq:appell} on physical grounds, provides a generating function representation of the Appell series. This might be interesting to study from a purely mathematical point of view.} We further show in Appendix \ref{sec:replica2} that the divergence is not due to spatially coincident operator insertions: it persists if the operators are separated by a finite amount in space. (See also Fig.\ \ref{fig:replicaplot} below, where the divergence is plotted for $\Delta = 1$ and various different values of spatial separation.)

\subsection{Relative entropy in thermal states with finite spatial separation}
\label{sec:replica2}

\begin{figure}
\begin{center}
\includegraphics[width=.8\textwidth]{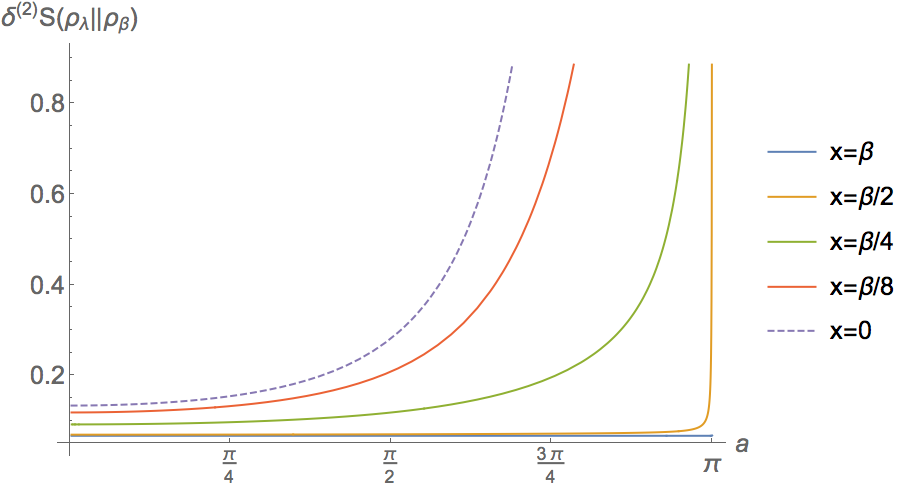}
\end{center}
\caption{Plot of $S(\rho_\lambda||\rho_\beta)$ at order $\lambda^2$ for dimension $\Delta=1$ operator insertions separated by a distance $a$ in Euclidean time, and a distance $x$ in the spatial direction. The divergence as $a\rightarrow \frac{\beta}{2} = \pi$ persists, though it does get milder as we increase $x$. For $x = \beta$ the divergence should be understood as localized to a point.}
\label{fig:replicaplot}
\end{figure}

We can generalize the results of \S\ref{sec:replica} to the case where the operators have some spatial separation $x$. In this case, the propagator between points on the cylinder separated by $a$ in the direction with period $\beta$ and by $x$ in the infinite direction is
\be
\Delta_\beta(a,x) = \left({2 \pi^2 \over \beta^2 \left( \cosh \left({2 \pi x \over \beta} \right) - \cos\left({2 \pi a \over \beta}\right) \right) }\right)^\Delta.
\ee
We need to calculate
\bea
\label{res1a}
\left[ {d \over d n} \tr(\rho_\lambda^n)|_{n \to 1} \right]_{\lambda^2} &=&  {1 \over 2} {d \over d n} \left[n \sum_{k=1}^{n-1} (\Delta_{n \beta}^2(k\beta,0) + \Delta_{n \beta}(k\beta + a,x) \Delta_{n \beta}(k\beta - a,x))\right]_{n \to 1} \cr
&=& {1 \over 2} {d \over d n} \left[n\, B^\Delta_n(\beta,0,0) + n\,B^\Delta_n(\beta,a,x) \right],
\eea
where
\be
\begin{split}
B^\Delta_n(\beta,a,x) &\equiv \sum_{k=1}^{n-1} \Delta_{n \beta}(k\beta + a,x) \Delta_{n \beta}(k\beta - a,x) \\
 &= \sum_{k=1}^{n-1} \left({\pi \over n \beta}\right)^{4 \Delta} {1 \over \left[(\sin^4(\pi k /n) + 2(\alpha^2 - \gamma^2 - 2 \alpha^2 \gamma^2) \sin^2(\pi k /n)  + (\alpha^2 +\gamma^2)^2 \right]^\Delta} \,,
 \end{split}
\ee
with
\be
\alpha = \sinh\left({\pi x \over n \beta}\right)\,, \qquad \qquad \gamma = \sin \left({\pi a \over n \beta}\right)\,.
\ee
We have applied some simplification using standard trigonometric identities to obtain the latter expression above. We can define a generating function for the integer operator dimension cases:
\bea
G_n(t) &\equiv&  \sum_{\Delta=1}^\infty t^{2 \Delta} \sum_{k=1}^{n-1} \Delta_{n \beta}(k\beta + a,x) \Delta_{n \beta}(k\beta - a,x) \cr
&=& \sum_{k=1}^{n-1} {T^2 \over \sin^4(\pi k /n) + 2(\alpha^2 - \gamma^2 - 2 \alpha^2 \gamma^2) \sin^2(\pi k /n)  + (\alpha^2 +\gamma^2)^2 - T^2} \cr
&=& {T^2 \over A_+-A_-} \sum_{k=1}^{n-1} {1 \over \sin^2(\pi k /n) - A_+} + {T^2 \over A_--A_+} \sum_{k=1}^{n-1} {1 \over \sin^2(\pi k /n) - A_-}\,,
\eea
where $T \equiv \frac{\pi^2}{(n\beta)^2}\, t$ and
\be
A_\pm = 2 \alpha^2 \gamma^2 + \gamma^2 - \alpha^2 \pm \sqrt{T^2 + 4 \alpha^2 \gamma^2 (\alpha^2 + 1)(\gamma^2 -1)} \; .
\ee
Then, using the formula \eqref{eq:sinsum}, we have
\be
G_n(t) =  - {T^2 \over A_+A_-} + {T^2 \over A_+ - A_-} \left[{n\cot(n \arcsin\sqrt{A_-}) \over \sqrt{A_-} \sqrt{1-A_-}} - {n\cot(n \arcsin\sqrt{A_+}) \over \sqrt{A_+} \sqrt{1-A_+}} \right].
\ee
The order $t^{2 \Delta}$ term in this expression gives the result for $B^\Delta_n(\beta,a,x)$. Using this, we can insert the result into (\ref{res1a}) to find the second order relative entropy. Carrying this out, we find that the expression still diverges as $a \to \beta/2$ for any value of $x$, though the divergence is milder as $x$ becomes larger. For instance, with $\Delta=1$ and various values of $x$, we find the behavior shown in Fig.\ \ref{fig:replicaplot}.

\section{Perturbative relative entropy at $k$-th order}
\label{app:kthorder}

In this appendix we provide a generalization of the discussion in \S\ref{sec:faulknermethod}. We will show the breakdown of perturbation theory at $k$-th order in the double-trace states.

The relative entropy is $S(\rho||\rho_0) = \text{Tr} (\rho \, \log \rho - \rho \, \log \rho_0 )$. At $k$-th order in $\delta \rho$ (for $k\geq 2$), this receives two contributions, both coming from the first term in the trace:
\begin{equation}\label{eq:Kint}
\begin{split}
& \delta^{(k)} S(\rho||\rho_0) \equiv \delta^{(k)} S(\rho||\rho_0)_I + \delta^{(k)} S(\rho||\rho_0)_{II} \\
 &\qquad \delta^{(k)} S(\rho||\rho_0)_I =  (-)^k  \int ds_1 \cdots ds_{k-1} \; {\cal K}_{k-1}(s_1,\ldots,s_{k-1}) \; \text{Tr} \left[ \rho_0^{\frac{1}{2}}\,\delta \rho_b\,\rho_0^{\frac{1}{2}} \prod_{r=1}^{k-1} \rho_0^{-\frac{i s_r}{2\pi}} \, \delta \rho_b \, \rho_0^{\frac{is_r}{2\pi}} \right] \\
 &\quad\;\;\delta^{(k)} S(\rho||\rho_0)_{II} = (-)^{k+1}\int ds_1 \cdots ds_{k} \; {\cal K}_{k}(s_1,\ldots,s_{k}) \; \text{Tr} \left[ \rho_0 \prod_{r=1}^{k} \rho_0^{-\frac{i s_r}{2\pi} } \, \delta \rho_b \, \rho_0^{\frac{is_r}{2\pi}} \right].
\end{split}
\end{equation}

First, let us simplify the second term, $\delta^{(k)} S(\rho||\rho_0)_{II}$. It can always be written as an integral over $k-1$ Schwinger parameters as in $\delta^{(k)} S(\rho||\rho_0)_{I}$. We start from the first line of \eqref{eq:Kint} and perform a change of variables according to
\begin{equation}
\begin{split}
  u &= s_1 +\ldots +  s_k \,,\qquad\quad v_r = s_r-s_{r-1} \quad (r=2,\ldots,k)\,.
\end{split}
\end{equation}
The trace in $\delta^{(k)} S(\rho||\rho_0)_{II}$ is independent of $u$, so we can simply perform the $u$-integral explicitly. We find
\begin{equation}
\begin{split}
\delta^{(k)} S(\rho||\rho_0)_{II}& =(-)^{k+1}  \int dv_2 \cdots dv_k \; \frac{(2\pi)^2}{(4\pi)^{k+1}} \frac{2\,i^{k-1} \,(v_2 + \ldots + v_k)}{\sinh \frac{v_2 + \ldots + v_k}{2} \; \prod_{r=2}^k \sinh \frac{v_r}{2} } \\
&\qquad\qquad\qquad\qquad\qquad\qquad \times  \text{Tr} \left[ \delta \rho_b\, \prod_{r=2}^{k} \left( \rho_0^{-\frac{i v_r}{2\pi}} \, \delta \rho_b \right)\rho_0^{\frac{i(v_2+\ldots+v_k)}{2\pi} +1} \right]\,.
\end{split}
\label{eq:Kint01}
\end{equation}
Now we change variables again in order to bring the kernel back to the form of ${\cal K}_{k-1}$:
\begin{equation}
 v_2 + \ldots + v_k = -\tilde{s}_1 
     \,,\qquad\quad v_r = \tilde{s}_r - \tilde{s}_{r-1} \quad (r=2,\ldots,k-1)\,.
\end{equation}
The integral becomes
\begin{equation}
\begin{split}
\delta^{(k)} S(\rho||\rho_0)_{II}
&=(-)^{k+1} \int d\tilde{s}_1 \cdots d\tilde{s}_{k-1} \; {\cal K}_{k-1}(\tilde{s}_1+ \pi i ,\ldots ,\tilde{s}_{k-1} + \pi i) \\
&\qquad\qquad\qquad\qquad \times
\frac{i \tilde{s}_1}{2\pi} \;\text{Tr} \left[\delta \rho_b \, \rho_0 \prod_{r=1}^{k-1}  \left( \rho_0^{-\frac{i \tilde{s}_r}{2\pi}} \, \delta \rho_b\, \rho_0^{\frac{i \tilde{s}_r}{2\pi}} \right) \right]\,.
\end{split}
\end{equation}
Shifting all contours as $\tilde{s}_r \rightarrow \tilde{s}_r - \pi i$, we recover the form of $\delta^{(k)} S(\rho||\rho_0)_{I}$:
\begin{equation}
\begin{split}
\delta^{(k)} S(\rho||\rho_0)_{II}
&=(-)^{k+1} \int d\tilde{s}_1 \cdots d\tilde{s}_{k-1} \; {\cal K}_{k-1}(\tilde{s}_1 ,\ldots ,\tilde{s}_{k-1} ) \\
&\qquad\qquad\qquad\qquad \times
\left(\frac{i \tilde{s}_1}{2\pi} + \frac{1}{2} \right) \;\text{Tr} \left[\rho_0^{\frac{1}{2}} \,\delta \rho_b \, \rho_0^{\frac{1}{2}} \prod_{r=1}^{k-1}  \left( \rho_0^{-\frac{i \tilde{s}_r}{2\pi}} \, \delta \rho_b\, \rho_0^{\frac{i \tilde{s}_r}{2\pi}} \right) \right]\,.
\end{split}
\end{equation}
From here, it follows immediately that
\begin{equation}
\begin{split}
\delta^{(k)} S(\rho||\rho_0)
&=(-)^{k} \int ds_1 \cdots ds_{k-1} \; {\cal K}_{k-1}(s_1 ,\ldots ,s_{k-1} ) \\
&\qquad\qquad\qquad\qquad \times
\left( \frac{1}{2} -\frac{i s_1}{2\pi}\right) \;\text{Tr} \left[\rho_0^{\frac{1}{2}} \,\delta \rho_b \, \rho_0^{\frac{1}{2}} \prod_{r=1}^{k-1}  \left( \rho_0^{-\frac{i s_r}{2\pi}} \, \delta \rho_b\, \rho_0^{\frac{i s_r}{2\pi}} \right) \right]\,.
\end{split}
\label{eq:SrelkApp}
\end{equation}
The manipulations leading to this expression were somewhat formal. In order for \eqref{eq:SrelkApp} to be sensible and well-defined, it needs to be complemented with a contour prescription for the modular $s$-integrals such that the Euclidean trace in the integrand is well-defined. In particular, if $\delta \rho_b$ consists of various operator insertions, we need to ensure that the Euclidean correlator is time-ordered. As we shall see momentarily, this is not always possible.

\subsubsection*{Divergences for double-trace states}

We use  \eqref{eq:SrelkApp} to write the relative entropy as
\begin{equation}
\begin{split}
&\delta^{(k)} S(\rho||\rho_0) \\
&\quad=(-)^{k}  \int dx_1dy_1 \cdots dx_kdy_k \, \source[2](x_1,y_1) \cdots \source[2](x_k,y_k) \, \int ds_1 \cdots ds_{k-1} \; {\cal K}_{k-1}(s_1 ,\ldots ,s_{k-1} ) \\
&\qquad\quad \times
\left( \frac{1}{2} -\frac{i s_1}{2\pi}\right) \;\text{Tr} \Big[ {\cal O}(x^0_1+is_1,\vec{x}_1) {\cal O}(y^0_1 + i s_1,\vec{y}_1) \cdots\\
&\qquad\qquad\qquad \cdots {\cal O}(x^0_{k-1}+is_{k-1},\vec{x}_{k-1}) {\cal O}(y^0_{k-1} + i s_{k-1},\vec{y}_{k-1})  \; {\cal O}(x_k^0-\pi,\vec{x}_k){\cal O}(y_k^0-\pi,\vec{y}_k)  \,\rho_0\Big] \,.
\end{split}
\label{eq:SrelkOOApp}
\end{equation}

\begin{figure}
\begin{center}
\includegraphics[width=.35\textwidth]{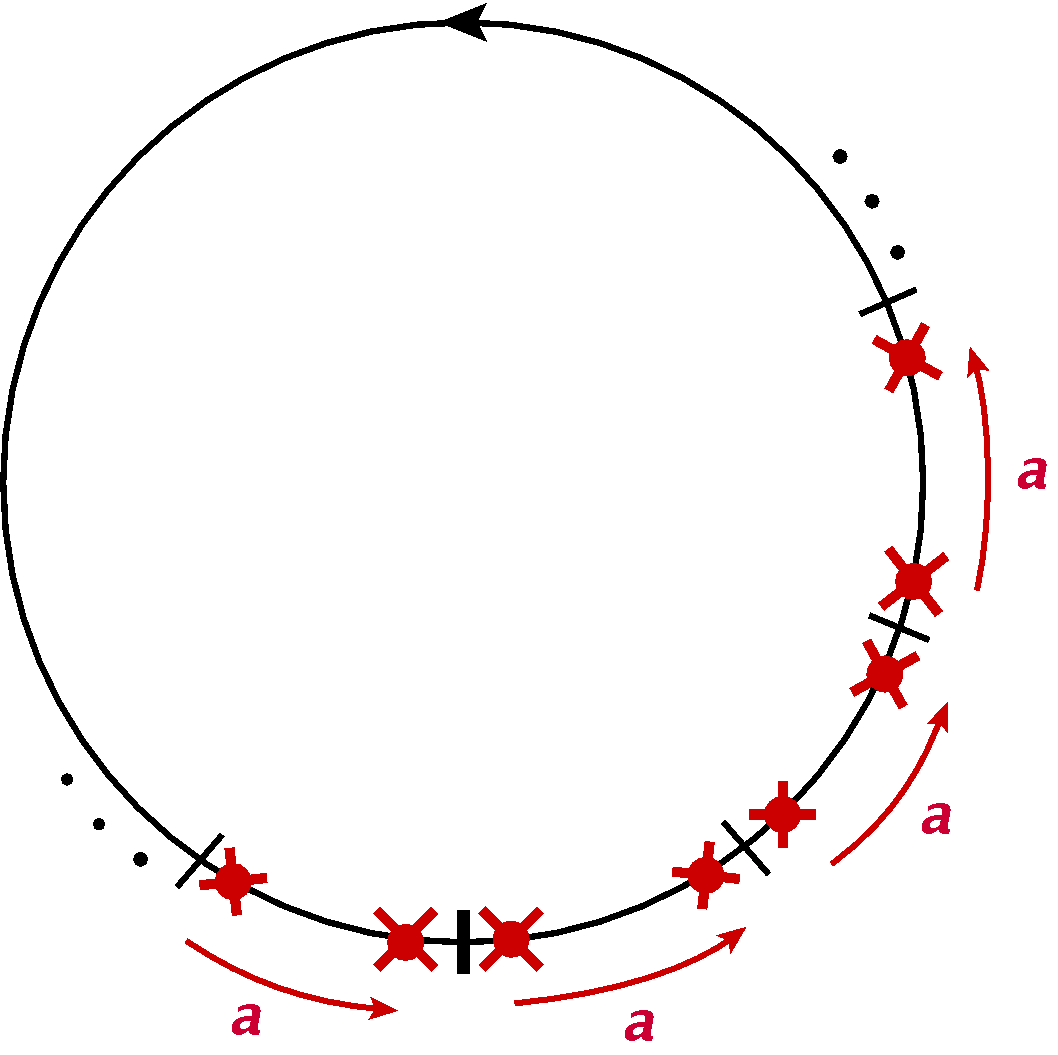}
\put(-133,37){$\frac{2\pi(k-1)}{k}$}
\put(-90,25){$0$}
\put(-60,38){$\frac{2\pi}{k}$}
\put(-40,65){$\frac{4\pi}{k}$}
\put(-45,100){$\frac{6\pi}{k}$}
\end{center}
\caption{Configuration of Euclidean insertion times in the correlator of \eqref{eq:SrelkOO2}. Operators (red crosses) connected by a double-trace source are separated by $a$ around the Euclidean circle. If $a > \frac{\beta}{k} = \frac{2\pi}{k}$, then the order of insertions in the correlator does not coincide with the time ordering along the Euclidean circle anymore. This leads to a breakdown of perturbation theory in $\source[2]$.}
\label{fig:kcontours}
\end{figure}
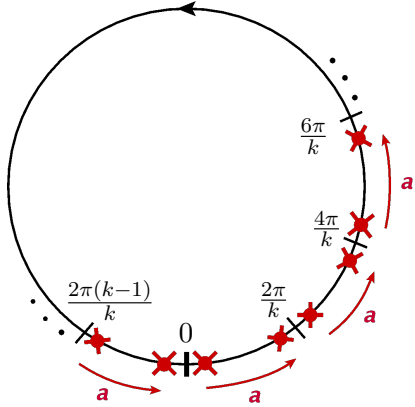

At leading order in $1/N$ the correlation function in \eqref{eq:SrelkOOApp} factorizes into $(2k-1)!!$ products of $k$ two-point functions. For the local sources \eqref{eq:sourcesEx}
we can time-order the correlator in \eqref{eq:SrelkOOApp} by defining the modular flow integrals to run along the contours lying between $-\pi i$ and $\pi i$ with maximal separation from each other, i.e., separation $\frac{2\pi}{k}$. This is illustrated in Fig.\ \ref{fig:kcontours}. It can be implemented by a shift of integration variables
\be
s_r \; \longrightarrow \; s_r - i\pi + \frac{2\pi i r}{k}
\ee
and the integration of all $s_r$ along the real line:
\begin{equation}
\begin{split}
&\delta^{(k)} S(\rho||\rho_0) =(-)^{k} \, \int_{-\infty}^\infty ds_1 \cdots ds_{k-1} \; {\cal K}_{k-1}\left(s_1- i\pi + \frac{2\pi i}{k} ,\ldots ,s_{k-1}- i\pi + \frac{2\pi i(k-1)}{k} \right) \\
&\qquad \times
\left( \frac{1}{k}-\frac{i s_1}{2\pi}\right) \;\text{Tr} \bigg[ {\cal O}\left(\tau_0+a+\frac{(k-1)2\pi}{k}+is_1,0\right){\cal O}\left( \tau_0+\frac{(k-1)2\pi}{k}+is_1,0\right) \cdots\\
&\qquad\qquad\quad \cdots {\cal O}\left(\tau_0+a+\frac{2\pi}{k}+is_{k-1},0\right){\cal O}\left( \tau_0+\frac{2\pi}{k}+is_{k-1},0\right)\;
{\cal O}\left(\tau_0+a,0\right){\cal O}\left( \tau_0,0\right) \,\rho_0\bigg],
\end{split}
\label{eq:SrelkOO2}
\end{equation}
where $\tau_0$ represents some overall shift which we are free to make use of due to translation invariance. This correlation function is time-ordered (hence finite) for small enough $a$. However, as $a$ approaches the critical value $\frac{\beta}{k} = \frac{2 \pi}{k}$, the Euclidean separations are no longer in time order. For similar observations and further technical insights, see also \cite{Sarosi:2017rsq,Lashkari:2018tjh}.

\bibliographystyle{JHEP}

\providecommand{\href}[2]{#2}\begingroup\raggedright\endgroup

\end{document}